\documentclass[a4paper,fleqn,usenatbib]{mnras}

\usepackage{newtxtext,newtxmath}

\usepackage[T1]{fontenc}
\usepackage{ae,aecompl}

\usepackage[]{graphicx}
\usepackage{psfig}
\usepackage{rotating}
\usepackage{epsfig}
\usepackage{fixltx2e}
\usepackage{appendix}
\usepackage{psfrag}
\usepackage{amsmath,amssymb}
\usepackage{threeparttable}
\usepackage{float}
\usepackage{longtable}
\usepackage{lscape}
\usepackage{afterpage}
\usepackage{url}
\usepackage{color}
\usepackage{verbatim}
\usepackage{enumitem}

\def \oiii {[O{\small~III}]}

\def \ha {H$\alpha$}

\def \Msol {$M_{\odot}$}

\def\apjl{ApJL}

\def\aap{A\&A}
\def\apjs{ApJS}

\title[SAMI: Origin of Gas in Galaxies]{The SAMI Galaxy Survey: Stellar and gas misalignments and the origin of gas in nearby galaxies}

\author[J. J. Bryant et al.]{J. J. Bryant$^{1,2,3}$\thanks{E-mail: jbryant@sydney.edu.au (JJB)}, S. M. Croom$^{1,3}$, J. van de Sande$^{1,3}$, N. Scott$^{1,3}$, L. M. R. Fogarty$^{1}$,      
 \newauthor
J. Bland-Hawthorn$^{1,2,3}$, J. V. Bloom$^{1}$, E. N. Taylor$^{7}$,  S. Brough$^{3,6}$,  A. Robotham$^{3,5}$,   
 \newauthor
L. Cortese$^{3,5}$, W. Couch$^{4}$,  M. S. Owers$^{8}$, A.M. Medling$^{9,10}$, C. Federrath$^{9}$,
\newauthor
K. Bekki$^{5}$, S. N. Richards $^{1,11}$, J.S. Lawrence$^{4}$, I. S. Konstantopoulos$^{12}$
 \\
$^{1}$ Sydney Institute for Astronomy (SIfA), School of Physics, The University of Sydney, NSW 2006, Australia \\
$^{2}$ Australian Astronomical Optics, AAO-USydney, School of Physics, University of Sydney, NSW 2006, Australia;\\
$^{3}$ ARC Centre of Excellence for All Sky Astrophysics in 3 Dimensions (ASTRO 3D); \\
$^{4}$ Australian Astronomical Optics, AAO-Macquarie, Faculty of Science and Engineering, Macquarie University, 105 Delhi Rd, North Ryde, NSW 2113, Australia;\\
$^{5}$ ICRAR, The University of Western Australia, Crawley WA 6009, Australia; \\
$^{6}$ School of Physics, University of New South Wales, NSW 2052, Australia\\
$^{7}$ Centre for Astrophysics \& Supercomputing, Swinburne University of Technology, Mail H29, PO Box 218, Hawthorn, VIC 3122, Australia\\
$^{8}$ Department of Physics and Astronomy, Macquarie University, NSW 2109, Australia\\
$^{9}$ Research School of Astronomy and Astrophysics, The Australian National University, Canberra, ACT 2611, Australia\\
$^{10}$ Cahill Center for Astronomy and Astrophysics California Institute of Technology, MS 249-17 Pasadena, CA 91125, USA\\
$^{11}$ SOFIA Operations Center, USRA, NASA Armstrong Flight Research Center, 2825 East Avenue P, Palmdale, CA 93550, USA\\
$^{12}$ Atlassian 341 George St Sydney, NSW 2000\\
}

\date{Accepted XXX. Received YYY; in original form ZZZ}

\pubyear{2017}

\begin{document}
\label{firstpage}
\pagerange{\pageref{firstpage}--\pageref{lastpage}}
\maketitle

\begin{abstract}

Misalignment of gas and stellar rotation in galaxies can give clues to the origin and processing of accreted gas.
Integral field spectroscopic observations of 1213 galaxies from the SAMI Galaxy Survey show that 11\% of galaxies with fitted gas and stellar rotation are misaligned by more than $30^{\circ}$ in both field/group and cluster environments. Using SAMI morphological classifications and S\'{e}rsic indices, the misalignment fraction is $45\pm6$\% in early-type galaxies, but only $5\pm1$\% in late-type galaxies. The distribution of position angle offsets is used to test the physical drivers of this difference. Slower dynamical settling time of the gas in elliptical stellar mass distributions accounts for a small increase in misalignment in early-type galaxies. However, gravitational dynamical settling time is insufficient to fully explain the observed differences between early- and late-type galaxies in the distributions of the gas/stellar position angle offsets. LTGs have primarily accreted gas close to aligned rather than settled from misaligned based on analysis of the skewed distribution of PA offsets compared to a dynamical settling model. Local environment density is less important in setting the misalignment fractions than morphology, suggesting that mergers are not the main source of accreted gas in these disks. 
Cluster environments are found to have gas misalignment driven primarily by cluster processes not by gas accretion.

\end{abstract}


\section{Introduction}
\label{intro}
The morphology-density relation shows how galaxies in different density environments have a different distribution of morphologies \citep{Dre80}. Both the environment and morphology affect how galaxies build up mass through accretion of gas. However, how gas collapses onto the different shaped galaxy gravitational potentials and then forms stars remains poorly understood. It may both be influenced by, and impact, the morphology of the galaxy and be controlled by the local (nearest neighbour), global (field, group or cluster) or large-scale structure environment. Galaxies with the bulk of their baryonic mass in a thin disk are rotationally supported (late-type galaxies, LTGs), while galaxies with a spherical stellar mass distribution were previously thought to be dispersion supported (early-type - elliptical and lenticular -  galaxies, ETGs). However, it is now known that not only do LTGs host gas disks, but $\sim40$\% of ETGs have molecular or atomic gas disks and/or ionised gas disks traced by emission lines \citep{Sar06,Kra08,Kra11, Ems11,Cap11,Dav11,Ser12,You14}. 
The dynamics of gas in these disks compared to stars can be used to trace the origins of the gas in both ETG and LTGs.

Accreted gas will be converted into stars within a depletion time, defined as the mass of the accreted gas divided by the rate at which stars are being formed. Typical depletion times are of order a few gigayears, but yet most LTG, and some ETGs are observed to be forming stars in the current universe and have been for a Hubble time \citep{Lar80,Ken94}. Accretion of new external gas 
is therefore necessary regularly over the life of a galaxy, at least every few Gyr, to provide fuel to sustain the star formation \citep{Bau10,Tac10,Put12, San14}.  For LTGs, estimated rates of gas accretion (based on high velocity clouds and merger rates) are $\sim 0.2$ solar masses per year \citep[e.g.][]{San08}, 
which is insufficient to sustain the typical rate of forming stars ($\sim 1$ solar mass per year). These statistics are highly dependent on galaxy types, different samples and measurement methods, but the discrepancy highlights the difficulty of measuring gas accretion because gas accreted through mergers has observational tracers, while other gas origins are more difficult to detect. 

Galaxies can increase their gas supply through either internal or external processes. Internal processes include stellar mass loss in which the resulting gas should have the same dynamics as the stars.  External stochastic accretion of gas in the $\Lambda$ cold dark
matter ($\Lambda$CDM) universe can come from accretion of cold gas from filaments \citep[e.g.][]{Ker05,Chu12,Ser14}, hot gas from the outer halo that then cools \citep{Lag14,Lag15}, or alternatively clumpy accretion from gas-rich mergers or interactions between gas-rich galaxies. Recycled gas may also appear to be accreted externally. Massive stars can eject hot gas from the galaxy. If that gas then cools it can be recycled back on to the galaxy in a galactic fountain and potentially fuel future star formation \citep{Sha76,Hou90}. 
Evidence for external accretion lies in the cold gas and the ionised gas having angular momentum axes that are decoupled from those of the stars. 

Not all externally accreted gas must accrete with an angular momentum axis misaligned to the stars, and equally, not all misaligned gas is recently accreted because gas can form stable misaligned orbits.
Galaxy mergers or interactions can lead to disruption of the velocity fields
that is 
transient until the dynamical interaction is complete \citep{Blo17a,Blo17b}. On the other hand, smooth accretion from filaments or cooling from the hot halo can be more continuous. Both accretion mechanisms lead to gas rotation that is aligned or misaligned from the stellar rotation.
If this accreted gas is in place before most of the stars form, then the stellar and gas velocities will align. However, if accretion occurs well after the bulk of the stars have formed, then accreted gas will settle after some dynamical time into an aligned or counter-rotating orbit compared to the stars, both of which are stable and long-lived. 
Counter-rotating gas must have been introduced from processes not connected with the galaxy formation because counter-rotating gas would have been unstable in the formation process. 
There is a clear difference between measuring misalignment that is due to a current interaction, and gas and stars in stable misaligned orbits. The latter may point to the origin of how the gas got into the galaxy, but the former can be used to test the timescales of mergers. 

The larger structure and halo that a galaxy resides in can have an impact on the gas accretion. Simulations have shown that misalignment of either two spatially-separate stellar disks (e.g. decoupled cores) or misalignment of stellar and gas disks can be accounted for without mergers but instead by accretion of gas from different filaments. Using high-resolution cosmological hydrodynamical simulations,  \citet{Bro09} show 70\% of the gas accreted from outside the virial radius has come from that galaxy's outer halo as either shocked or un-shocked gas, while less than $30$\% has come from the halo of a different galaxy (a merger or interaction). They find that cold accretion of un-shocked gas occurs in disk galaxies less than L* up to current day because filaments penetrate the hot halo, feeding cold gas well inside the virial radius to the central galaxy and fuelling star formation.  While feeding of cold gas through filaments is important at high redshift, it continues to play a role up to the local universe \citep{Ker05,Ker09,Mas07,Ocv08,Sal12}.

Depending on the local environment and the shape of the galaxy's gravitational potential and stellar mass distribution, the timescales over which gas is accreted and then settles into stable orbits within a galaxy can be vastly different. In gas-rich accretion events, if gas coming into a galaxy from another interacting galaxy has sufficiently high momentum then it will not initially be pulled down onto the disk. At some point, the source of high-angular-momentum external gas is exhausted and then gravitational torques will eventually dominate as the gas loses momentum and is pulled down to align with the rotation of the stellar disk. \citet{vdV15} simulated such an interaction in a $10^{11}$ M$_{\odot}$ galaxy merger and found that once the incoming gas supply is cut off, the system goes through a warped stage, before ending up aligned. The timescale for dynamical settling of the gas into the disk in the central regions of the galaxy ($\sim$1\,kpc) after introduction of new gas, was $\sim 5-10$  dynamical times, but could reach $80-100$ dynamical times in ETGs when there is continual gas accretion. The longer the accretion timescale, the larger the fraction of galaxies are expected to be misaligned at any point in time.  

\citet{Dav16} compared the distribution of stellar and gas rotation position angle (PA) misalignment in ETGs from the ATLAS$^{3D}$ Survey, to distributions generated based on different models of dynamical times for mergers. 
The fraction of counter-rotating objects (misalignment angle $\sim180^{\circ}$) is the biggest discriminant between models. The fraction aligned cannot be used to estimate external gas accretion because it will include not only systems where the gas has external origins but will be contaminated with galaxies where aligned gas has come from internal processes. 
In order to match the distribution of PA misalignments seen in the ATLAS$^{3D}$ ETGs they required a relaxation time of ~80 dynamical times, substantially larger than standard values of $\sim 5-10$ dynamical times but in keeping with the timescales found in \citet{vdV15}. Their sample consisted of 260 ETGs of which $\sim200$ had fitted dynamical PAs for the stars and gas and only $\sim20$ of those were in cluster environments (rather than group/field galaxies). In order to test these simulations, the distribution of the PA misalignment angles is required for extended samples that encompass a range of galaxy morphologies as well as a range of environments.

In LTGs, on the other hand, which are typically more gas-rich than ETGs, accreted gas that is misaligned will precess to become aligned or counter-aligned with the existing disk within a few dynamical times \citep{Toh82,vdV15}. Semi-analytic models from \citet[{\sc dark sage};][]{Ste16} trace evolution of the angular momentum of gas disks which are initially offset from the stellar rotation in galaxies. They have shown that the distribution of PA offsets between the gas and stellar angular momentum vectors is remarkably different when the model allows gas precession. With gas precession, by redshift zero, the bulk of disk-dominated galaxies have aligned PAs and 7.25\% have counter-rotating gas compared to stars. Again the fraction of galaxies with counter-rotating stars and gas is the crucial discriminator because without gas precession in the models, these aligned and misaligned peaks in the galaxy distribution disappear.  They point out that there is to-date no published data capable of testing this model.  

The test of these ETG and LTG simulations requires stellar and gas kinematic data from galaxies across all morphological types. Very few papers have compared resolved gas and stellar dynamics because it requires spatially-resolved integral field spectroscopy (IFS) for many galaxies. Some  have measured stellar and ionised gas dynamics in small samples such as $\leq 30$ LTGs using the \oiii\, line \citep{Gan06, Martinsson13} but not addressed misalignment rates or gas accretion models. Others have measured \ha\, kinematics in spiral galaxies in different environments without the stellar dynamics \citep{Fat09, Epi08}. On the other hand several individual systems with signatures of mergers have been investigated in detail \citep[e.g.][]{Eng10, wil14} but statistically larger samples are needed to test theoretical predictions of gas accretion. The largest two surveys have focussed on just ETGs \citep{Dav11} or LTGs \citep{Bar14,Bar15} alone. 

\citet{Dav11} used 260 ETGs from the ATLAS$^{3D}$ survey to find that 36\% of ETGs have gas misaligned by more than $30^{\circ}$ to the stars and hence concluded that gas is externally acquired. Furthermore, they found a significant difference between the (mis)alignment of gas in galaxies in the Virgo cluster and the field with essentially no misalignment detected in the cluster ETGs.
The largest study to-date of stellar/gas misalignment in LTGs focussed on the impact of interactions. Firstly \citet{Bar14} used the CALIFA survey \citep{San12,Wal14} to look at misalignments between ionised gas and stellar rotations in spiral galaxies that were not interacting. They found that both the stellar and gas components have a strong tendency to follow the gravitational potential of the disc even if the spiral is strongly barred, and 90\% of their galaxies had stellar and ionised gas rotations aligned within $16^{\circ}$.
However, \citet{Bar15} investigated the misalignments in a sample of 66 interacting galaxies from the CALIFA Survey and found that 18\% had stellar/gas rotation misalignments above $30^{\circ}$, significantly higher than the non-interacting sample. We note that while the gas/stellar misalignment was not compared to morphology in those papers \citep{Bar14,Bar15}, the noninteracting sample was strongly biased towards LTGs but the interacting sample has a distribution from ETGs to LTGs. Their conclusion that misalignments are substantially higher in interacting galaxies may therefore be influenced by the morphological classifications of their interacting and non-interacting groups. Instead they investigate the different stages of mergers and show that misalignment is higher during the merger and the remnant stage and that the gas dynamics are more affected by the merger than the stellar dynamics. 

Previous work has hinted at a morphological trend with misalignment. For example several works have found that S0 galaxies are more likely to have gas counter-rotating compared to stars, while LTGs have lower counter-rotating fractions \citep{Bur06,Kan01,Piz04}. The largest survey to-date that included a range of morphological types is \citet{Jin16} using the Mapping Nearby Galaxies at Apache Point Observatory \citep[MaNGA;][]{Bun15} survey. Their sample has 66 galaxies with misaligned gas and stellar rotation and they found a lower fraction of misalignment in galaxies with fewer neighbours. Having only 4 galaxies in cluster environments they note the dependence on environment was not significant and a larger sample spanning a broader range in environment is necessary. Their results have a trend with stellar mass such that the misalignment fraction peaks at $\sim10^{10.5}\,M_{\odot}$ and the fraction of misaligned galaxies increases with lower SFRs, but the analysis unfortunately does not constrain the counter-rotating fractions required to test the theoretical predictions discussed above.
So far there has not been an IFS survey of galaxies covering a sufficient number of galaxies across a broad range in morphological types, stellar masses and environments with both stellar and ionised gas dynamics, to disentangle the impact of morphology and environment on the accretion processes and timescale for gas in all galaxies.
 
The Sydney-AAO Multi-object Integral field spectrograph (SAMI) Galaxy Survey \citep{JB2015,Cro12} provides IFS data that is ideal to investigate the misalignment of gas and stellar kinematics because it can extend the \citet{Dav11,Dav16} and \citet{Bar14,Bar15} work to include a broader morphological and environment parameter space. The SAMI Galaxy Survey v0.9 includes $\sim1213$ galaxies with $z<0.1$ and has the advantage for this science of IFS data covering a broad range in stellar mass, environment (clusters, groups, field galaxies), and morphologies, with substantial auxiliary information including a range of environment metrics.

Using the SAMI Galaxy Survey data, this paper aims to consider the galaxies that {\it have both gas and stellar rotation} in order to use PA misalignment as a test of the accretion and dynamical timescales of that gas.  Therefore galaxies without gas or stellar rotation are not considered.
The outline of this paper is as follows: Section~\ref{SurveyandData} presents the data from the SAMI Galaxy Survey and the measurements of gas and stellar rotation and misalignment; Section~\ref{Results} compares misalignment measures with regard to galaxy properties; Section~\ref{Discussion} discusses observed misalignments in terms of the shape of the galaxy's baryonic mass distribution (morphology) and timescales for accretion and gas settling, as well as environment and the implications for gas accretion in galaxies. A summary of our results is in Section~\ref{Conclusion}.

Throughout this paper, we adopt the concordance cosmology:
($\Omega_{\Lambda}$,$\Omega_{\rm m}$, h) $=$ (0.7, 0.3, 0.7) \citep{Hin09}.
Colour versions of all figures appear in the online version.

\section{SAMI Galaxy Survey} 
\label{SurveyandData}
\subsection {Survey structure and observations}

The SAMI Galaxy Survey \citep{JB2015} is currently underway with the SAMI instrument \citep{Cro12} on the 3.9\,m Anglo-Australian Telescope (AAT). The SAMI instrument uses imaging fibre bundles called `hexabundles' \citep{JB2014, JB2012, JB2011, JBH2011}. Each hexabundle has a very high filling-factor of 75\%. Twelve hexabundles are positioned across a $1^{\circ}$ diameter field to each simultaneously observe a different galaxy.  61-fibre cores are each 1.6\,arcsec in diameter, giving a 15\,arcsec diameter field-of-view. The median effective radius (R$_{e}$) of galaxies in the sample is 4.4\,arcsec (see Fig.~\ref{Re_ellip}). Each galaxy is observed in either 6 or 7 dithered pointings of 1800s each. SAMI feeds into the AAOmega spectrograph \citep{Sau04, Smi04, Sha06} which gives a median resolution of FWHM$_{blue} = 2.65$\AA\, from 3700-5700\AA\, and FWHM$_{red} = 1.61$\AA\, from 6300-7400\AA\, \citep[see][]{vdS17}.

\begin{figure*}
\includegraphics[width=9.0cm]{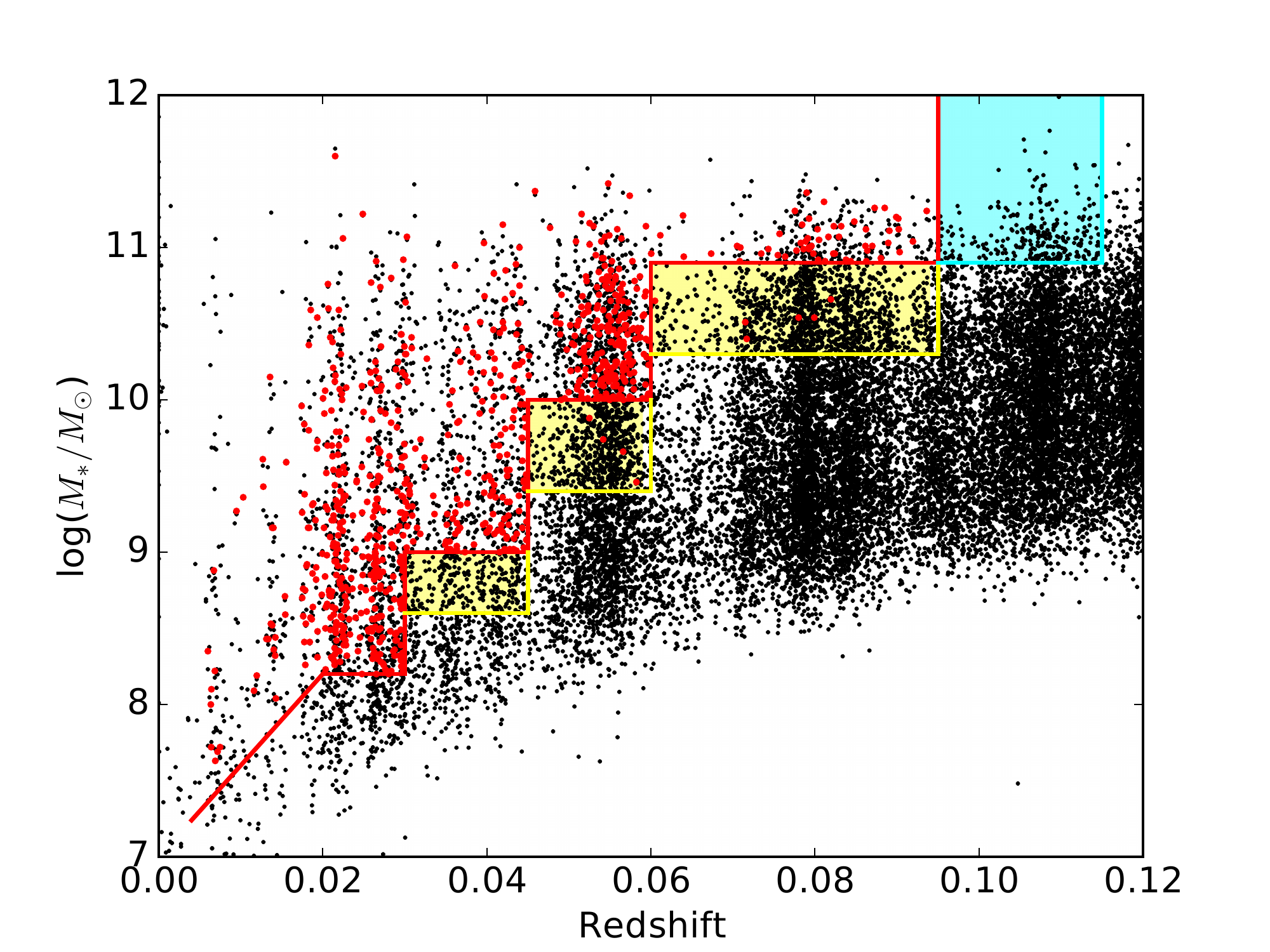}
\includegraphics[width=10.0cm]{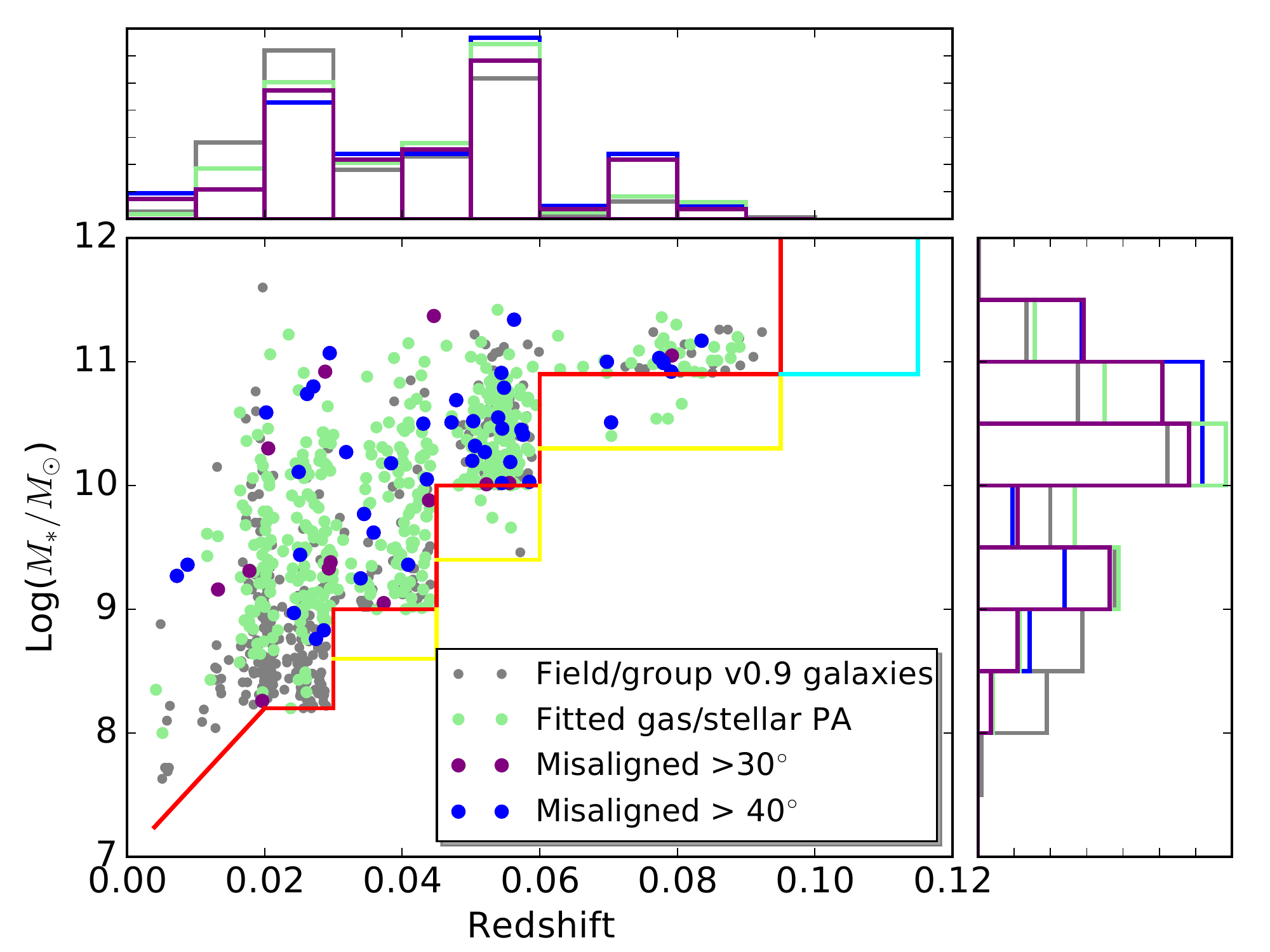}
\includegraphics[width=10.0cm]{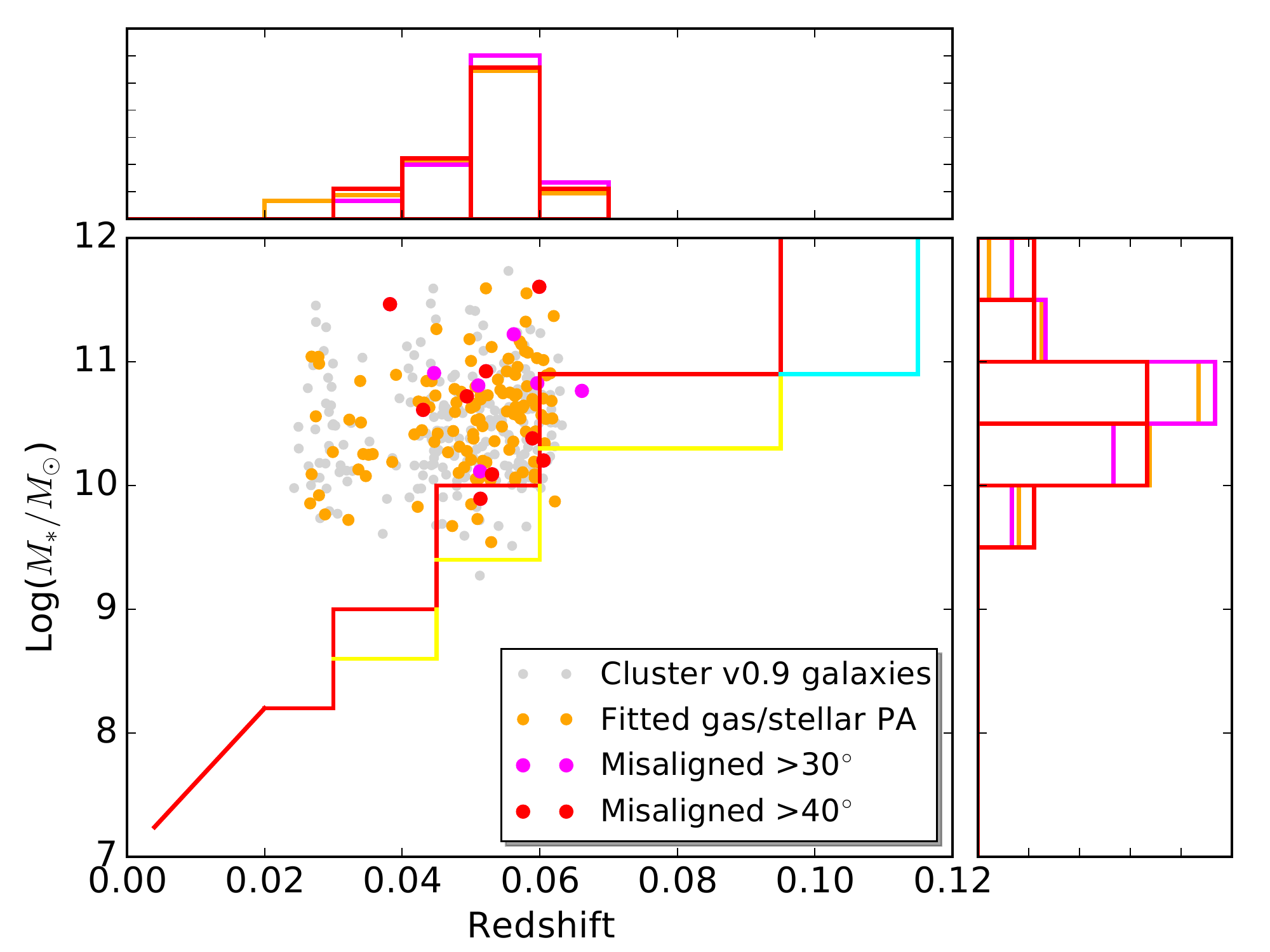}
\vspace*{3mm}
\caption{Top: Redshift versus stellar mass for galaxies in the GAMA parent sample (black), and the galaxies selected from this sample, lying above the red line are the primary targets for the SAMI Survey. Galaxies in the yellow and cyan regions are the SAMI Survey filler targets \citep[for details see][]{JB2015}. Galaxies in internal data release v0.9 used in this work (red) are a subset of the SAMI Survey selection. Middle (field/group) and lower (clusters): Redshift versus stellar mass for the v0.9 galaxies (grey), the galaxies for which both stellar and gas PAs  could be fitted (field/group sample: light green; clusters: orange), misaligned (PA$>30^{\circ}$) galaxies (field/group sample: purple; clusters: magenta), misaligned (PA$>40^{\circ}$) galaxies (field/group sample: blue; clusters: red). This point colour scheme is used consistently throughout all plots in this paper.
}
\label{zvsMstar}

\end{figure*}

The survey aims to observe a total of $\sim3000$ $z<0.1$ galaxies including $\sim800$ cluster galaxies in 8 clusters and $\sim2200$ galaxies from the GAMA survey \citep{Dri11} G09, G12 and G15 regions, including field and group galaxies.  Therefore, a full range of global environments are represented from the field to clusters.  Throughout this paper we refer to the GAMA-region galaxies as the field/group sample. Full details of the survey target selection for the SAMI survey are given in \citet{JB2015}, and the cluster galaxy selection is described further in \citet{Owe17}. The selection is based on redshift and stellar mass cut-offs forming four volume-limited samples. This analysis uses galaxies from the internal data release v0.9, and Fig.~\ref{zvsMstar} shows the distribution of those 1213 objects compared to the full survey selection. These galaxies were observed from mid-2013 to mid-2015 and include 833 GAMA-region (field/group sample) galaxies and 380 cluster galaxies from all 8 clusters.

\subsubsection{Ancillary data}

Supporting data from the GAMA survey are available for the field and group galaxies. These data include two environmental measures, the GAMA galaxy group catalogue (G$^3$C) \citep{robotham11} and the 5th nearest neighbour local surface density \citep{Bro13}. The G$^3$C is an adaptive friends-of-friends group catalogue which provides group mass and membership including central group galaxy. We note that the field galaxies are those in which there were no detected group members in the G$^3$C but this may mean they are the central galaxy of a low mass halo where the satellites were undetected. The 5th nearest neighbour local surface density values were recalculated for SAMI following the same method as used in \citet{Bro13}, but with a brighter absolute magnitude limit of M$_{r}=-18.5$ mag. Several quantities measured by the GAMA survey have been incorporated in the SAMI catalogue including stellar mass \citep{Tay11}, aperture-matched colours \citep{Hil11, Lis15}, semi-major axis effective radius from $r$-band S\'{e}rsic fits (R$_{e}$) and ellipticity from $r$-band S\'{e}rsic fits \citep{Kel12}.

5th nearest-neighbour surface densities $\Sigma_{5}$ have also been calculated for galaxies in the cluster regions using the same method as for the GAMA regions \citep[see][for details]{Bro13,Bro17}. All surrounding galaxies brighter than M$_{r}=-18.5$ mag within $\pm1000$\,km\,s$^{-1}$ contribute to the surface density estimate, which is then divided by the survey completeness in that region. 109 field/group-region and 4 cluster-region galaxies for which the surface density measurement is unreliable have been excluded from the $\Sigma_{5}$ analysis. The unreliability is because the 5th nearest neighbour is outside the boundary of the survey region.

\subsection{Data reduction}
The SAMI galaxy observations were reduced by the SAMI survey reduction pipeline which is detailed in \citet{Sha15}, \citet{All15} and \citet{Gre17}. Briefly, this includes data reduction using the 2{\sc dfdr} pipeline \citep{Cro04,Sha10} to produce wavelength-calibrated, row-stacked spectra that have night sky emission subtracted. Correction for atmospheric dispersion and removal of telluric absorption features is done using the secondary standard stars that are observed in one hexabundle in each field and a primary spectrophotometric standard which is observed each night. The reduced row-stacked spectra for each of the 7-point dither observations are then combined into the final flux-calibrated IFS cubes.  The cubes have $50\times50$ spaxels (spatial pixels) each of which are $0.5'' \times 0.5''$.  

\subsection{Data analysis}
\subsubsection{Emission line position angles}
The reduced data cubes were processed through the emission-line fitting software called {\sc LZIFU} \citep{Ho16b}, which subtracts the stellar continuum using the penalised pixel-fitting routing {\sc PPXF} \citep{Cap04}, then fits 1, 2 or 3-component Gaussians to each line using a non-linear least-squares fit. For this work the 1-component fits only are used. The resultant 2D H$\alpha$ velocity maps had the axis of rotation fit using the {\sc fit\_kinematic\_pa} code developed by \citet{Cap07} and \citet{Kra11} based on the methods presented in \citet{Kra06}. Only spaxels with H$\alpha$ S/N$>5$ were included in the fits. The number of galaxies with sufficient S/N for the PA to be fitted was 665 out of 833 and 143 out of 380 in the field/group regions and cluster regions respectively.

As the aim here is to fit the global gas PA, these fits are immune to higher-order perturbations such as galactic winds which affect only a small fraction of the gas. We note that of the 15 wind-dominated galaxies from the SAMI sample that are presented in \citet{Ho16a}, none are misaligned by our measure. 

\subsubsection{Stellar rotation position angles}
Using a similar method to that described in \citet{Fog2014}, the PA of the stellar rotation was measured from 2D stellar kinematic maps using {\sc fit\_kinematic\_pa}, which is base on the method described in Appendix C of \citet{Kra06}. A detailed description of the SAMI survey stellar kinematics fitting is presented in \citet{vdS17}. In summary, the blue and red spectra are re-binned and combined onto a common velocity scale. The penalised pixel fitting code \citep[{\sc pPXF};][]{Cap04} is used to extract the line-of-sight velocity distribution, and is run in a multi-step process that removes emission line regions, accounts for noise in the spectrum, identifies the best templates from binned spectra then fits individual spaxels in each image. The MILES library stellar template spectra \citep{San06} were used to derive the optimal template. Line-of-sight velocities were measured for all spaxels and those with maximum velocity uncertainty $>30$\,km\,s$^{-1}$ were excluded. This gave 586 out of 833 field/group and 354 out of 380 cluster galaxies with fitted PAs.

\subsubsection{Measurement of position angle offsets}
\label{PAmeasures}

486 out of 833 of the field/group galaxies and 136 out of 380 galaxies in the 8 clusters have PAs measurable for both gas and stars and therefore could have a PA offset measured.  
The PAs of the stellar and gas rotation axes were measured to be the counter-clockwise angle from north to the line perpendicular to the axis of rotation on the side where the rotation is receding. The PA vector along the velocity field is perpendicular to the line-of-sight angular momentum vector. 

\begin{figure*}
\begin{minipage}[t]{0.02\textwidth}
\centerline{(a)}
\end{minipage}%
\begin{minipage}[]{0.18\textwidth}
\includegraphics[width=2.5cm]{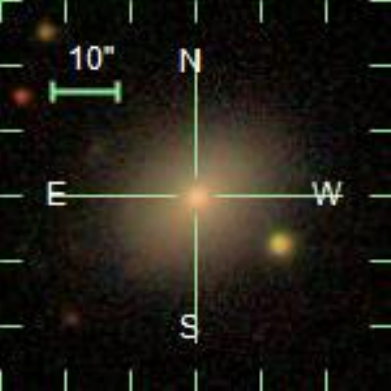}
\end{minipage}%
\hspace*{-3mm}
\begin{minipage}[]{0.2\textwidth}
\includegraphics[width=3.6cm]{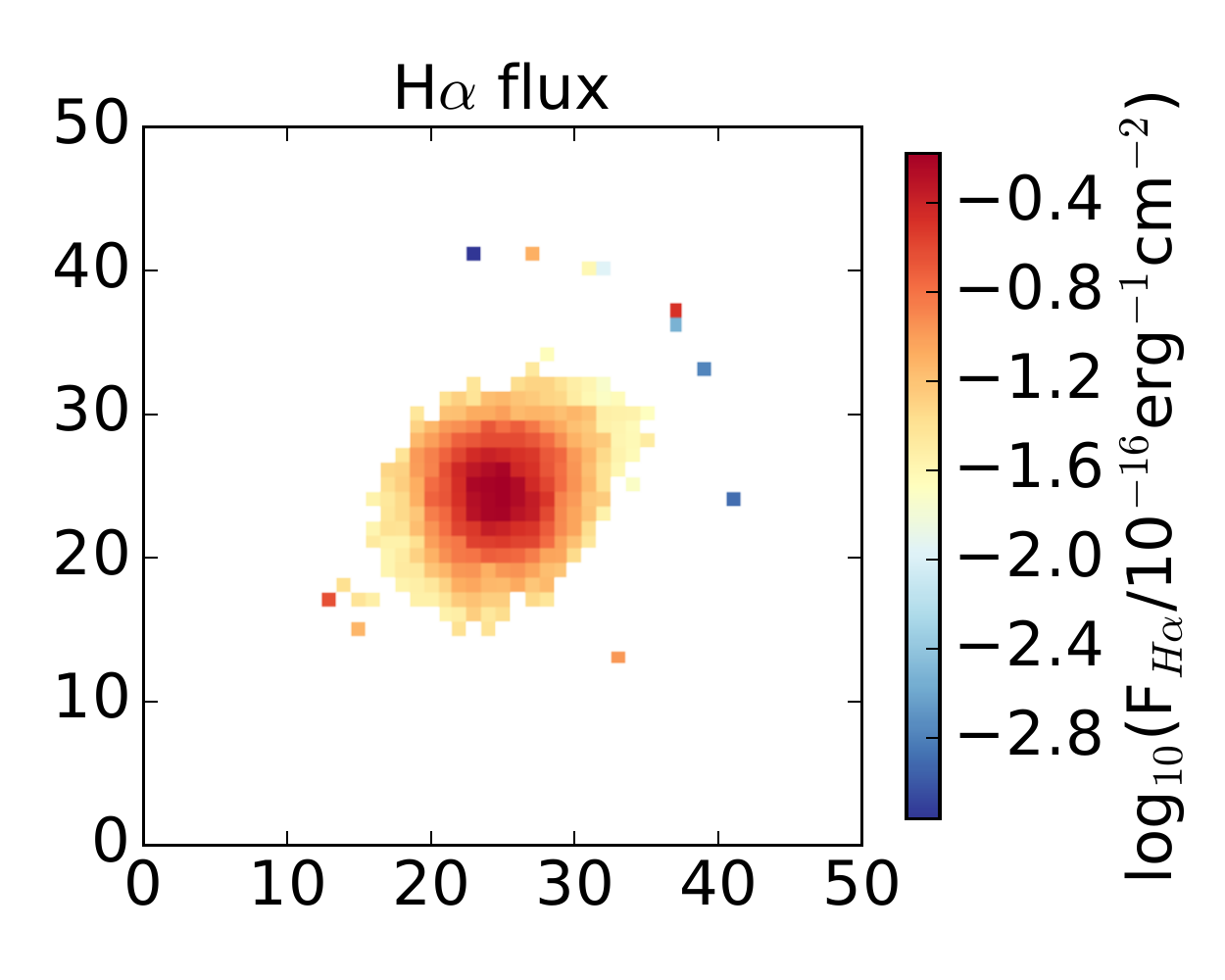}
\end{minipage}%
\begin{minipage}[]{0.2\textwidth}
\includegraphics[width=3.6cm]{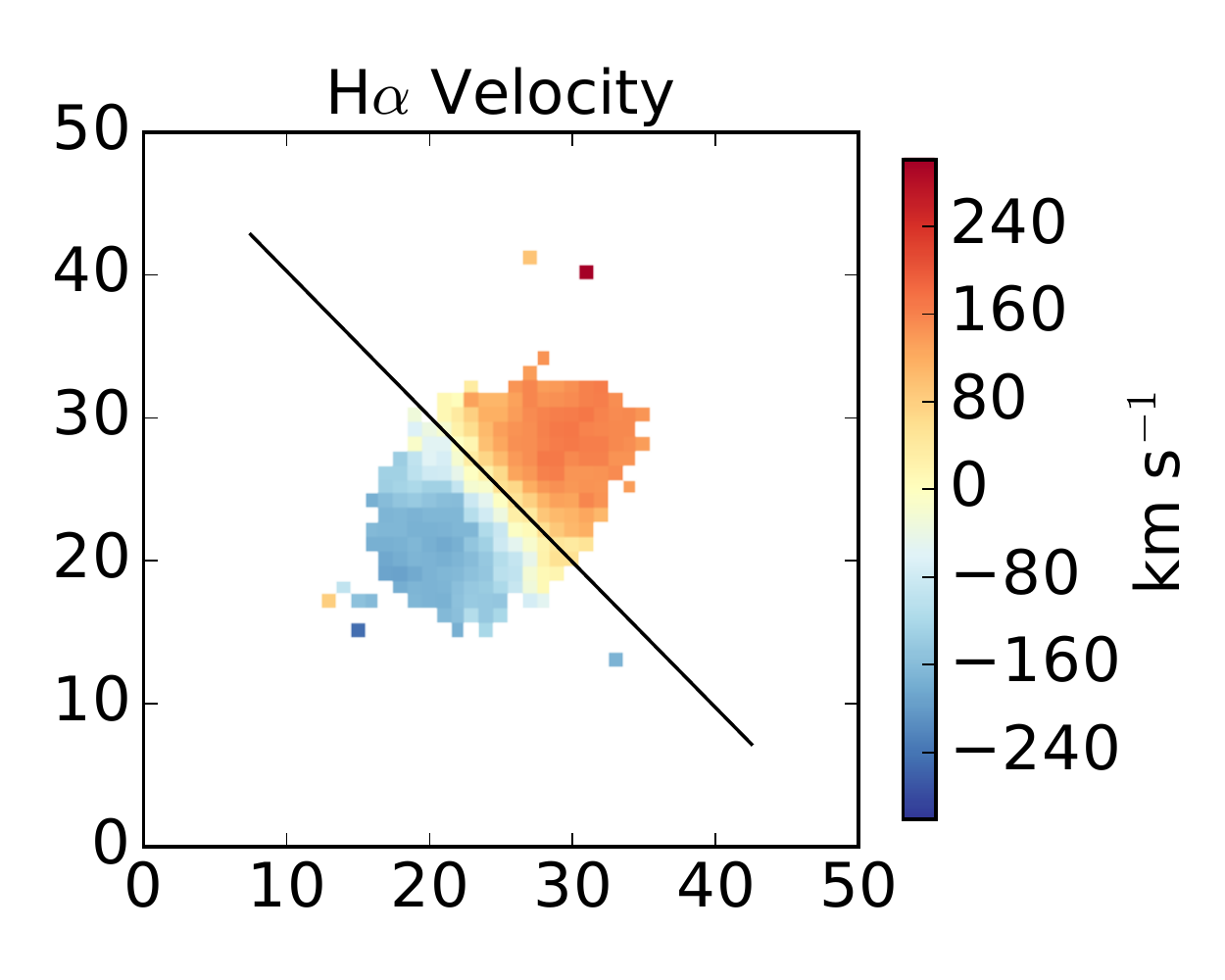}
\end{minipage}%
\begin{minipage}[]{0.2\textwidth}
\includegraphics[width=3.6cm]{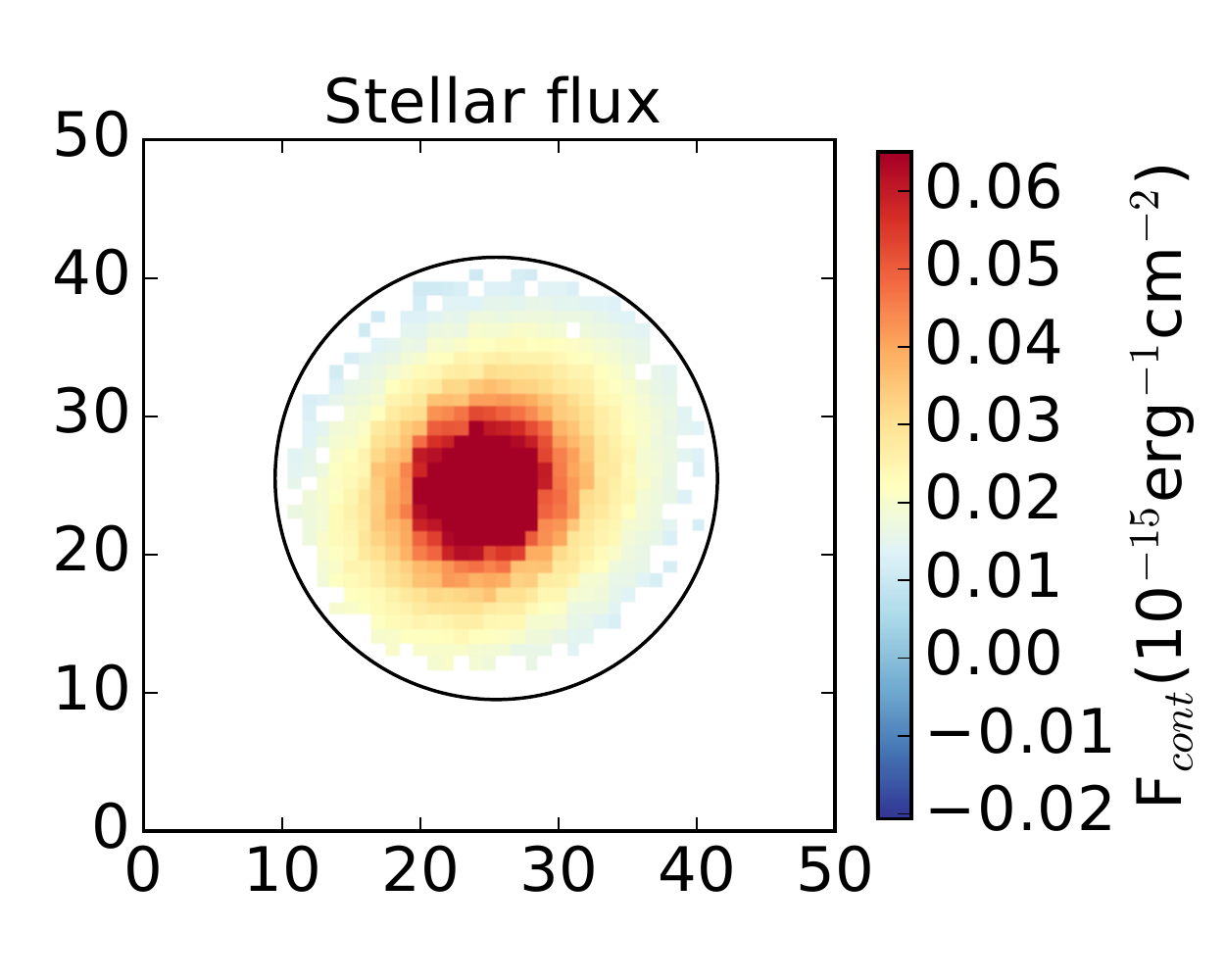}
\end{minipage}%
\hspace*{-3mm}
\begin{minipage}[]{0.22\textwidth}
\includegraphics[width=3.2cm]{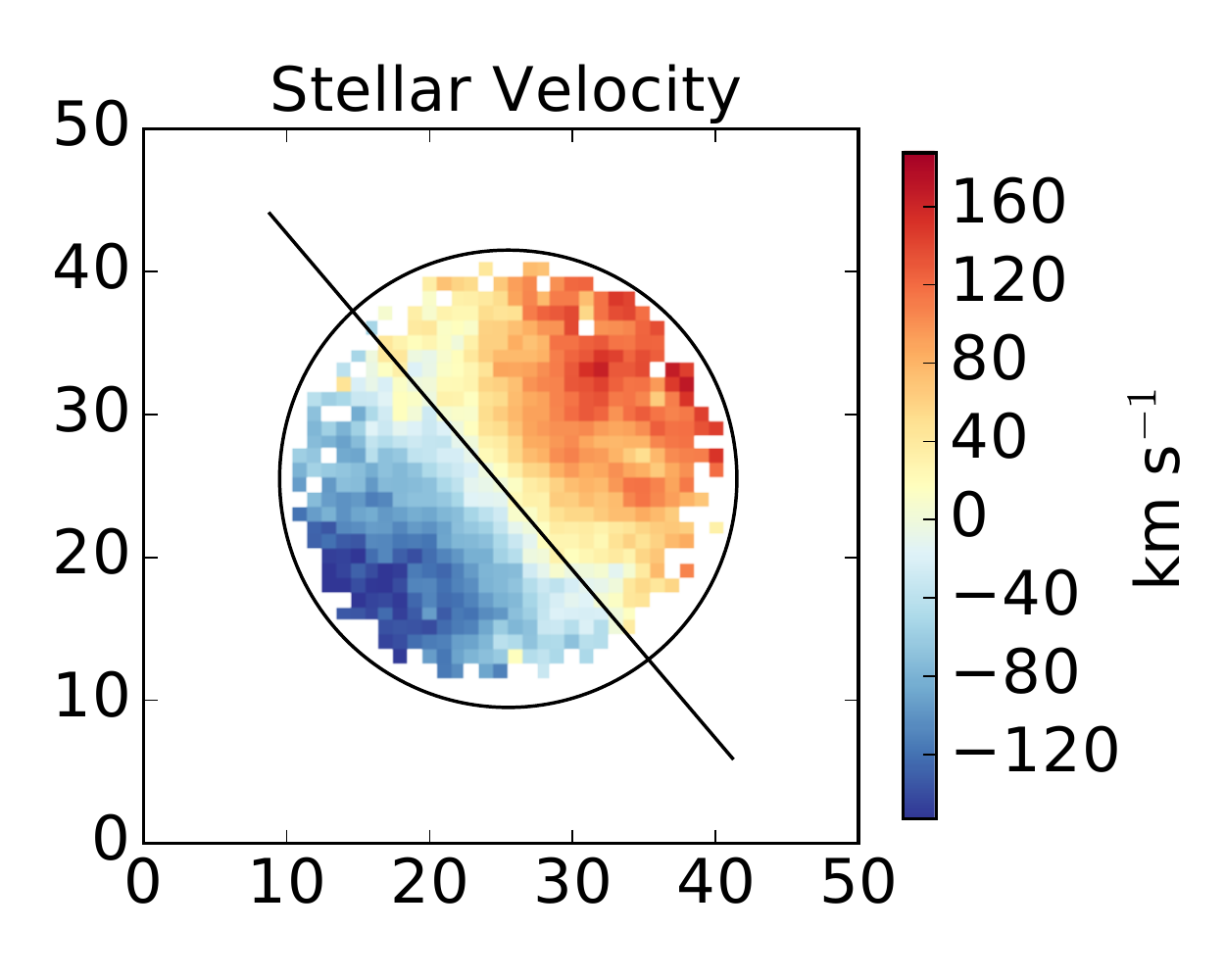}
\end{minipage}%
\vspace*{1mm}
\begin{minipage}[t]{0.02\textwidth}
\centerline{(b)}
\end{minipage}%
\begin{minipage}[]{0.18\textwidth}
\includegraphics[width=2.5cm]{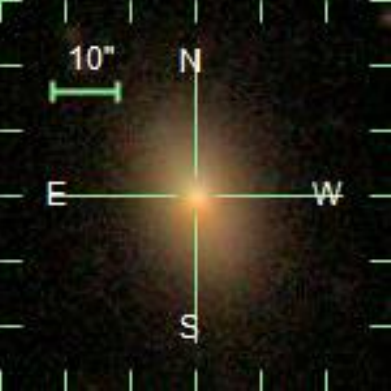}
\end{minipage}%
\hspace*{-3mm}
\begin{minipage}[]{0.2\textwidth}
\includegraphics[width=3.6cm]{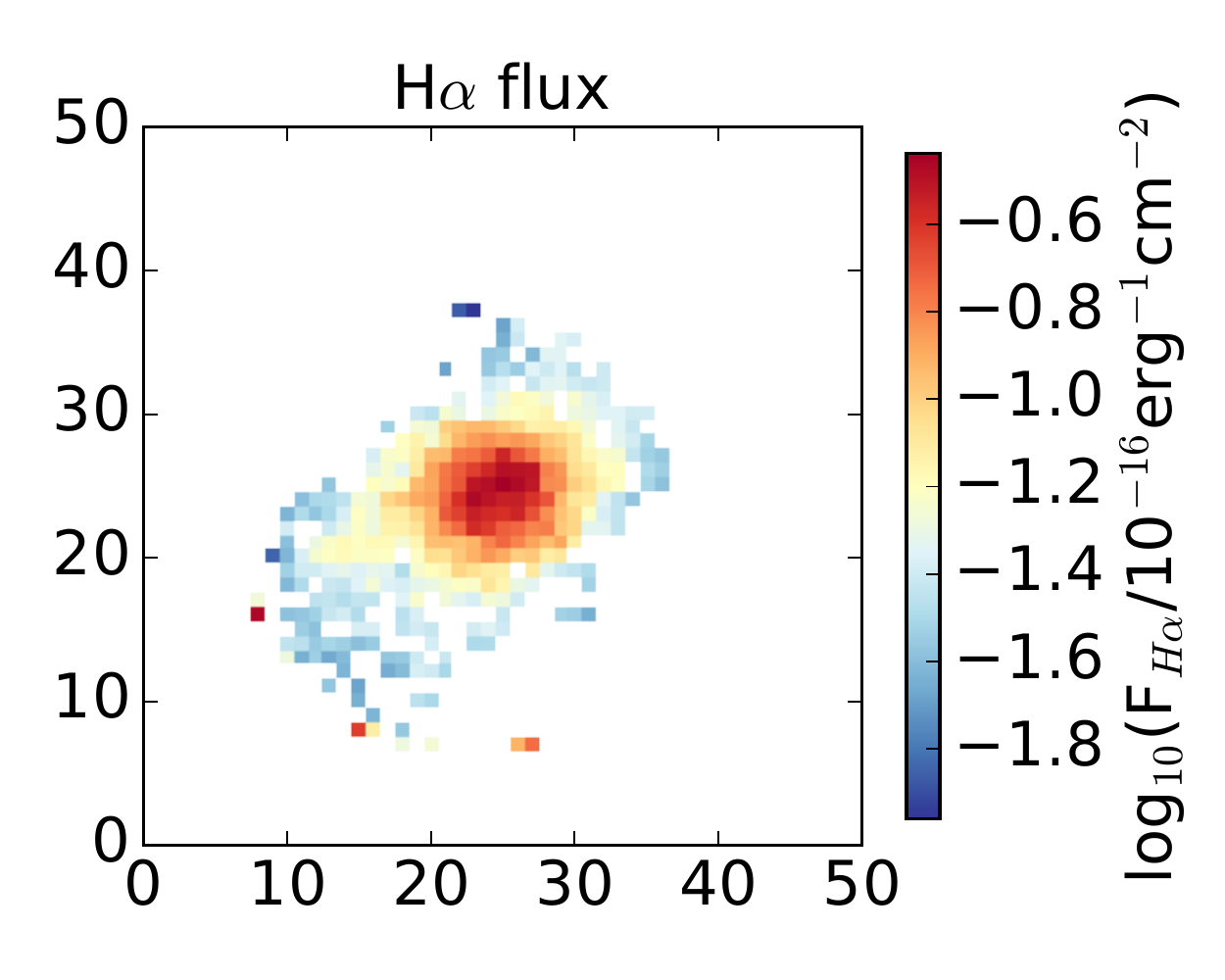}
\end{minipage}%
\begin{minipage}[]{0.2\textwidth}
\includegraphics[width=3.6cm]{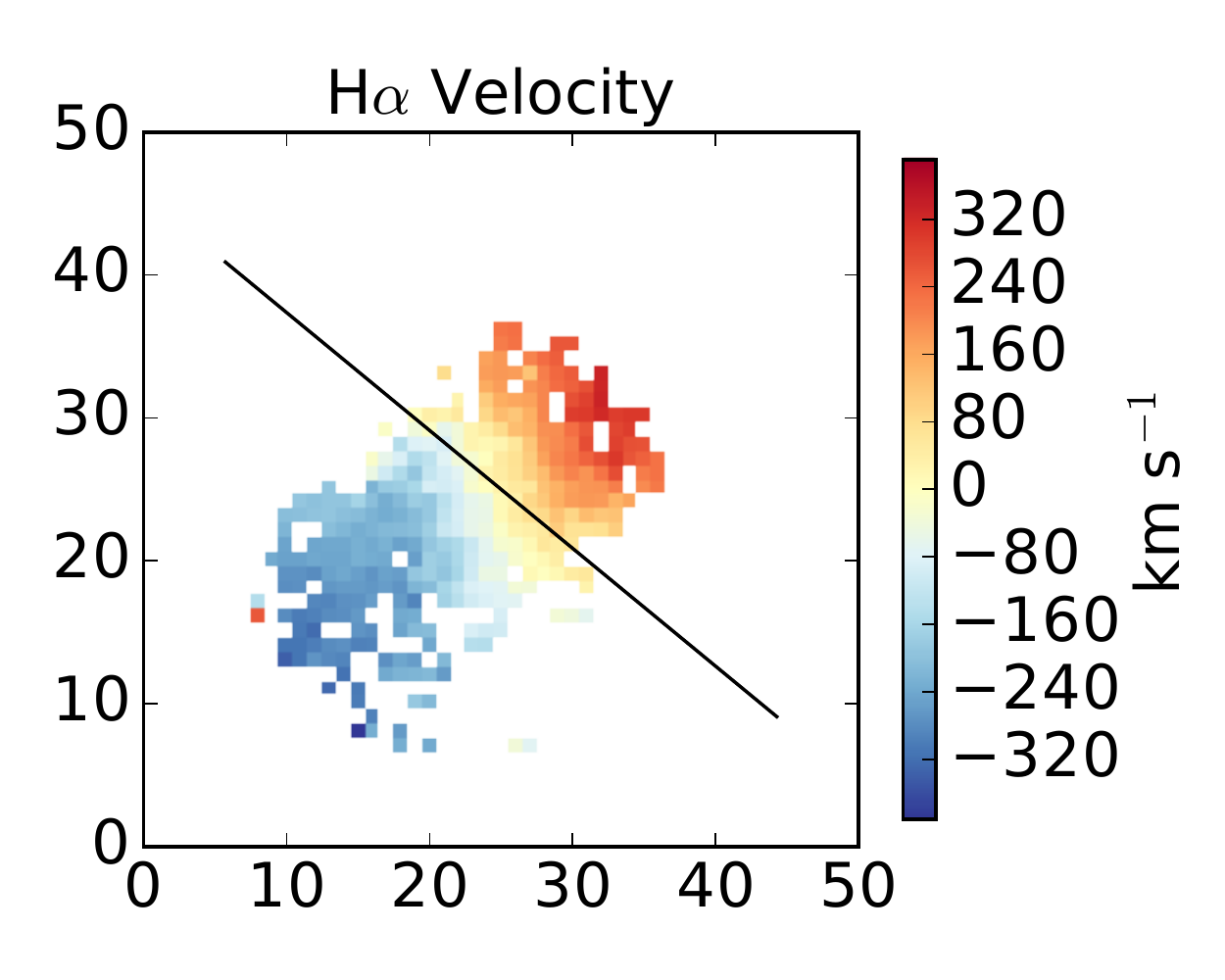}
\end{minipage}%
\begin{minipage}[]{0.2\textwidth}
\includegraphics[width=3.6cm]{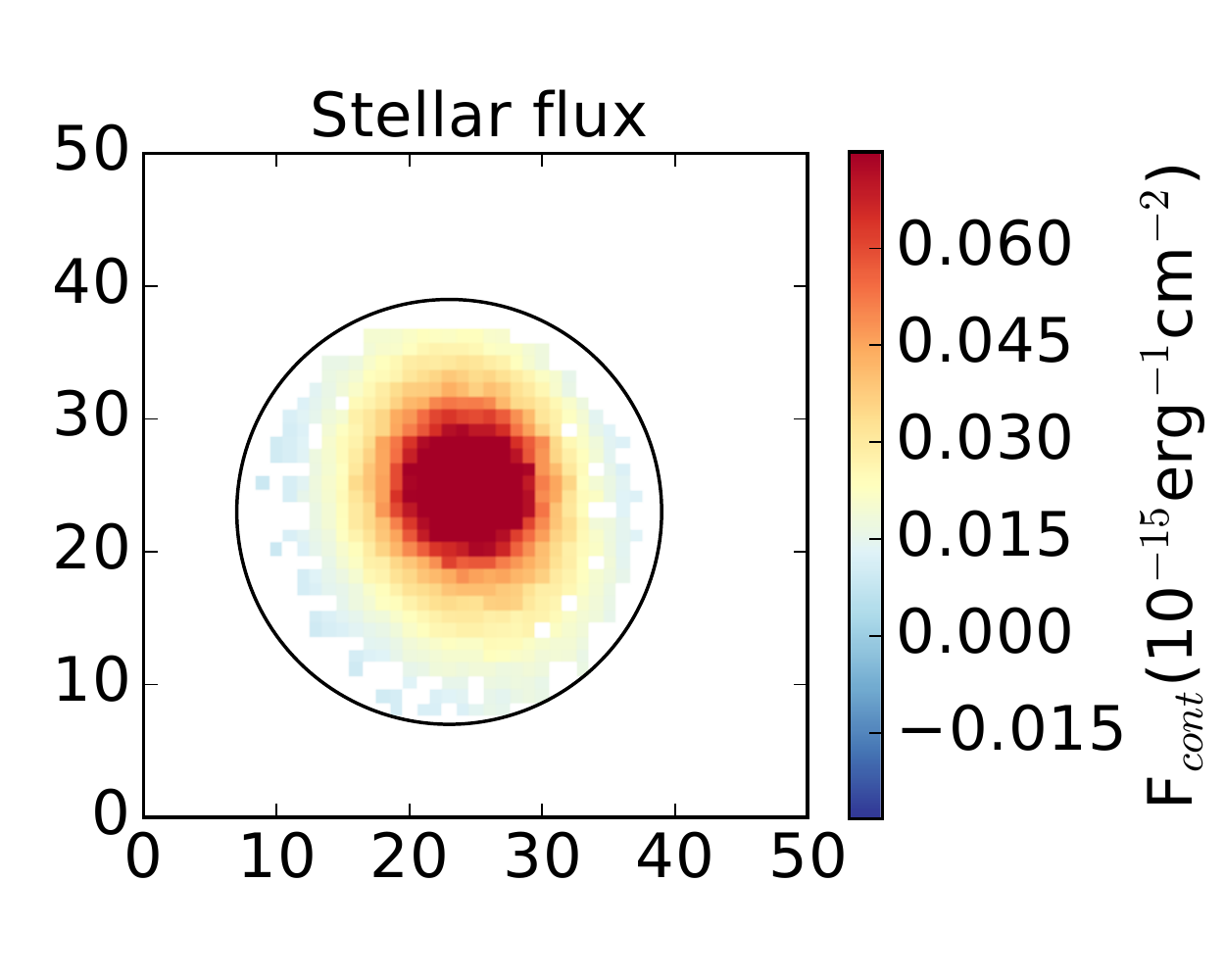}
\end{minipage}%
\hspace*{-3mm}
\begin{minipage}[]{0.22\textwidth}
\includegraphics[width=3.2cm]{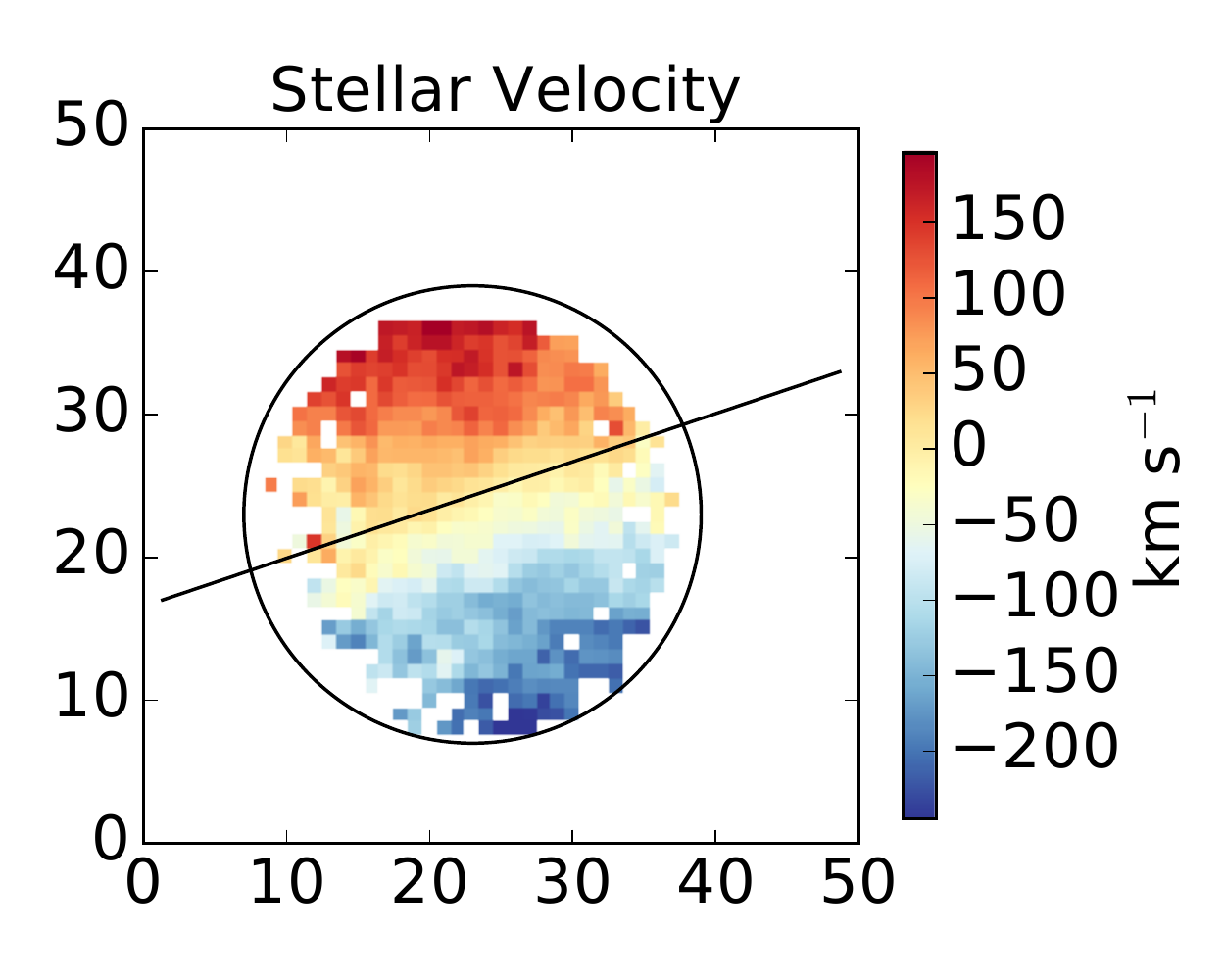}
\end{minipage}%
\vspace*{1mm}
\begin{minipage}[t]{0.02\textwidth}
\centerline{(c)}
\end{minipage}%
\begin{minipage}[]{0.18\textwidth}
\includegraphics[width=2.5cm]{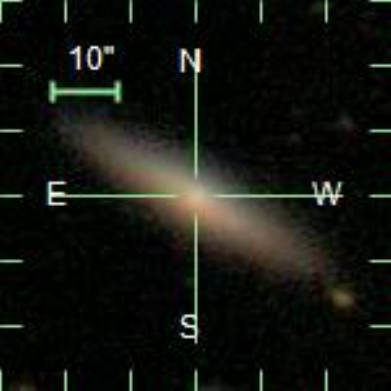}
\end{minipage}%
\hspace*{-3mm}
\begin{minipage}[]{0.2\textwidth}
\includegraphics[width=3.6cm]{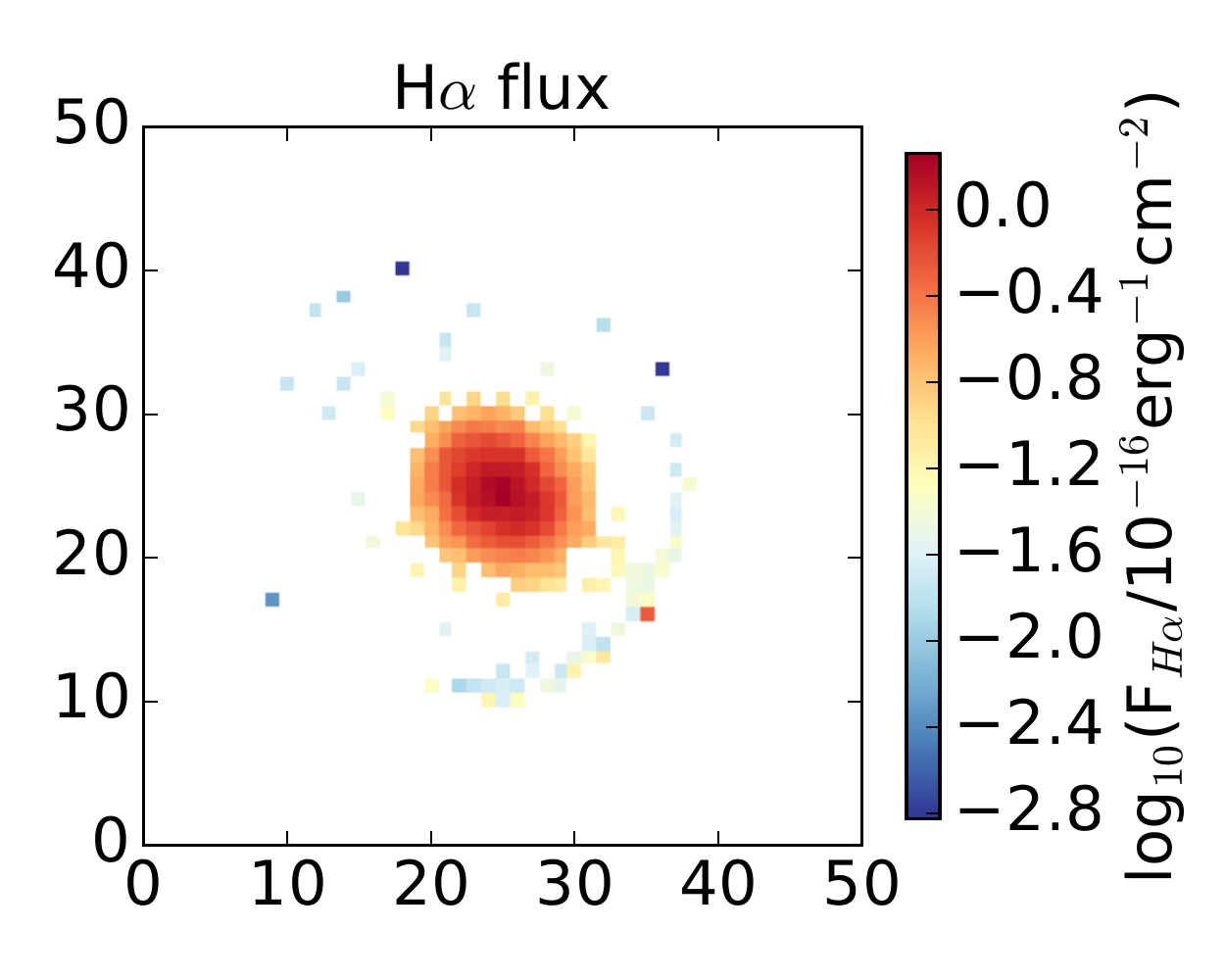}
\end{minipage}%
\begin{minipage}[]{0.2\textwidth}
\includegraphics[width=3.6cm]{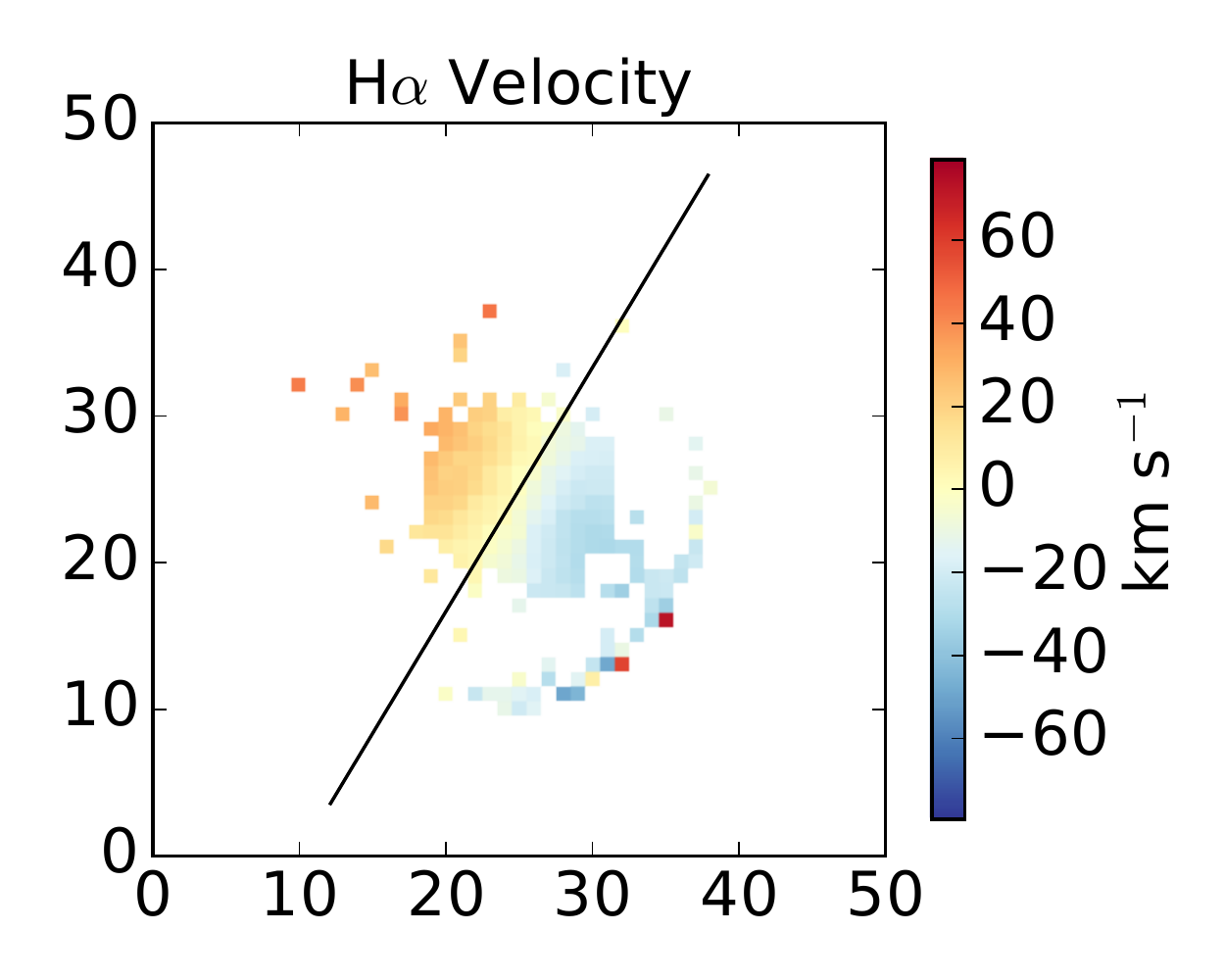}
\end{minipage}%
\begin{minipage}[]{0.2\textwidth}
\includegraphics[width=3.6cm]{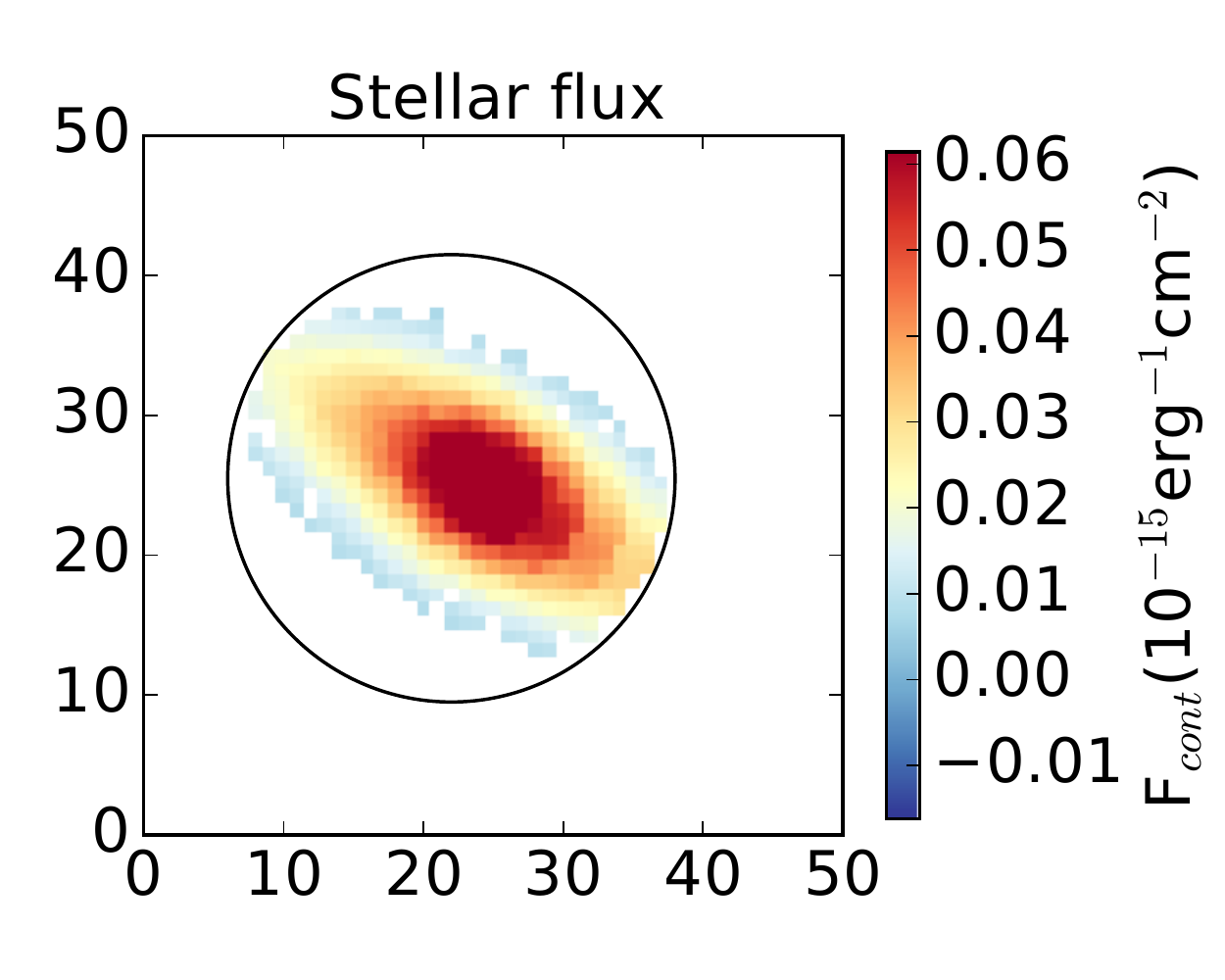}
\end{minipage}%
\hspace*{-3mm}
\begin{minipage}[]{0.22\textwidth}
\includegraphics[width=3.2cm]{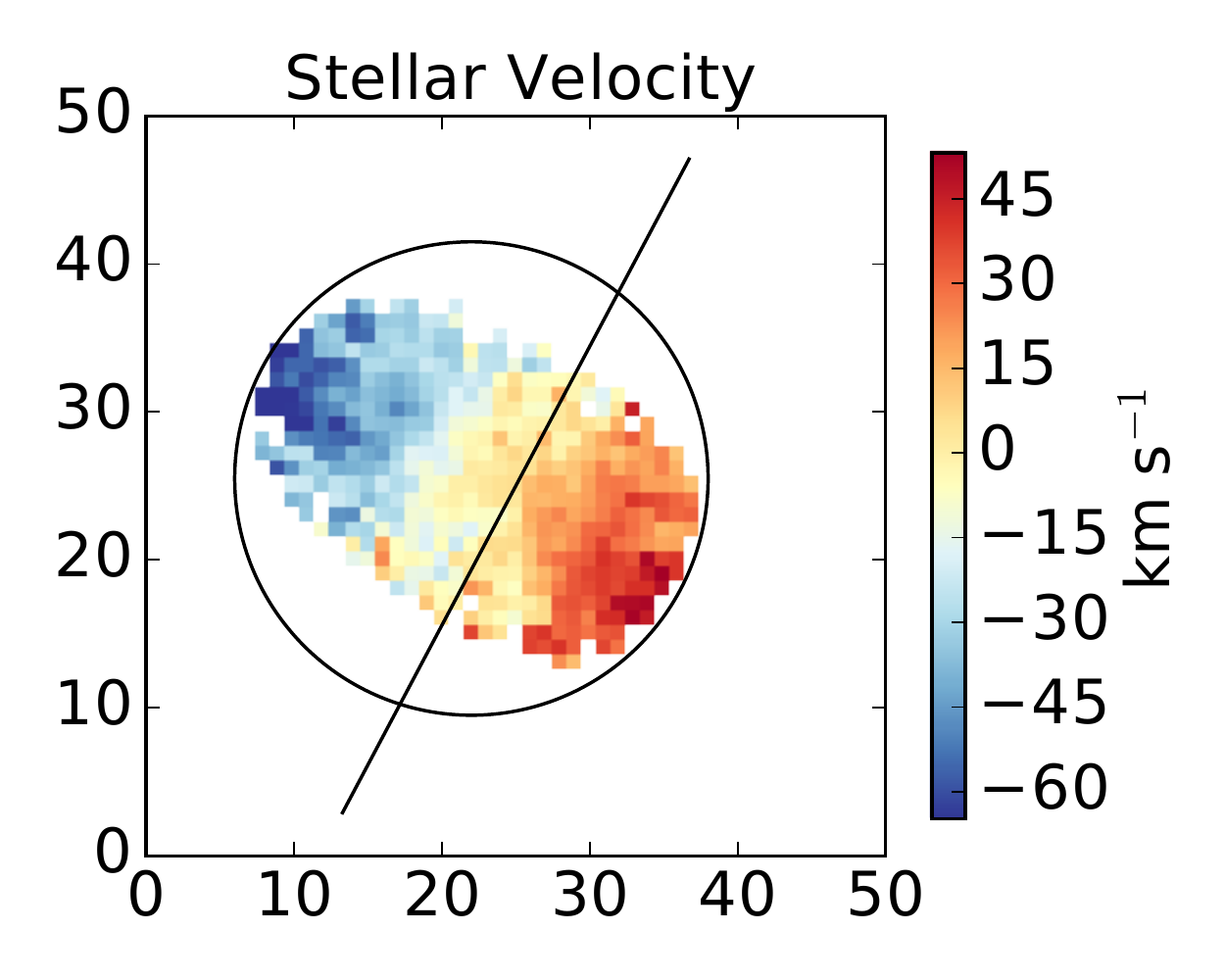}
\end{minipage}%
\vspace*{1mm}
\begin{minipage}[t]{0.02\textwidth}
\centerline{(d)}
\end{minipage}%
\begin{minipage}[]{0.18\textwidth}
\includegraphics[width=2.5cm]{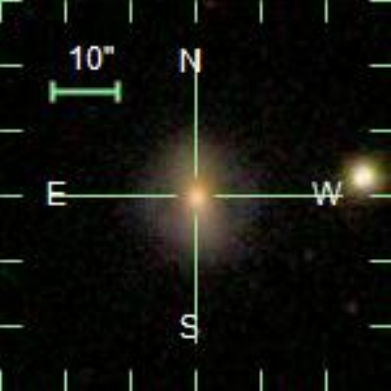}
\end{minipage}%
\hspace*{-3mm}
\begin{minipage}[]{0.2\textwidth}
\includegraphics[width=3.6cm]{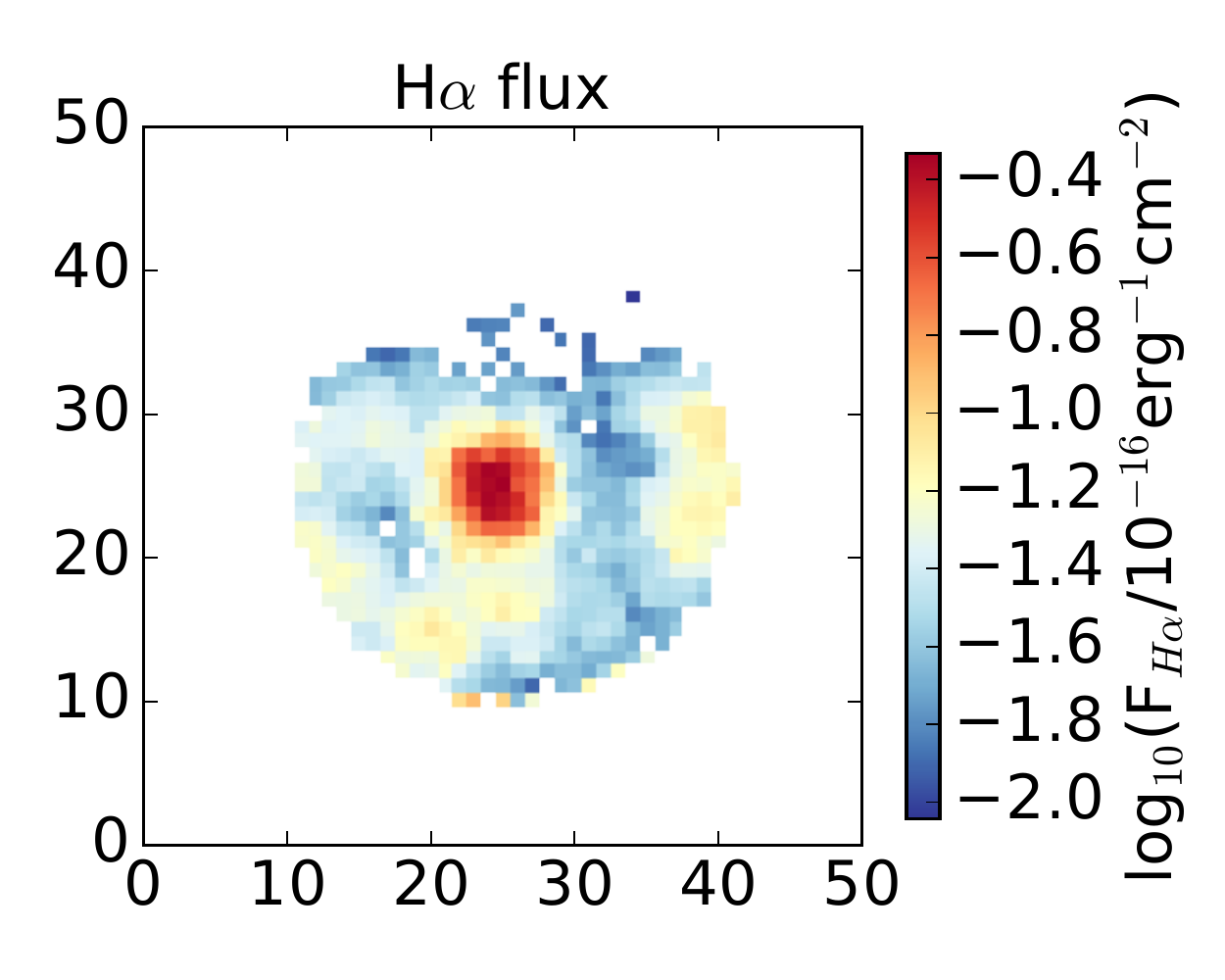}
\end{minipage}%
\begin{minipage}[]{0.2\textwidth}
\includegraphics[width=3.6cm]{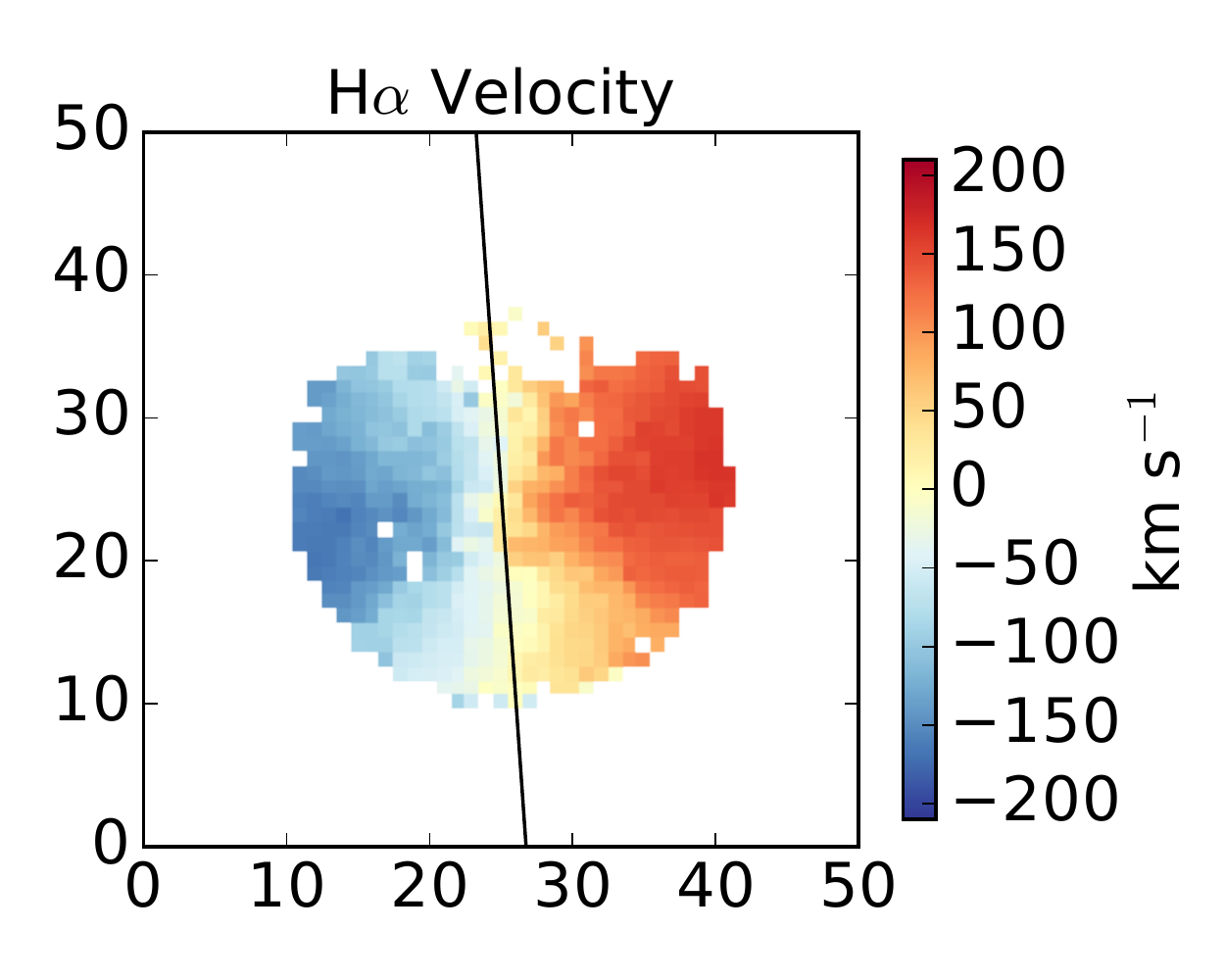}
\end{minipage}%
\begin{minipage}[]{0.2\textwidth}
\includegraphics[width=3.6cm]{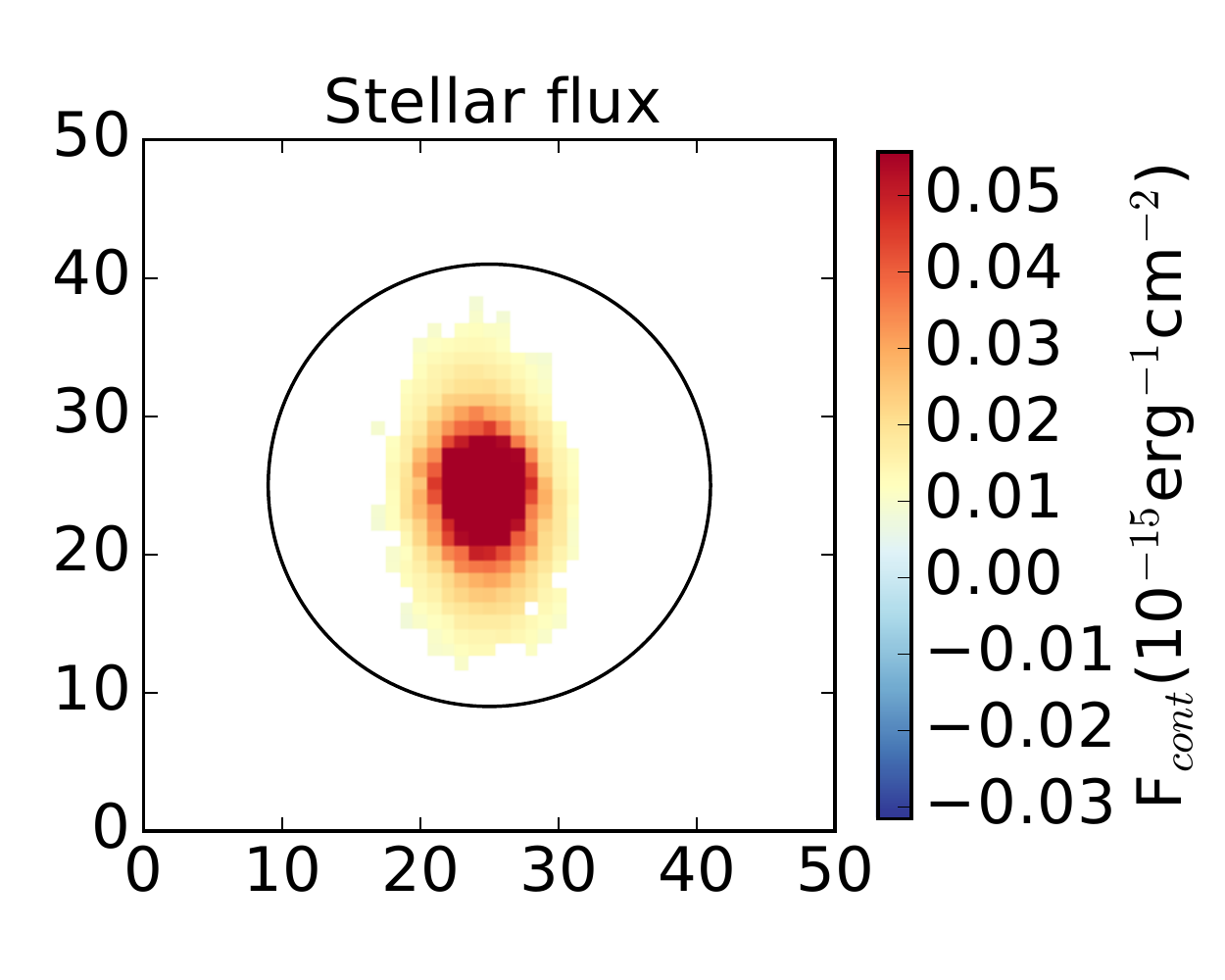}
\end{minipage}%
\hspace*{-3mm}
\begin{minipage}[]{0.22\textwidth}
\includegraphics[width=3.2cm]{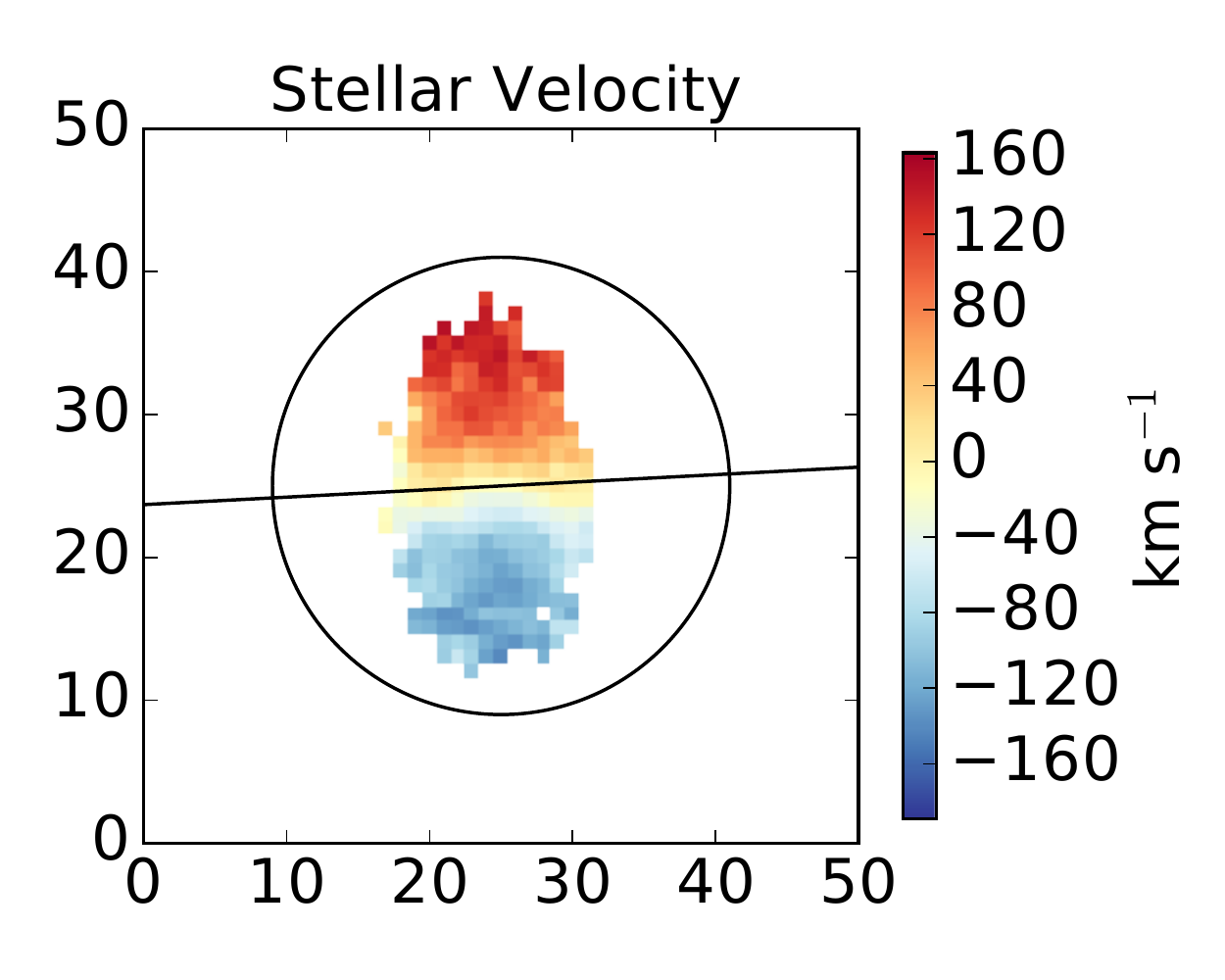}
\end{minipage}%
\vspace*{1mm}
\begin{minipage}[t]{0.02\textwidth}
\centerline{(e)}
\end{minipage}%
\begin{minipage}[]{0.18\textwidth}
\includegraphics[width=2.5cm]{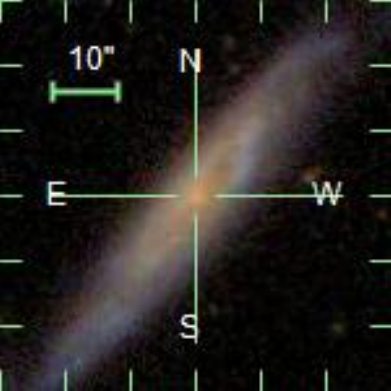}
\end{minipage}%
\hspace*{-3mm}
\begin{minipage}[]{0.2\textwidth}
\includegraphics[width=3.6cm]{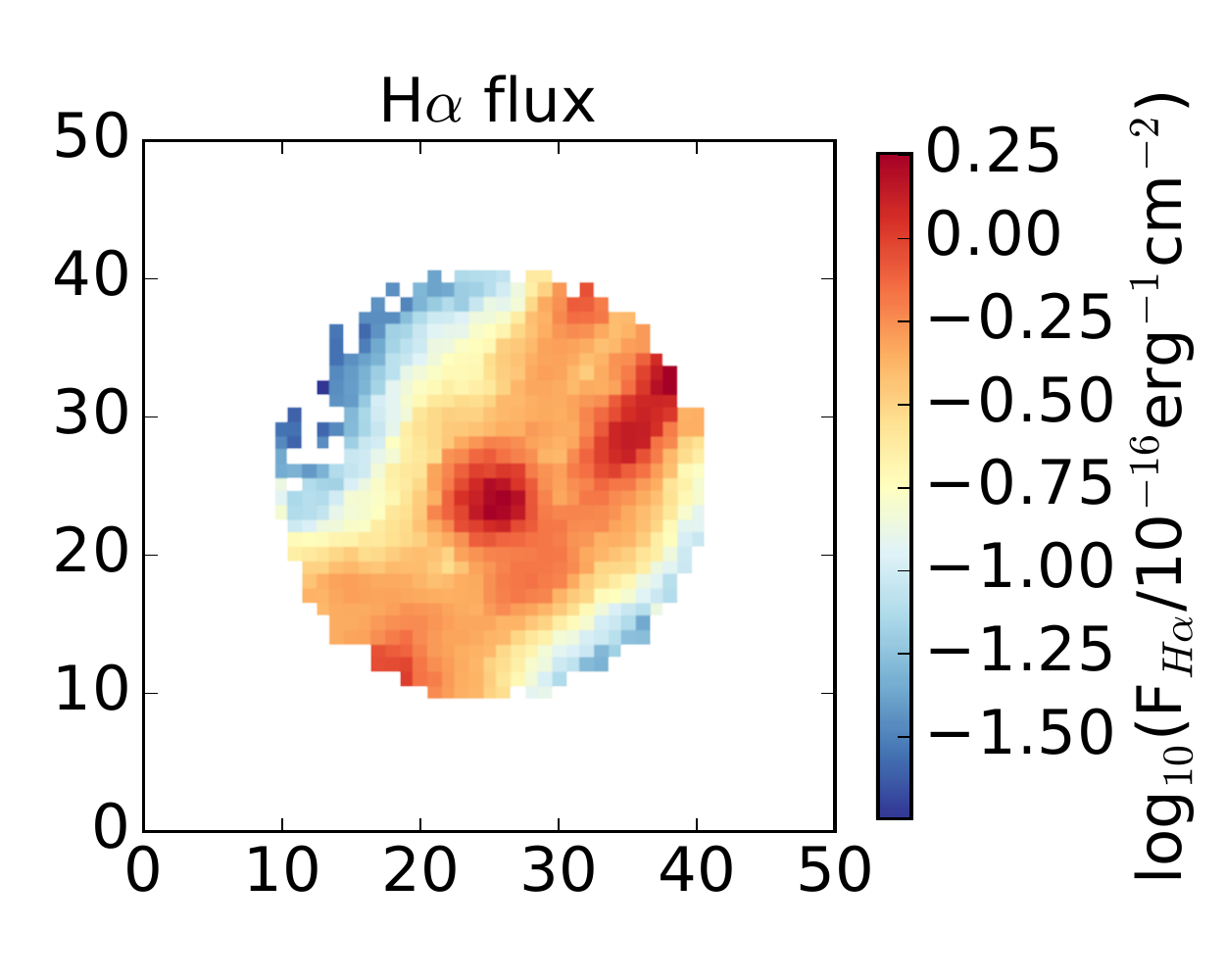}
\end{minipage}%
\begin{minipage}[]{0.2\textwidth}
\includegraphics[width=3.6cm]{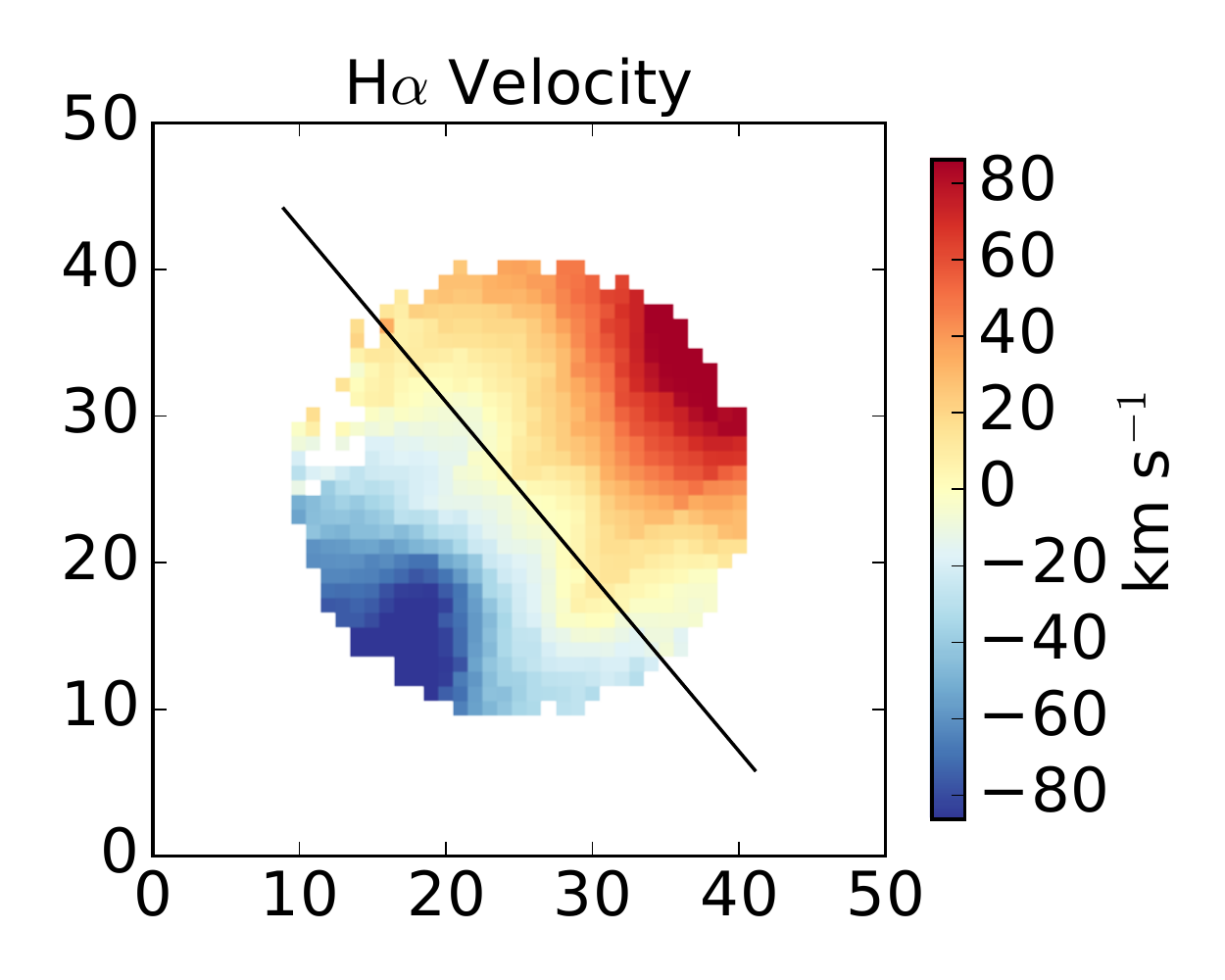}
\end{minipage}%
\begin{minipage}[]{0.2\textwidth}
\includegraphics[width=3.6cm]{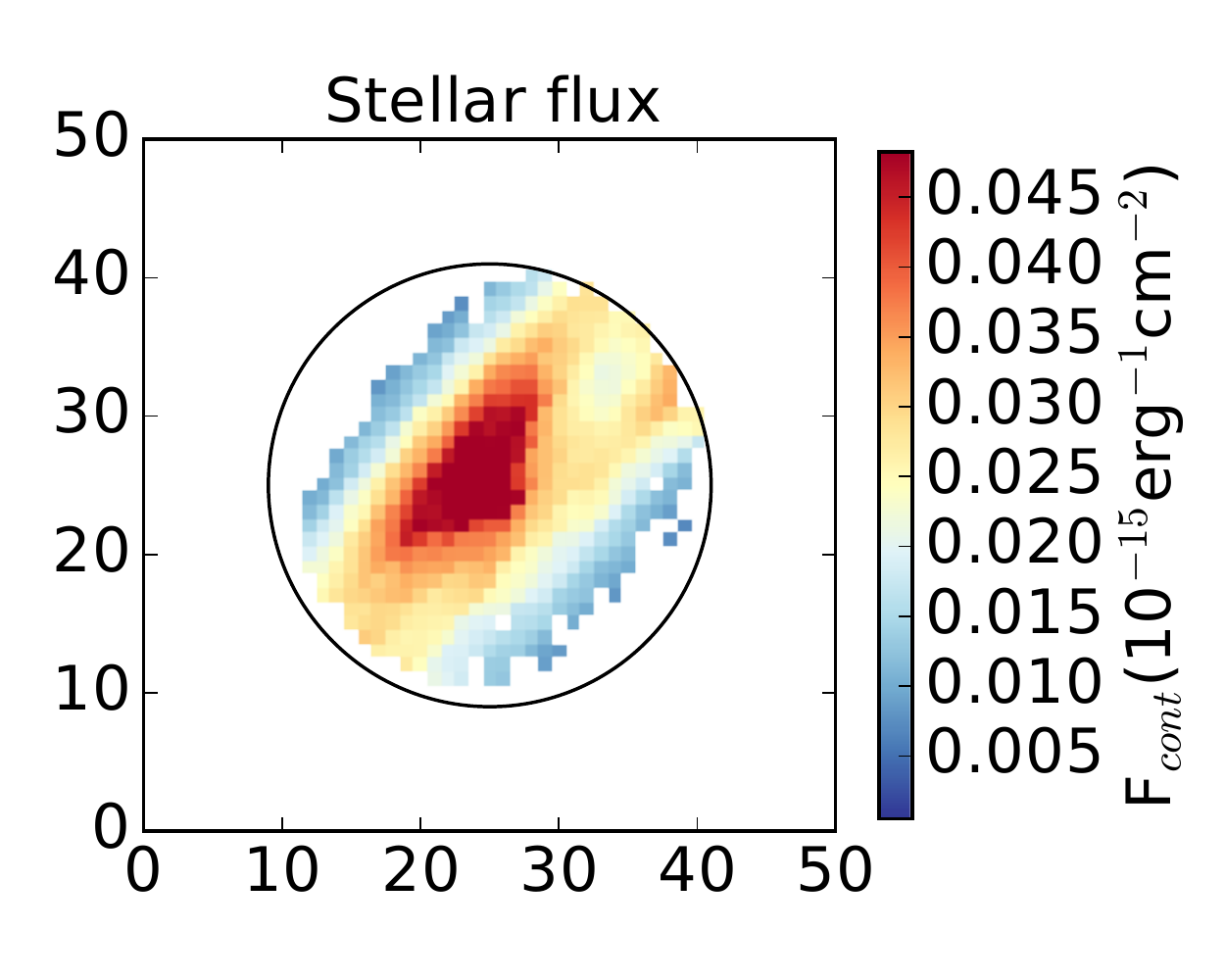}
\end{minipage}%
\hspace*{-3mm}
\begin{minipage}[]{0.22\textwidth}
\includegraphics[width=3.2cm]{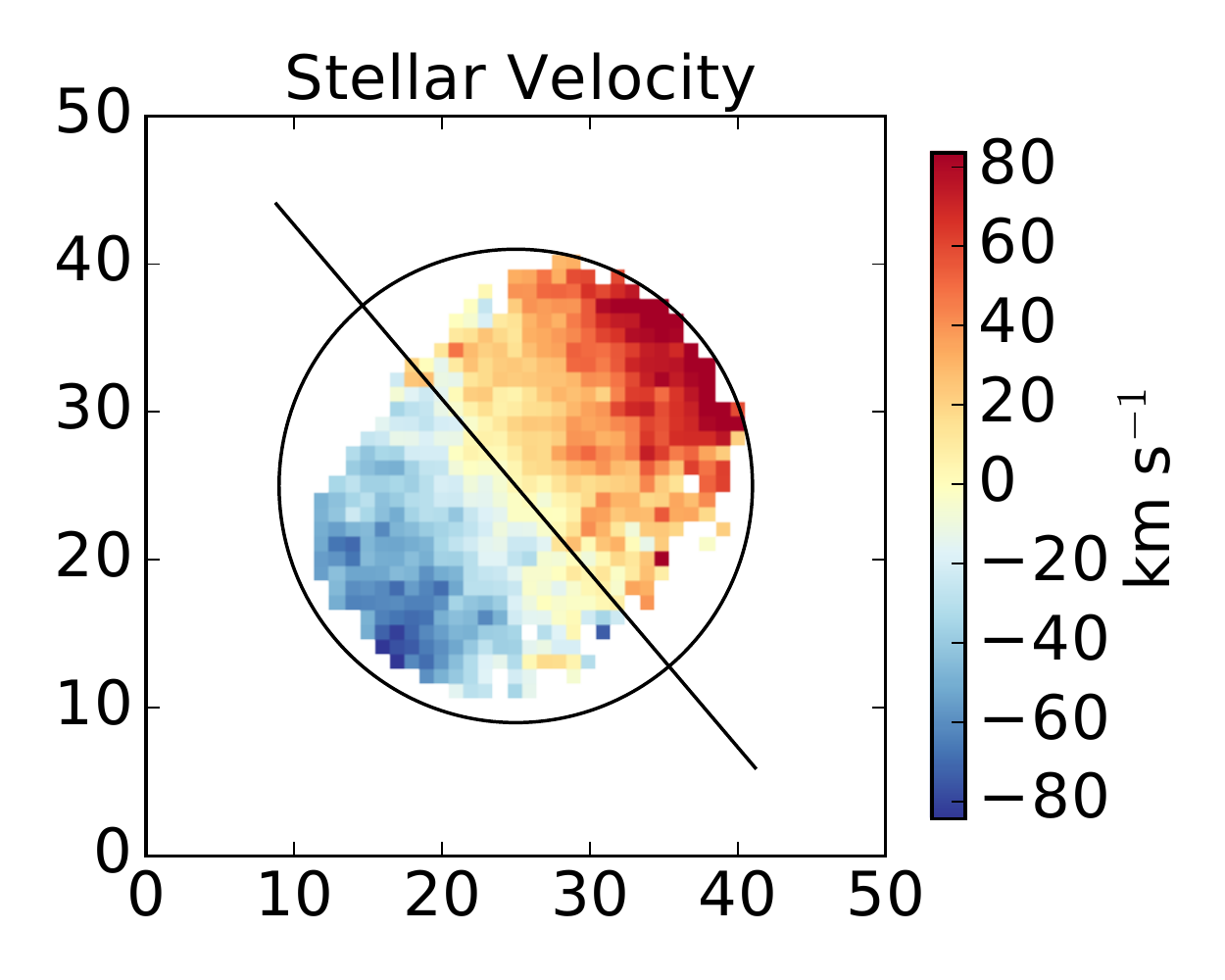}
\end{minipage}%
\vspace*{1mm}
\begin{minipage}[t]{0.02\textwidth}
\centerline{(f)}
\end{minipage}%
\begin{minipage}[]{0.18\textwidth}
\includegraphics[width=2.5cm]{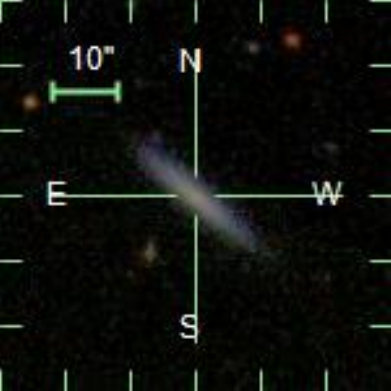}
\end{minipage}%
\hspace*{-3mm}
\begin{minipage}[]{0.2\textwidth}
\includegraphics[width=3.6cm]{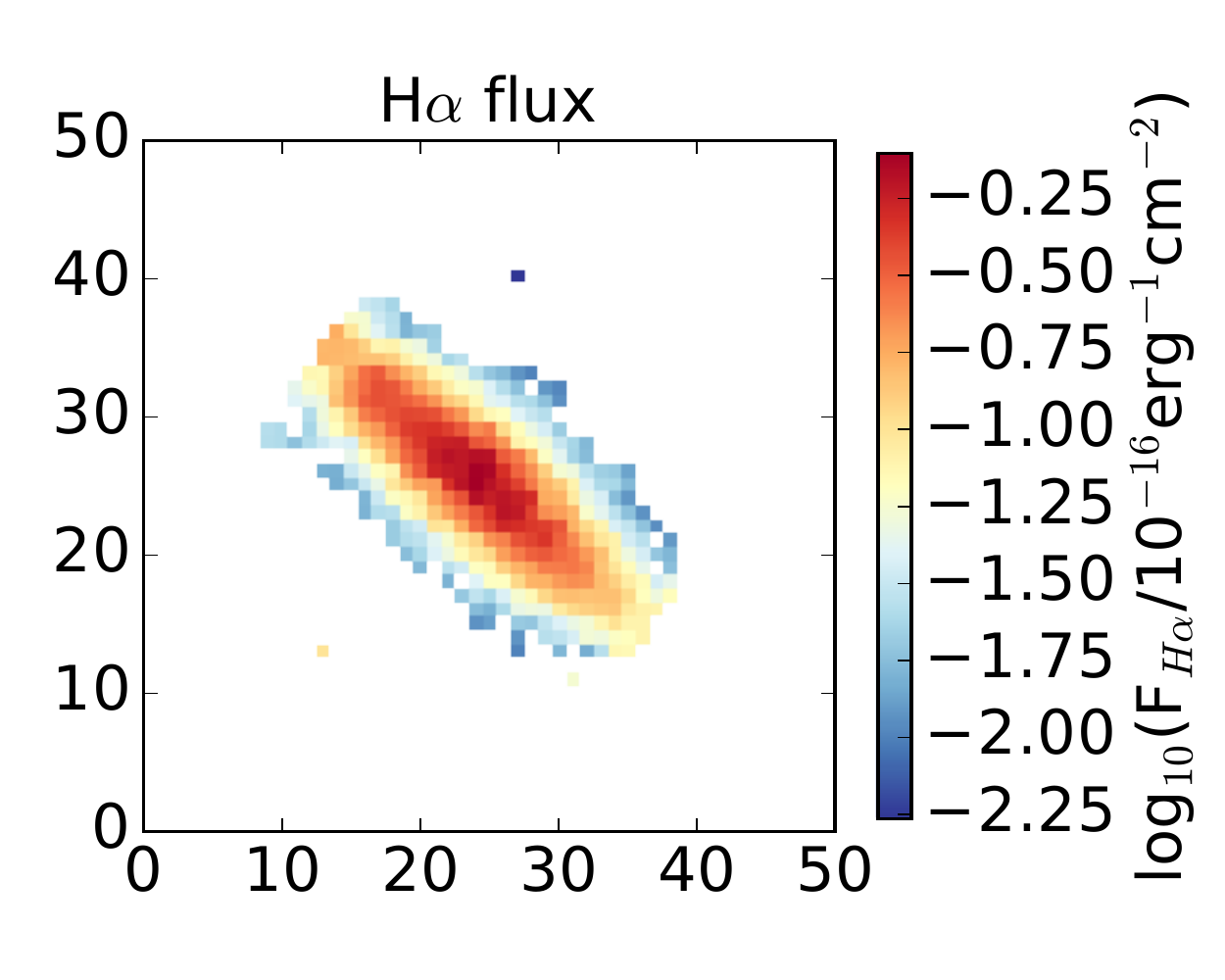}
\end{minipage}%
\begin{minipage}[]{0.2\textwidth}
\includegraphics[width=3.6cm]{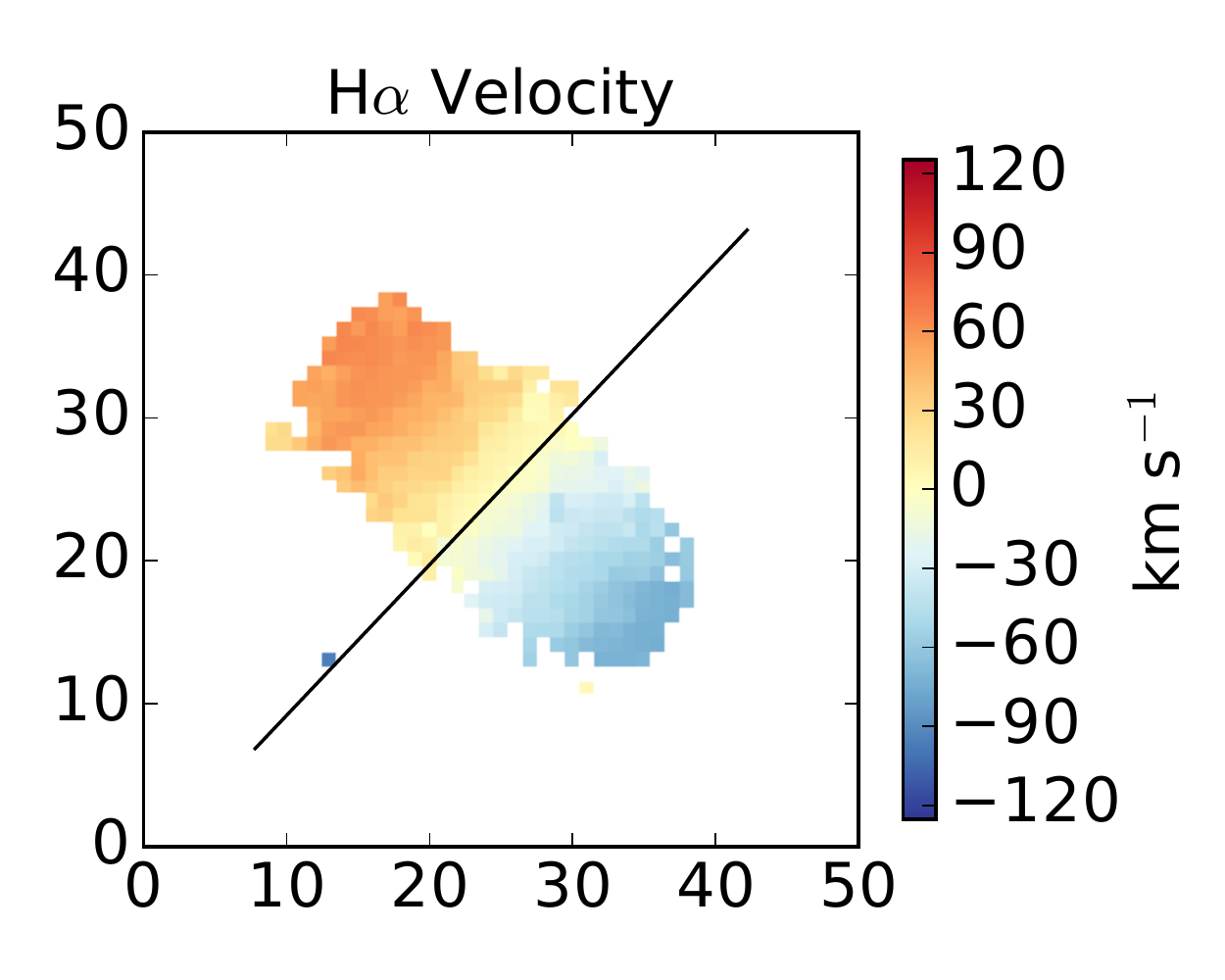}
\end{minipage}%
\begin{minipage}[]{0.2\textwidth}
\includegraphics[width=3.6cm]{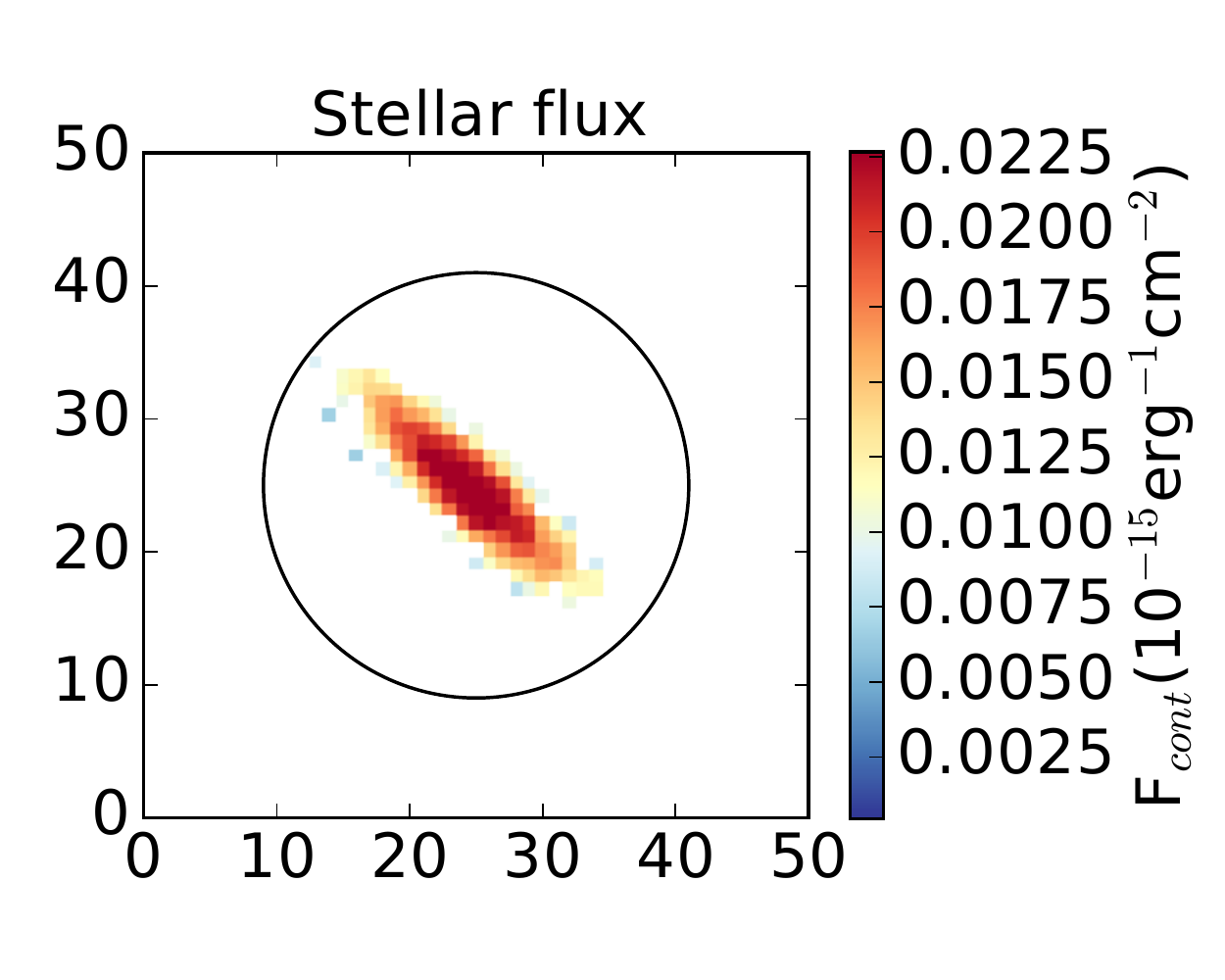}
\end{minipage}%
\hspace*{-3mm}
\begin{minipage}[]{0.22\textwidth}
\includegraphics[width=3.2cm]{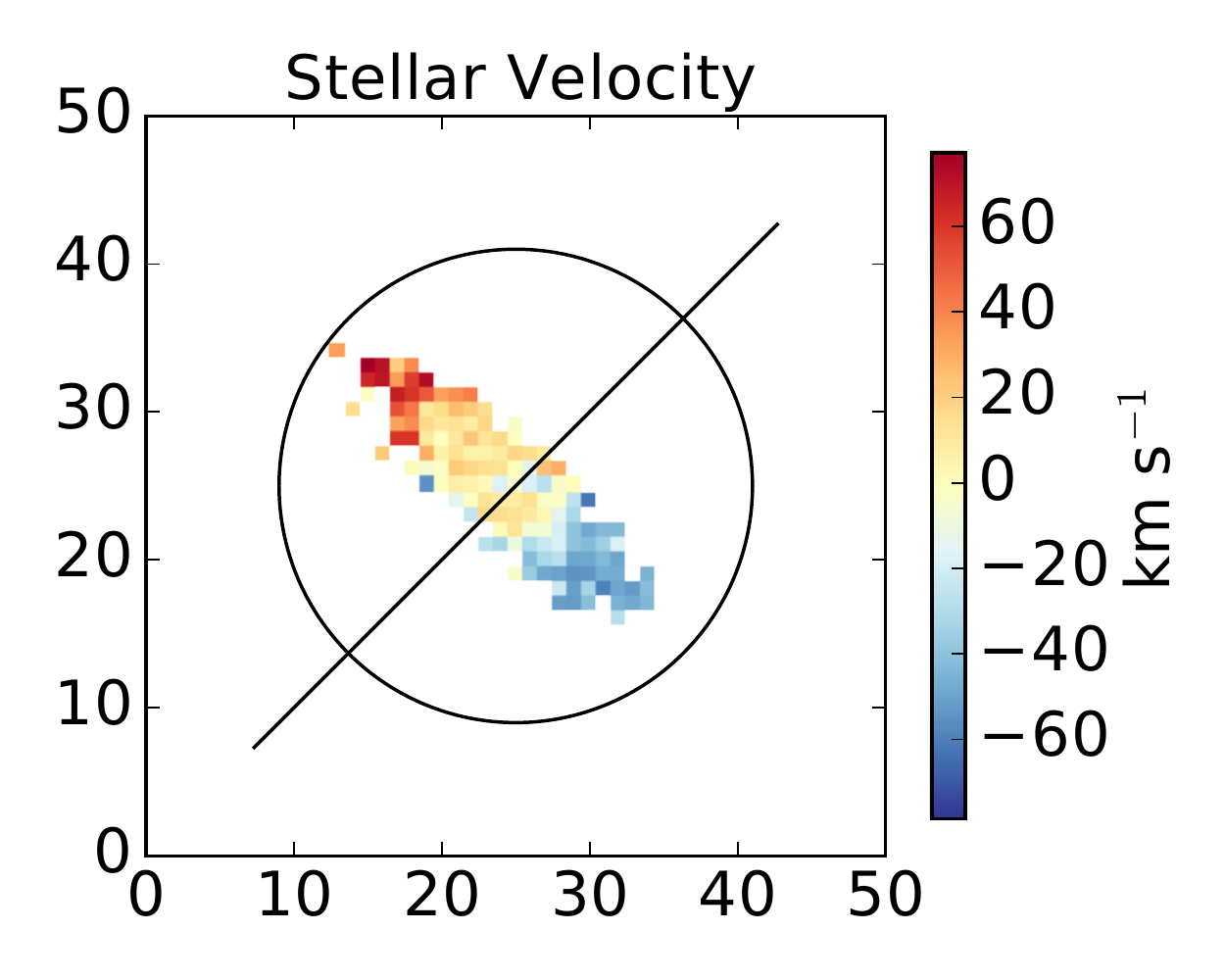}
\end{minipage}%
\vspace*{1mm}
\vspace*{3mm}
\caption{Examples of different types of fitted position angles. For each galaxy (row) the panels are the SDSS image, the log$_{10}$(H$\alpha$ gas flux) and H$\alpha$ velocity, then the stellar flux and stellar velocity. The lines mark the axis of rotation and the circle on the stellar images indicates the $15''$ size of the SAMI hexabundle. The SDSS images have a bar marking $10''$ and are not shown on the same scale as the SAMI images. One example of each of the following types of galaxies is shown (from top to bottom):
(a) Aligned ETG (91963);
(b) Misaligned ETG (551505), with a PA offset of $58^{\circ}$;
(c) Counter-rotating (511863);
(d) Polar ring galaxy with position angles close to perpendicular (570206);
(e) LTG with well aligned stellar/gas rotation even though the dynamics of the spiral arms distort the gas velocities (144239);
(f) Galaxy where the fitting failed and had to be fit by eye (302994). The fitting code gave a stellar rotation axis of $93.5\pm30^{\circ}$ while by eye we measure $135+\-10^{\circ}$ which aligns with the gas. 
}
\label{pics}
\end{figure*}

PA offsets were then calculated from the difference in these fitted stellar and gas velocity PAs when both velocities could be fitted with a reliable PA. The resultant value is the projected misalignment. Fig.~\ref{pics} shows examples of aligned, misaligned and counter-rotating galaxies.
In order to determine if the fitted PAs were reliable, all of the stellar and gas velocity maps with fitted PAs marked, were checked by eye. There were cases where the number of velocity data points in the map was sufficient to return a PA measurement, but the measurement could not be trusted because the fit PA was not aligned with the main velocity field but the correct PA could be clearly determined by eye. This is a failure of the fitting routine, and while rare, in those cases, a PA estimate by eye was recorded instead. The error on the PAs fitted by eye is estimated to be $\pm10^{\circ}$. 
An example is
shown in Fig.~\ref{pics} (row 6). There were 10 galaxies (1.6\%) fit by eye for their stellar PAs, and 24 (3.9\%) for the gas PAs. 

Spiral arms can give clear distortions in the gas kinematics for near-face-on galaxies, where the gas dynamics wrap around with the spiral arms.
However, it was confirmed by careful inspection that the global PA fitting is insensitive to these distortions due to spiral arms because they are small compared to the bulk disk rotation and serve only to cause kinks around the fitted gas rotation. Such gas distortions are therefore not indicative of merger events. We note that if fitting higher order kinemetry of such galaxies, the kinematic disturbance detected may be indistinguishable from mergers. An example of such a galaxy is shown in Fig.~\ref{pics} (row 5).

\begin{figure}
\begin{minipage}[]{0.5\textwidth}
\includegraphics[width=8.0cm]{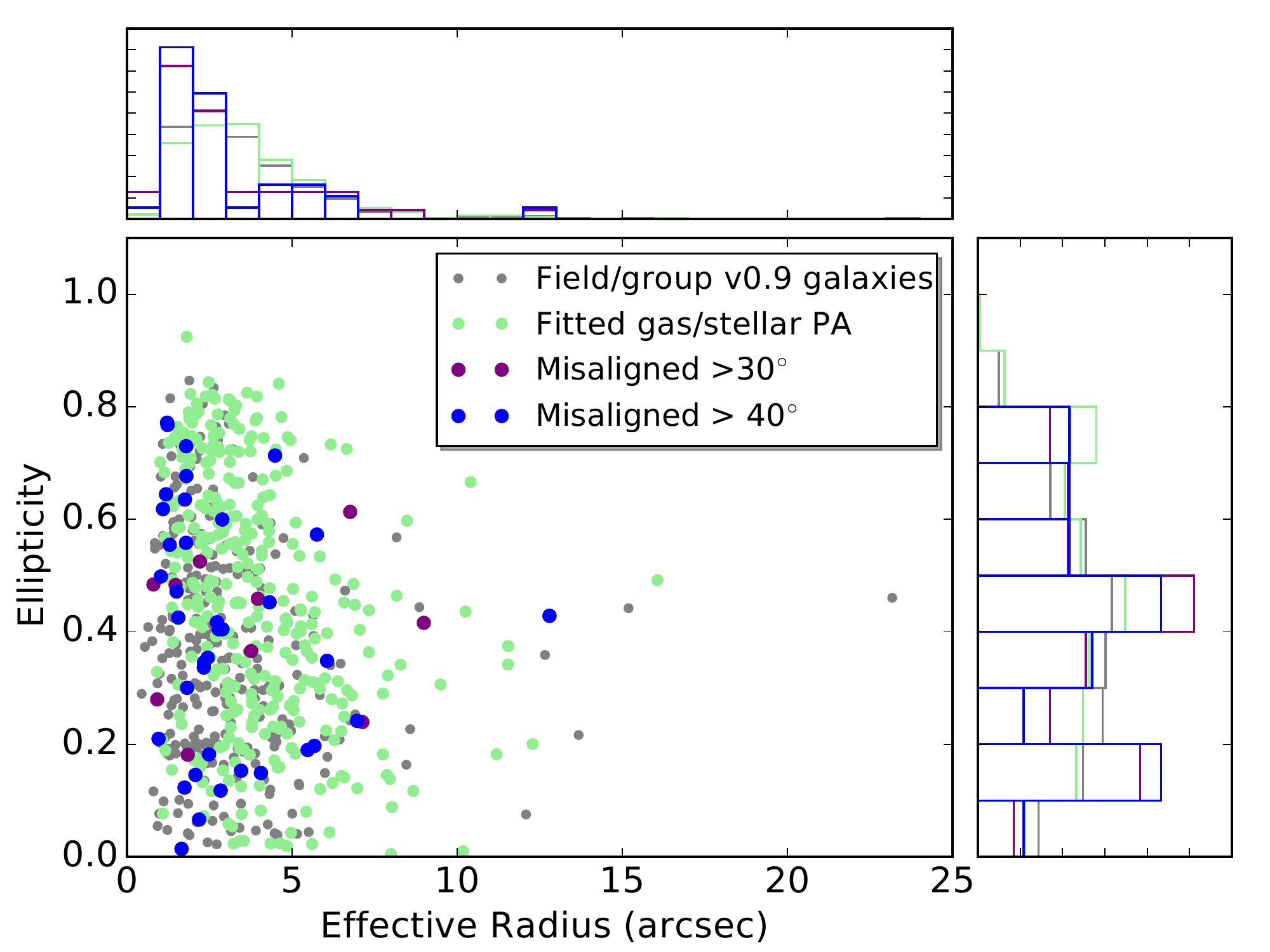}
\end{minipage}%
\caption{Ellipticity versus effective radius ($R_{e}$) for field/group sample galaxies in the v0.9 data release (grey), those for which the gas and stellar PAs could be fitted (light green), misaligned by $>30^{\circ}$ (purple) and misaligned by $>40^{\circ}$ (blue). 
}
\label{Re_ellip}
\end{figure}

Statistical tests were run on stellar mass and redshift (selection space for the SAMI survey) in order to test if the method for measuring PA offsets results in any bias in the population that can be fitted in both stellar and gas PAs, compared to the full v0.9 data release. The distributions of fitted and un-fittted galaxies are statistically different in the field/group sample in both redshift (KS test statistic=0.12, p-value=$1.9\times10^{-4}$) and stellar mass (KS test statistic=0.18, p-value=$5.7\times10^{-9}$) as shown in Fig.~\ref{zvsMstar}. There is a slight bias against fitting both PAs in the lowest stellar mass and redshift galaxies since the noise is higher in the stellar continuum for those galaxies \citep[see][]{vdS17}.  There is no such bias for cluster galaxies because they are selected to have stellar masses greater than $10^{9.5}$\Msol.
In addition biases in ellipticity and $R_{e}$ distributions were checked.  
Fig.~\ref{Re_ellip} shows that the distribution of galaxies for which both gas and stellar PAs could be measured (light green), is not representative of the whole v0.9 sample (grey) in ellipticity (KS test statistic=0.096, p-value=0.003) or in $R_{e}$ (KS test statistic=0.25, p-value=$2.5\times10^{-25}$). This is because there is a bias against measuring PAs for the lowest ellipticity galaxies, which are likely to be early-types and can have very little gas, and a bias against measuring PAs for the lowest  $R_{e}$ galaxies which are typically the low stellar mass dwarfs in which the stellar continuum has lower S/N.

\begin{table*}
\begin{center}
\caption{Statistics on the numbers and fractions of galaxies in the field/group sample and cluster samples that are misaligned by different PA offset ranges. }
\label{Misalign_stats}
\begin{tabular}{lccccccccc}
\hline
PA  & \multicolumn{6}{c}{GAMA regions}  & & \multicolumn{2}{c}{Cluster regions} \\ 
\cline{2-7}
\cline{9-10}
offset  & \multicolumn{2}{c}{All} & \multicolumn{2}{c}{Field} & \multicolumn{2}{c}{Groups} & & \multicolumn{2}{c}{} \\ 
 range & Number & Fraction & Number & Fraction & Number & Fraction & & Number & Fraction \\ 
\hline
$0-180^{\circ}$ & 486 & 1.00 & 192 & 1.00 & 294 & 1.00 & &136 & 1.00\\ 
 $>30^{\circ}$ & 55 & 0.11$\pm$0.01 & 14 & 0.07$\pm$0.02 & 41 & 0.14$\pm$0.02 & &15 & 0.11$\pm$0.03\\ 
 $>40^{\circ}$ & 42 & 0.09$\pm$0.01 & 11 & 0.06$\pm$0.02 & 31 & 0.11$\pm$0.02 & &9 & 0.07$\pm$0.02\\ 
 $30-150^{\circ}$ & 38 & 0.08$\pm$0.01 & 12 & 0.06$\pm$0.02 & 26 & 0.09$\pm$0.02 & &11 & 0.08$\pm$0.02\\ 
 $40-140^{\circ}$ & 24 & 0.05$\pm$0.01 & 8 & 0.04$\pm$0.01 & 16 & 0.05$\pm$0.01 & &5 & 0.04$\pm$0.02\\ 
 $>140^{\circ}$ & 18 & 0.04$\pm$0.01 & 3 & 0.02$\pm$0.01 & 15 & 0.05$\pm$0.01 & &4 & 0.03$\pm$0.01\\ 
 $>150^{\circ}$ & 17 & 0.03$\pm$0.01 & 2 & 0.01$\pm$0.01 & 15 & 0.05$\pm$0.01 & &4 & 0.03$\pm$0.01\\ 
 \hline
\end{tabular}

\end{center}
\end{table*}

\begin{table*}
\begin{center}
\caption{Statistics on the numbers and fractions of galaxies in the field/group sample (GAMA) and cluster samples that are misaligned by different PA offsets in morphological groups (E=Ellipticals, ESpirals=Early Spirals, LSpirals=Late spirals, NA=No agreement). }
\label{Misalign_stats_morph}
\begin{tabular}{lccccccccc}
\hline 
Description & E & E - S0 & S0 & S0 - ESpirals & ESpirals & ESpirals - LSpirals & LSpirals & Unknown & NA\\
\hline
All GAMA & 17 & 16 & 29 & 48 & 102 & 57 & 202 & 3 & 12\\
 GAMA $>30^{\circ}$ & 9 & 8 & 11 & 5 & 4 & 2 & 12 & 0 & 3\\
 FracGAMA $>30^{\circ}$  & 0.53 & 0.50 & 0.38 & 0.10 & 0.04 & 0.04 & 0.06 & 0.00 & 0.25\\
 GAMA $>40^{\circ}$  & 8 & 7 & 9 & 4 & 3 & 2 & 6 & 0 & 2\\
 FracGAMA $>40^{\circ}$  & 0.47 & 0.44 & 0.31 & 0.08 & 0.03 & 0.04 & 0.03 & 0.00 & 0.17\\
 GAMA $30-150^{\circ}$  & 6 & 5 & 9 & 2 & 2 & 1 & 11 & 0 & 2\\
 FracGAMA $30-150^{\circ}$  & 0.35 & 0.31 & 0.31 & 0.04 & 0.02 & 0.02 & 0.05 & 0.00 & 0.17\\
 GAMA $40-140^{\circ}$  & 5 & 4 & 7 & 1 & 1 & 1 & 4 & 0 & 1\\
 FracGAMA $40-140^{\circ}$  & 0.29 & 0.25 & 0.24 & 0.02 & 0.01 & 0.02 & 0.02 & 0.00 & 0.08\\
 \hline
AllClusters  & 19 & 10 & 16 & 21 & 31 & 15 & 19 & 0 & 1\\
 Clus $>30^{\circ}$  & 6 & 1 & 0 & 3 & 2 & 2 & 0 & 0 & 0\\
 FracClus $>30^{\circ}$  & 0.32 & 0.10 & 0.00 & 0.14 & 0.06 & 0.13 & 0.00 & 0.00 & 0.00\\
 Clus $>40^{\circ}$  & 3 & 1 & 0 & 2 & 0 & 2 & 0 & 0 & 0\\
 FracClus $>40^{\circ}$  & 0.16 & 0.10 & 0.00 & 0.10 & 0.00 & 0.13 & 0.00 & 0.00 & 0.00\\
 Clus $30-150^{\circ}$  & 5 & 1 & 0 & 2 & 2 & 1 & 0 & 0 & 0\\
 FracClus $30-150^{\circ}$  & 0.26 & 0.10 & 0.00 & 0.10 & 0.06 & 0.07 & 0.00 & 0.00 & 0.00\\
 Clus $40-140^{\circ}$  & 2 & 1 & 0 & 1 & 0 & 1 & 0 & 0 & 0\\
 FracClus $40-140^{\circ}$  & 0.11 & 0.10 & 0.00 & 0.05 & 0.00 & 0.07 & 0.00 & 0.00 & 0.00\\
 \hline
All GAMA + Clusters  & 36 & 26 & 45 & 69 & 133 & 72 & 221 & 3 & 13\\
 All $>30^{\circ}$  & 15 & 9 & 11 & 8 & 6 & 4 & 12 & 0 & 3\\
 FracAll $>30^{\circ}$  & 0.42 & 0.35 & 0.24 & 0.12 & 0.05 & 0.06 & 0.05 & 0.00 & 0.23\\
 All $>40^{\circ}$  & 11 & 8 & 9 & 6 & 3 & 4 & 6 & 0 & 2\\
 FracAll $>40^{\circ}$  & 0.31 & 0.31 & 0.20 & 0.09 & 0.02 & 0.06 & 0.03 & 0.00 & 0.15\\
 All $30-150^{\circ}$  & 11 & 6 & 9 & 4 & 4 & 2 & 11 & 0 & 2\\
 FracAll $30-150^{\circ}$  & 0.31 & 0.23 & 0.20 & 0.06 & 0.03 & 0.03 & 0.05 & 0.00 & 0.15\\
 All $40-140^{\circ}$  & 7 & 5 & 7 & 2 & 1 & 2 & 4 & 0 & 1\\
 FracAll $40-140^{\circ}$  & 0.19 & 0.19 & 0.16 & 0.03 & 0.01 & 0.03 & 0.02 & 0.00 & 0.08\\
 \hline

\end{tabular}
\end{center}
\end{table*}

\subsubsection{Definition of misalignment angle}
\label{nature}

For direct comparison to the literature \citep[e.g.][]{Dav16, Lag15} we nominally assume that galaxies with PA offsets less than $30^{\circ}$ are aligned, but note that this cut-off is somewhat arbitrary. This takes into account the following errors: Firstly, the PA measurements for aligned galaxies with strong stellar continuum and H$\alpha$ flux is typically $\pm4.9^{\circ}$. However, galaxies with less gas or lower continuum S/N can have much larger errors. Secondly, since the PAs are projected values, the correct 3-dimensional misalignment will be further affected by orientation, resulting in an uncertainty in the PA offset that is larger than the measurement fitting error. If aligned galaxies are taken to have PA offsets less than $30^{\circ}$ then galaxies with a PA offset within $30^{\circ}$ of $180^{\circ}$ ($150-180^{\circ}$) could be considered to be counter-rotating. 
However, the histogram of PA offsets (e.g. Figs ~\ref{misalign_hist} and ~\ref{PAhistMorph} later) clearly does not have a break or truncation at $30^{\circ}$, and instead the tail of the aligned distribution of galaxies extends beyond $30^{\circ}$. 

Careful inspection of galaxies with PA offsets between $30-40^{\circ}$ shows that some appear genuinely misaligned while for some, the errors on the fits mean that they are also in agreement with a PA offset below $30^{\circ}$. This is particularly apparent in the cluster sample (shown later in Fig.~\ref{cluster_radius}; top left) where there is a higher error in the gas PA fits for galaxies closer to the centre of clusters. Such galaxies have less gas and become increasingly difficult to fit, leading to a trend where the `aligned' population is below $\sim30^{\circ}$ at higher cluster radii but scatters up above $\sim30^{\circ}$ towards lower cluster radius. Therefore, by {\it also} considering a PA offset cut-off of $40^{\circ}$ for all galaxies (field/group and clusters) the impact of this contamination can be accounted for. The statistics are then less impacted by galaxies that are scattered up to higher offsets by these effects.   

It is clear that there is not a single value for stellar and gas misalignment that can separate physical reasons for misaligned gas and therefore statistics are presented for misaligned galaxies with both $30^{\circ}$ and $40^{\circ}$ definitions. 
The exact values of these cut-offs do not impact the main conclusions of this paper in most cases, exceptions are specifically discussed.

The statistics for aligned, misaligned and counter-rotating galaxies in the field, groups and cluster samples are given in Table~\ref{Misalign_stats}.

\subsubsection{Morphological classification}
\label{morph_intro}

All galaxies within the SAMI field/group sample and cluster regions have been morphologically classified by eye in batches, by a group of 8-12 SAMI team members. The classifications are 
0=ellipticals (no disk), 1= S0 (with disk), 2= Early spirals (with bulge; Sa-Sb), 3= Late spirals (no bulge), 5 and 6 are for galaxies where the classifying team did not agree or where they were undetermined. If 66\% of the classifying team members select the same morphological class then it was set. If not, then adjacent votes are combined into intermediate classes of 0.5=E/S0, 1.5=S0/Early-type spiral, 2.5=Early/Late-type spiral, and if the 66\% threshold is still not reached they are classified as 6= ``no agreement". Full details of the SAMI visual morphological classification can be found in \citet{Cor16}. 

The field/group galaxies were classified based on 3-colour SDSS imaging while 3-colour VST imaging was used to classify the cluster morphologies. The consistency between morphological classifications using these two different imaging sets was checked with a subsample of $\sim 15$ galaxies that had both SDSS and VST imaging. The classifications of those galaxies from SDSS and VST had 100\% agreement. Therefore we do not expect any strong bias in the morphological classification between the field/group galaxies and the cluster galaxies.

Table~\ref{Misalign_stats_morph} lists the numbers of galaxies and fractions misaligned in each morphological class.

\section{Results}
\label{Results}

Fig.~\ref{misalign_hist} shows the distribution of the difference in the fitted stellar and ionised gas PAs for both the field/group galaxies and the clusters. For initial comparison with literature samples, here we show misaligned galaxies as those with a PA offset of greater than $30^{\circ}$. Of the 486 field/group galaxies for which both stellar and gas rotation PAs could be measured, 55 or $11.3\pm1.4$\% are misaligned by more than $30^{\circ}$. For the 8 clusters, 15 out of 136 or $11.0\pm2.7$\% of galaxies are misaligned. 
While these misaligned fractions are similar in both field/groups and clusters, this is no longer the case when morphology is taken into account.

\subsection{Morphology, S\'{e}rsic index and misalignment}

\begin{figure*}
\centering
\includegraphics[width=14.0cm]{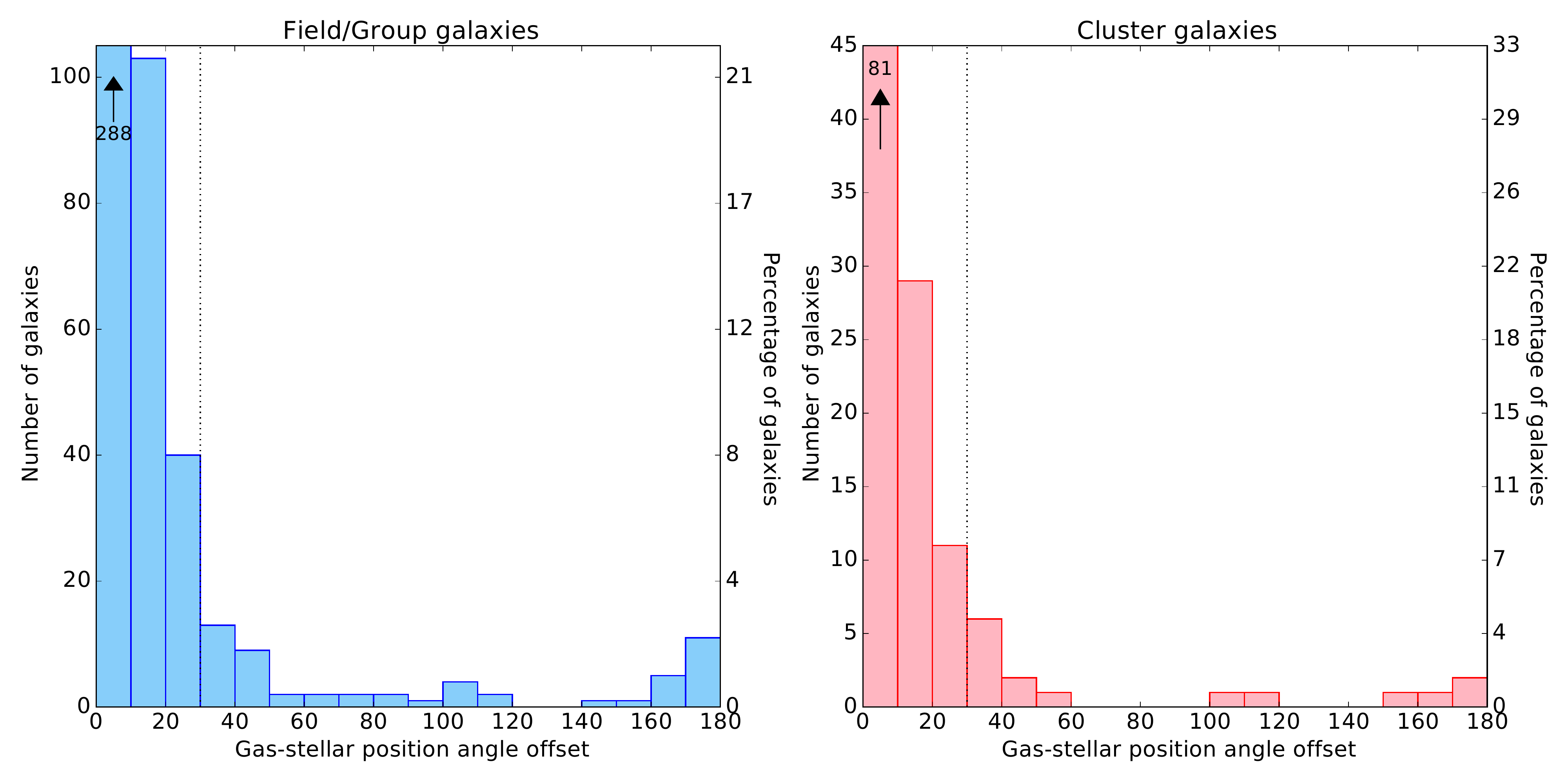}
\vspace*{3mm}
\caption{Distribution of PA offsets between the rotation axis of the stars and ionised gas for the field/group sample (left) and the combined sample of 8 clusters (right). The field/group galaxies have $11\pm1$\% (55/486) misaligned $>30^{\circ}$ (dotted line) while the clusters have $11\pm3$\% (15/136) misaligned. 
}
\label{misalign_hist}
\end{figure*}

\begin{figure*}
\centering
\includegraphics[angle=270, width=14.0cm]{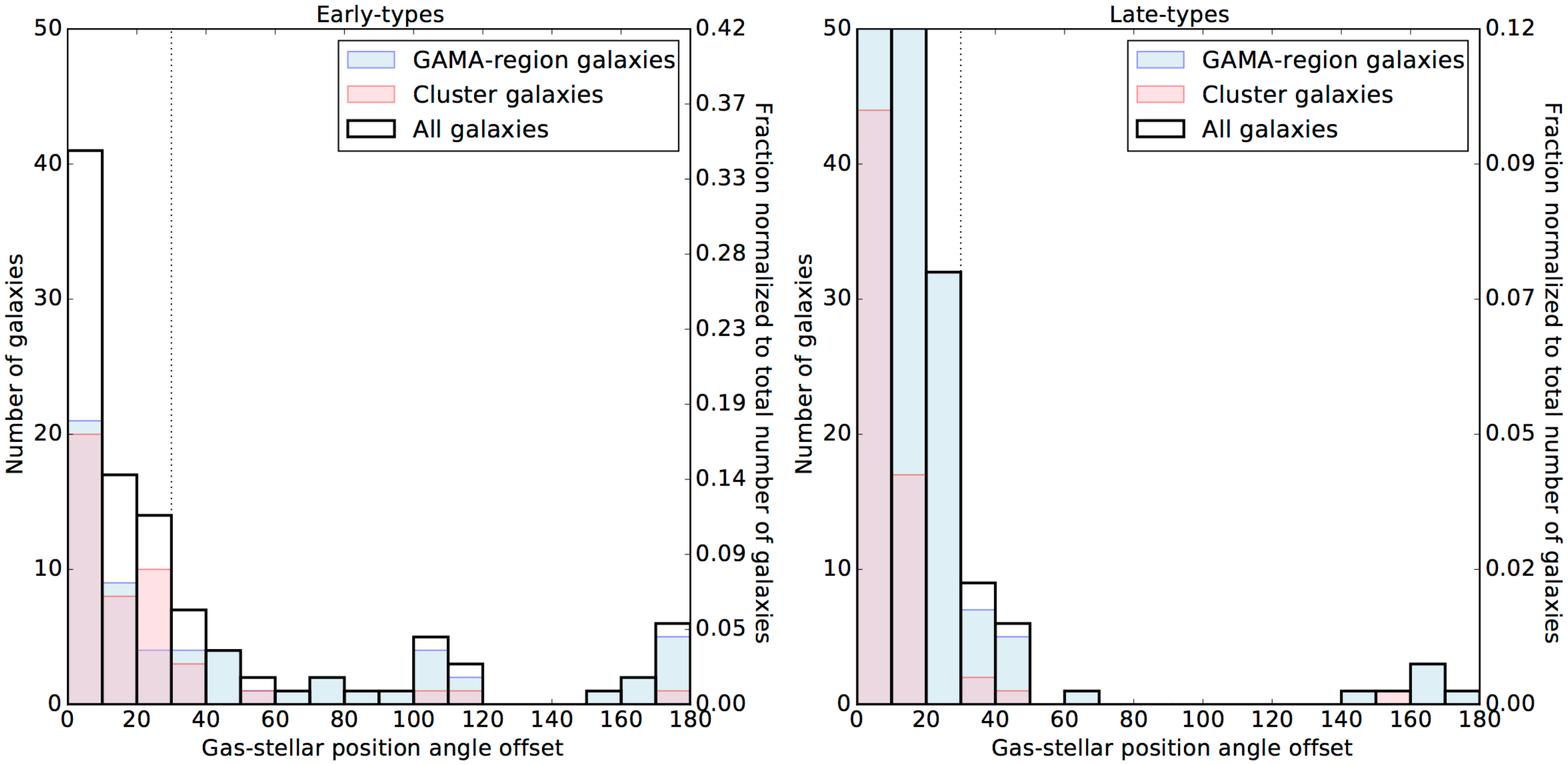}
\vspace*{3mm}
\caption{Distribution of PA offsets between the rotation axis of the stars and ionised gas for the field/group (blue), cluster (red) and combined (black) samples separated into ETGs (left; morphology index from Fig.~\ref{morphtype} $<1.5$) and LTGs (right; index $>1.5$). 
}
\label{PAhistMorph}
\end{figure*}

The PA offset distribution from Fig.~\ref{misalign_hist} was divided into 
an ETG (with morphological classification $<1.5$; including E to S0 galaxies) and LTG (with morphological classification $>1.5$; including early-spirals to late spiral galaxies) sample as shown in Fig.~\ref{PAhistMorph}.  The blended category with morphological classification 1.5, in between S0 and early spirals, has intentionally not been included to make a cleaner separation between LTGs and ETGs.  E and S0 galaxies are not separated at this point because this paper investigates how rotating gas is accreted and dynamically processed within galaxies, and the galaxies for which gas and stellar rotation can be measured, have a stellar disk. Therefore, it is important to note that {\it E-type galaxies without such a disk are not included in our sample}.

\begin{figure*}
\begin{minipage}[]{0.3\textwidth}
\includegraphics[width=6.0cm]{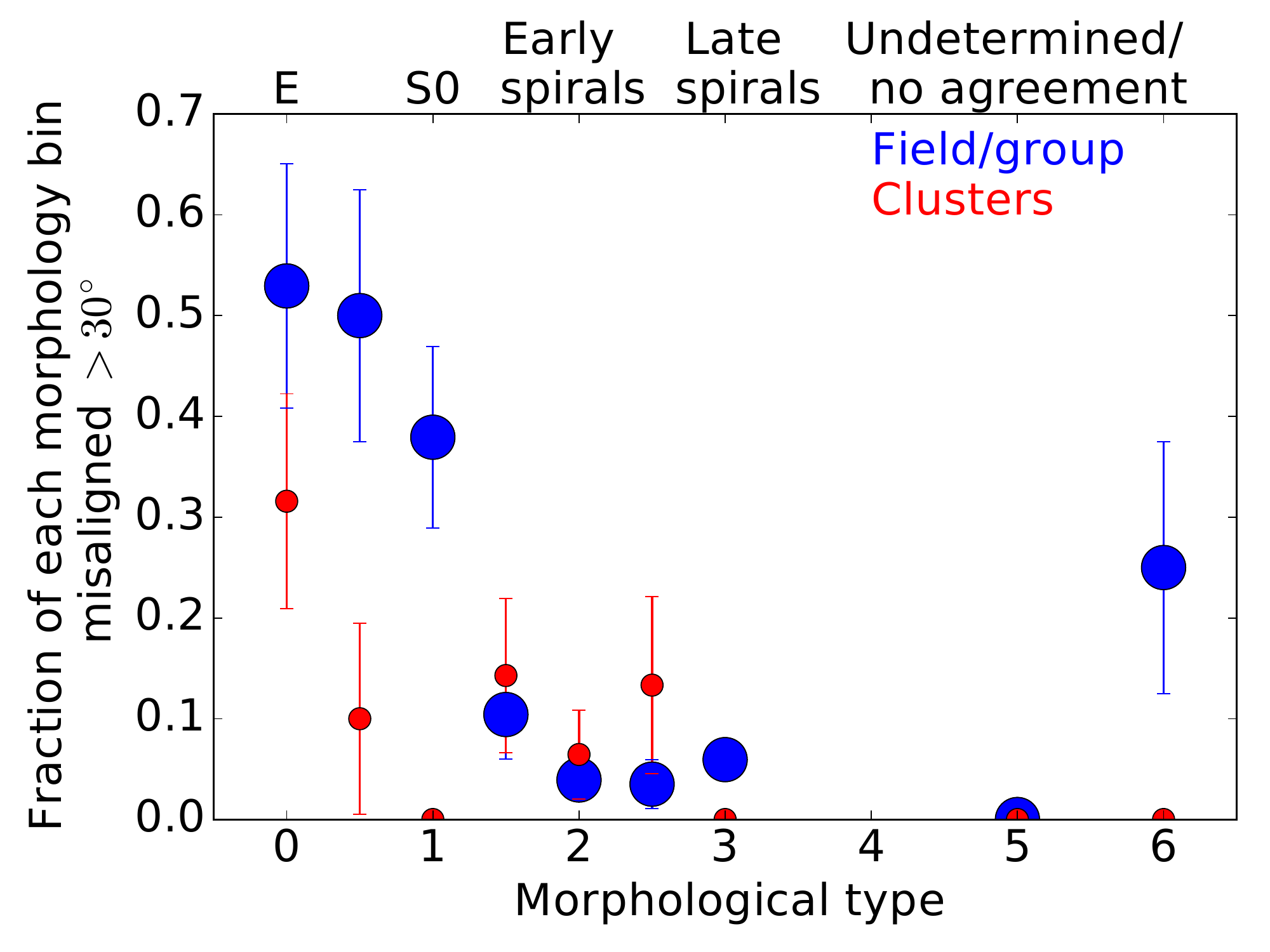}
\end{minipage}%
\hspace*{5mm}
\begin{minipage}[]{0.3\textwidth}
\includegraphics[width=6.0cm]{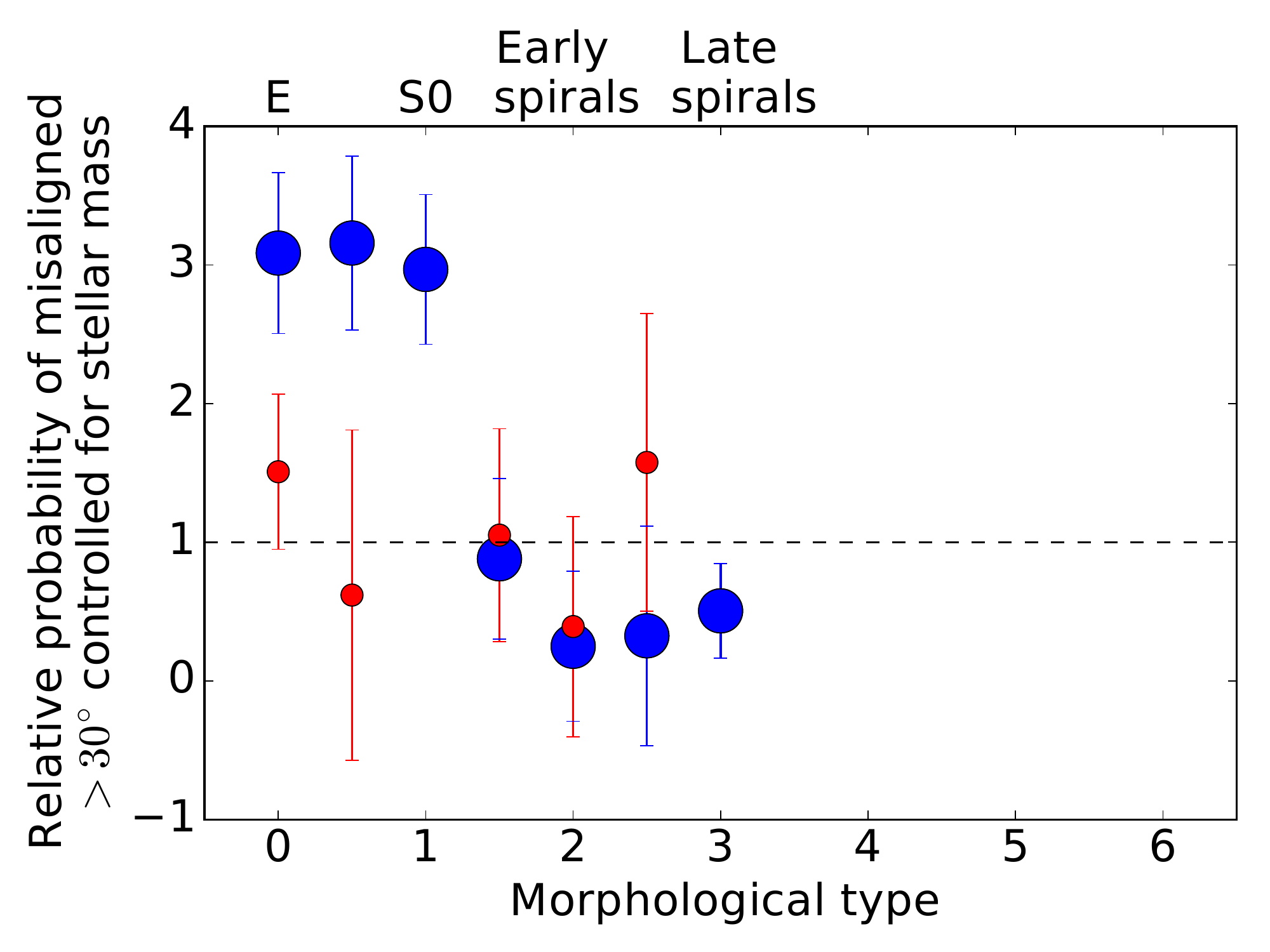}
\end{minipage}%
\vspace*{0.1mm}
\begin{minipage}[]{0.3\textwidth}
\includegraphics[width=6.0cm]{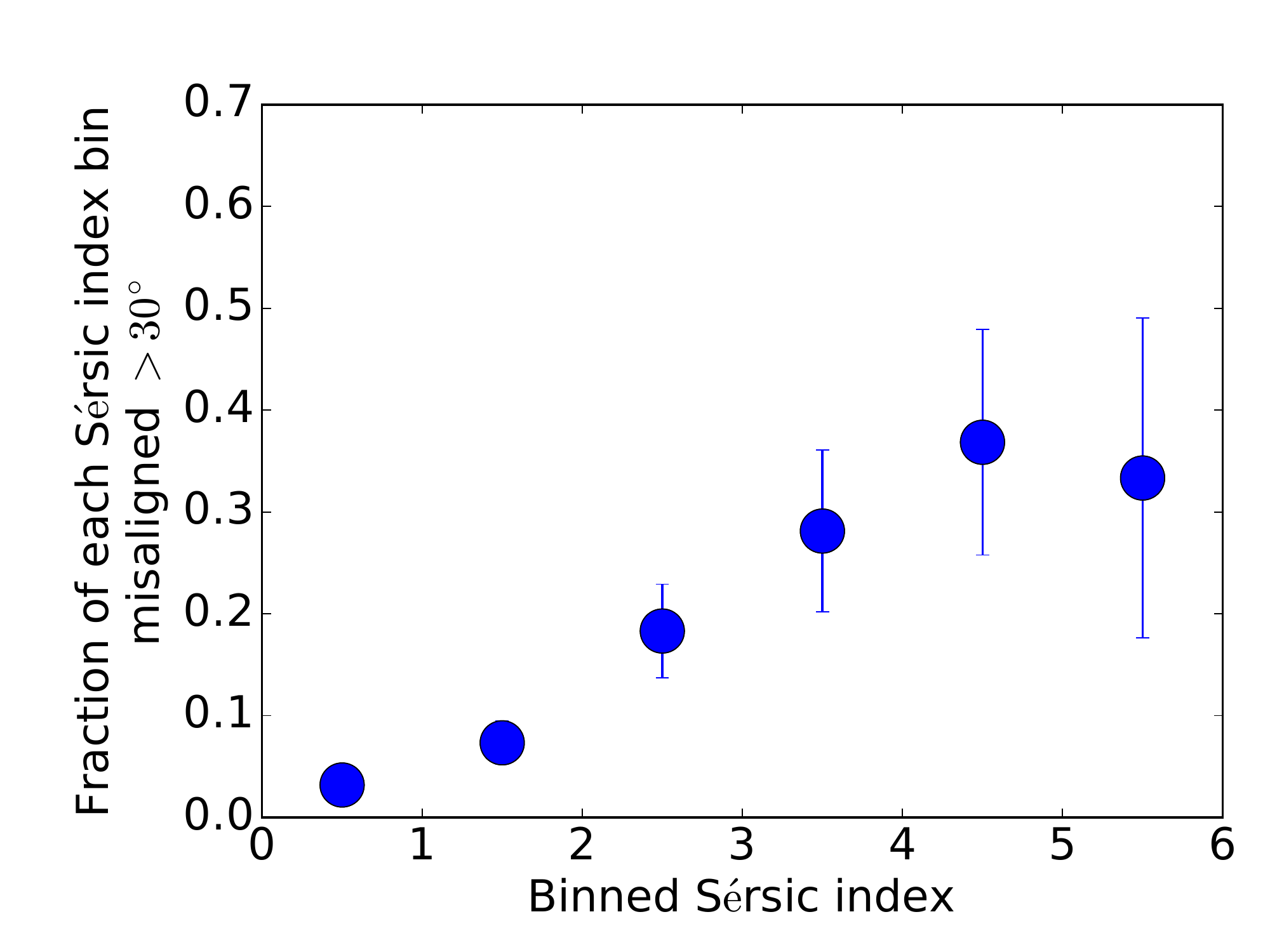}
\end{minipage}%
\hspace*{5mm}
\begin{minipage}[]{0.3\textwidth}
\includegraphics[width=6.0cm]{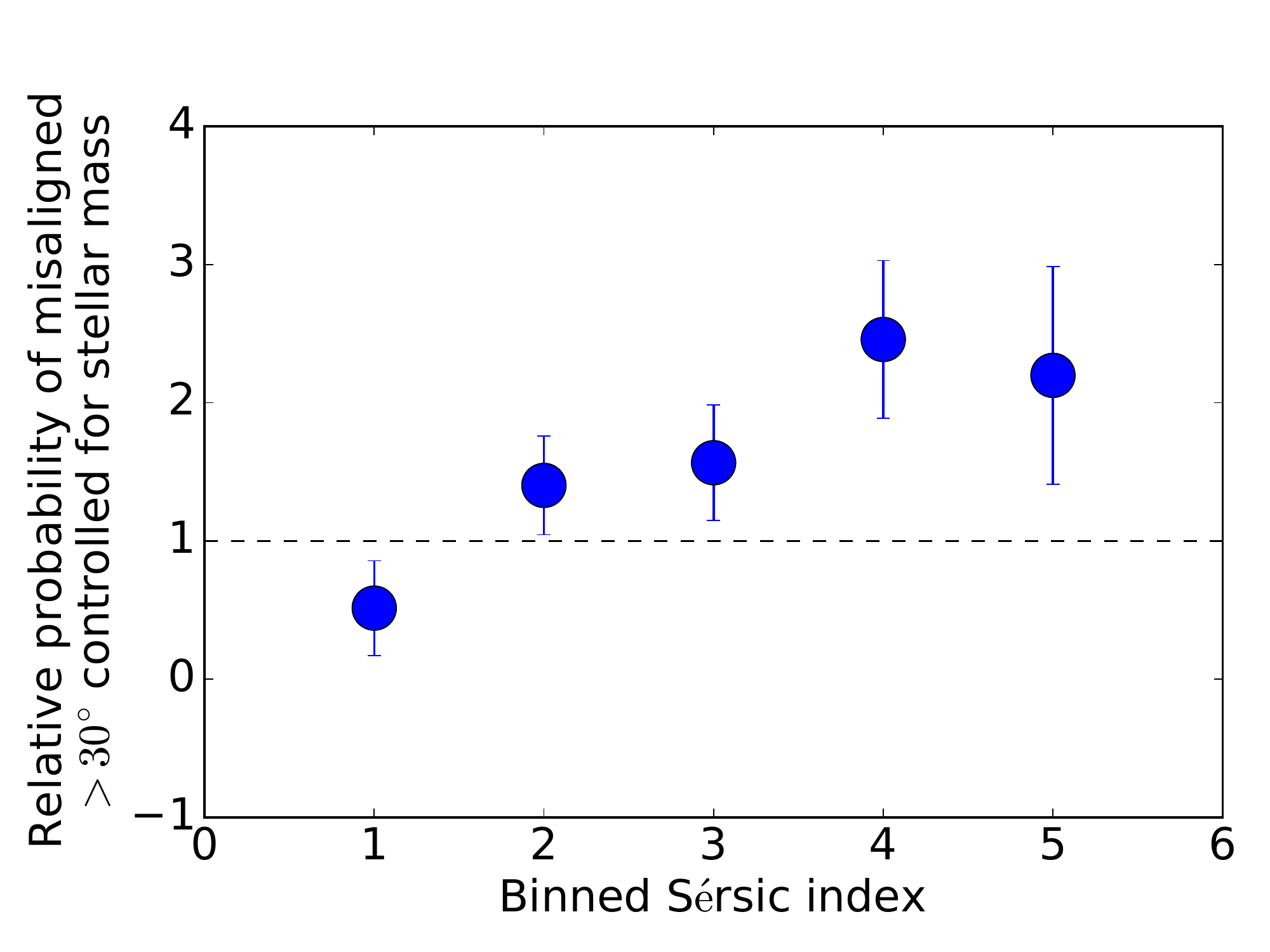}
\end{minipage}%
\vspace*{0.1mm}
\begin{minipage}[]{0.3\textwidth}
\includegraphics[width=6.0cm]{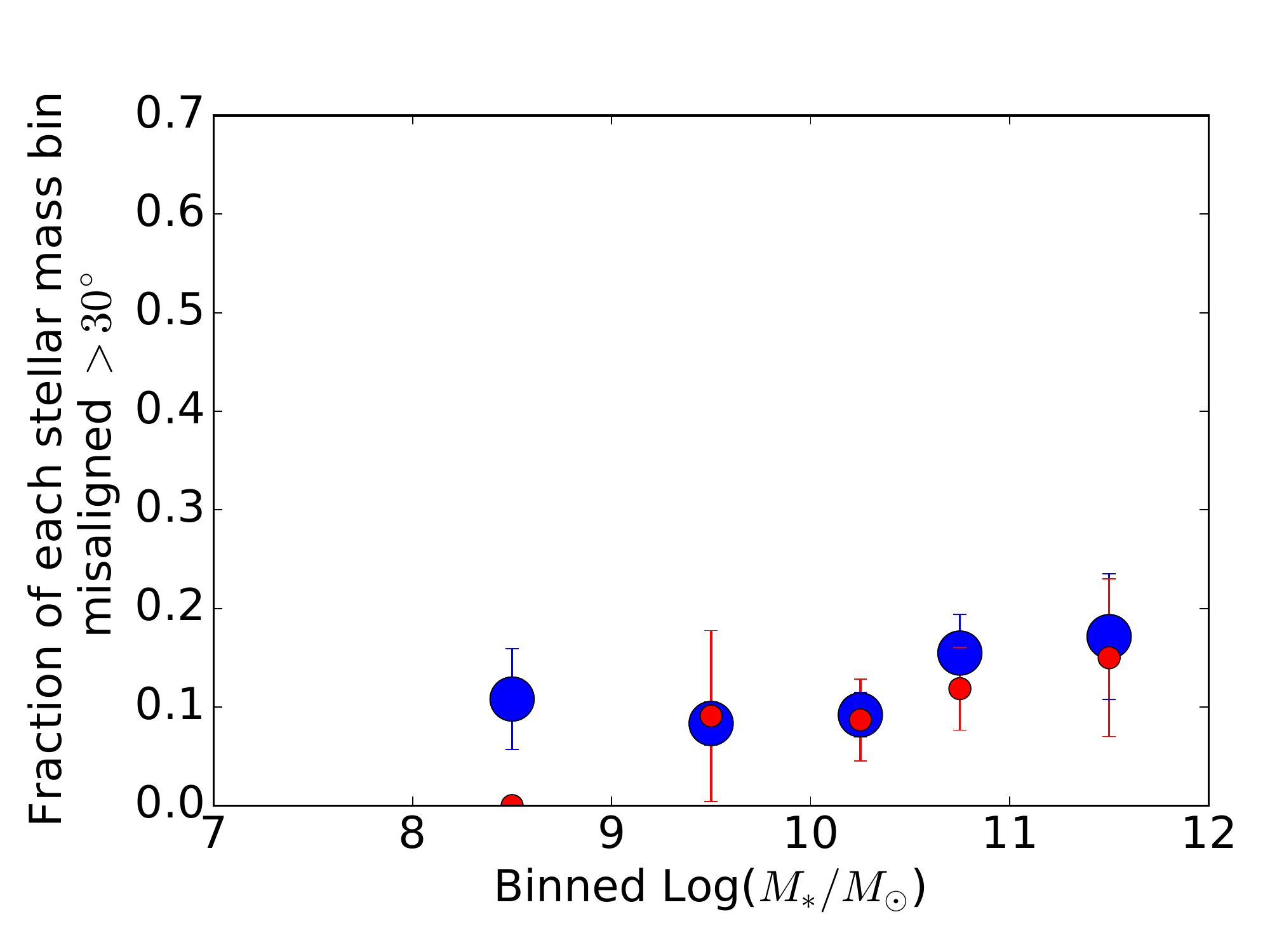}
\end{minipage}%
\hspace*{5mm}
\begin{minipage}[]{0.3\textwidth}
\hspace{5.3cm}
\end{minipage}%
\vspace*{0.1mm}
\begin{minipage}[]{0.3\textwidth}
\includegraphics[width=6.0cm]{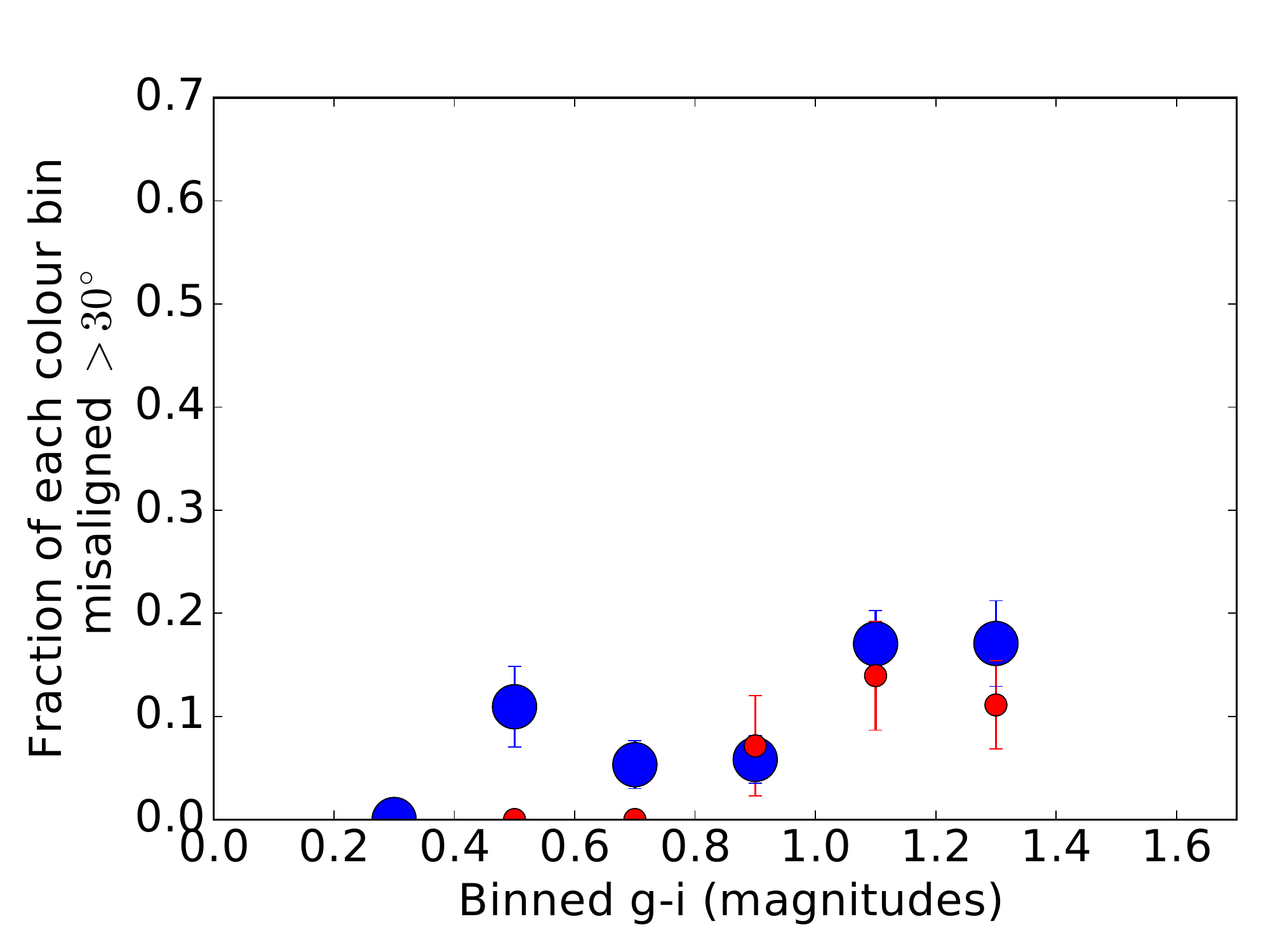}
\end{minipage}%
\hspace*{5mm}
\begin{minipage}[]{0.3\textwidth}
\includegraphics[width=6.0cm]{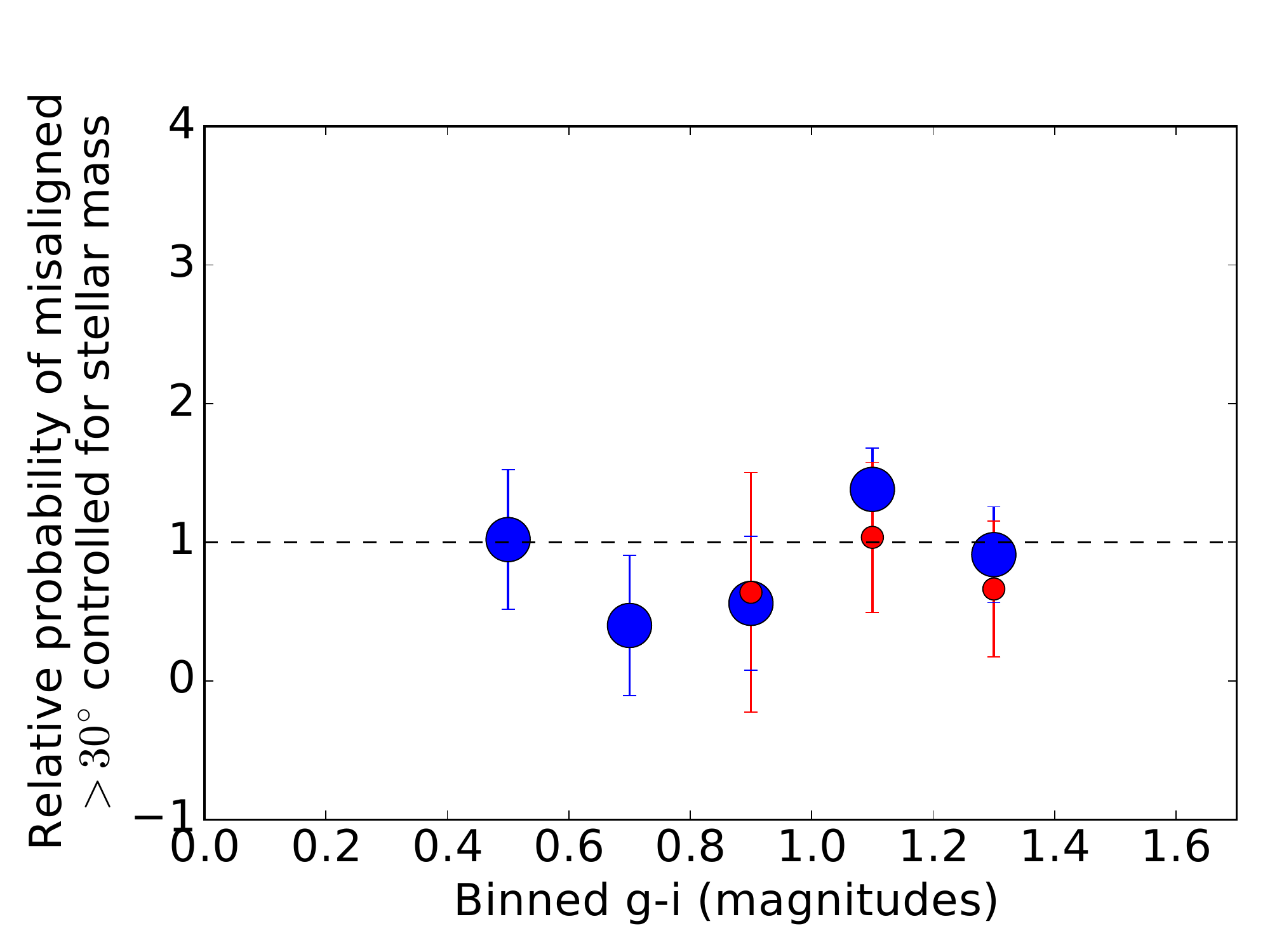}
\end{minipage}%
\vspace*{1mm}
\caption{Fraction of galaxies that are misaligned by greater than $30^{\circ}$ (left column) in bins of morphology (top row), S\'{e}rsic index (2nd row), stellar mass (3rd row) and g-i colour (4th row). 
Morphological types are 0= Elliptical, 1= S0, 2= Early spirals 3=Late spirals, 5 and 6 are galaxies that were undetermined due to complex structure or a lack of consensus among the classification team members (see Section~\ref{morph_intro}). Blended categories are represented between these main bin values (e.g. E/S0 = 0.5). Blue points refer to the field/group galaxies while cluster galaxies are in red. S\'{e}rsic indices are not available for the cluster galaxies.
The largest change in misalignment fraction is with morphology. The changes in misalignment fraction with galaxy colour and stellar mass are very much less. The same trends are seen for both a $>30^{\circ}$  and $>40^{\circ}$ misalignment definition (shown in Appendix 1). The right column is the result of controlling the plots in the left column for stellar mass to account for the dependence of morphology, S\'{e}rsic index and colour on stellar mass. A value above $1.0$ in the controlled plots indicates how much more likely that bin is to be misaligned compared to other galaxies in the same stellar mass range. Therefore, if the dependence is stellar mass driven, then the stellar mass controlled plots would have all points at 1.0. The morphology and S\'{e}rsic index plots instead show that the ETGs are up to $\sim 3$ times more likely to be misaligned (LTGs are less likely to be misaligned) than galaxies at the same stellar mass, confirming that the misalignment dependence on morphology and S\'{e}rsic index in the first column is stronger than the stellar mass dependence. The galaxies in the higher mass bins are a mix of ETGs and LTGs as shown in Fig~\ref{Mstarmorph} (e.g. 54\% of the galaxies in the highest stellar mass bin are ETGs) and therefore the stellar mass plot is not expected to follow the ETG plot.
}
\label{morphtypehist}
\end{figure*}

In Fig.~\ref{morphtypehist}, the fraction of galaxies with PA offset more than $30^{\circ}$ ($40^{\circ}$ version is shown in Appendix 1) is separated by morphological type (as defined in Section~\ref{morph_intro}), S\'{e}rsic index, stellar mass and $g-i$ colour. Misalignment fractions ($>30^{\circ}$) have a strong dependence on morphological type and S\'{e}rsic index, much more so than with stellar mass or $g-i$ colour (as discussed further in Sections~\ref{Mstar} and~\ref{MorphMisalign}). The fraction of field/group galaxies misaligned by $>30^{\circ}$ is 53, 50 and 38\% in the three morphological types from E to S0 (in Fig.~\ref{morphtypehist}) or  $45\pm6$\% for all ETGs in these three categories combined. This is consistent with the average 42\% found for fast-rotating early-types in ATLAS$^{3D}$ \citep{Dav11}. On the other hand, the LTGs (in the three morphological types from early spirals to late spirals) in the SAMI sample have only 4, 3.5 and 6\% of galaxies misaligned, bringing the overall misalignment fraction across all morphologies down to $11\pm1$\% (for PA offset$>30^{\circ}$). 

It is particularly notable that both morphology and S\'{e}rsic index show a striking trend with misalignment. Higher S\'{e}rsic indices and ETGs have higher misalignment fractions. While ETGs and LTGs can have an overlapping range of S\'{e}rsic indices, both morphology and S\'{e}rsic index track the underlying shape of the galaxy stellar mass distribution. In this paper we address the impact of that distribution on the accretion and dynamical processing of gas. The S\'{e}rsic  index trends agree with that from the morphological classifications. However, his work focusses on the galaxy morphology rather than S\'{e}rsic index firstly because S\'{e}rsic indices have not yet been accurately fit for the cluster sample, and secondly, 105 of our galaxies are flagged as bad fits in the GAMA S\'{e}rsic catalogue for the field/groups region and can not be used. 

While overall the misalignment fraction in the clusters is the same ($11\pm3$\%) as in the field/group environments,  the misalignment fraction for ETGs is vastly different in the clusters from the field/groups. The misalignment fraction in the cluster ETGs is $16\pm5$\% which agrees within errors with the fraction found by ATLAS$^{3D}$ for ETGs in the Virgo cluster (2 of the 20 or 10+/-7\%). 
Section~\ref{clusters} will discuss how measured misalignment in cluster galaxies can be attributed to different physical processes compared to field/group galaxies and therefore a direct comparison of these misalignment ratios can be misleading.

\subsection{Stellar mass and misalignment}
 \label{Mstar}

\begin{figure*}
\centering
\begin{minipage}[]{0.5\textwidth}
\includegraphics[width=9.0cm]{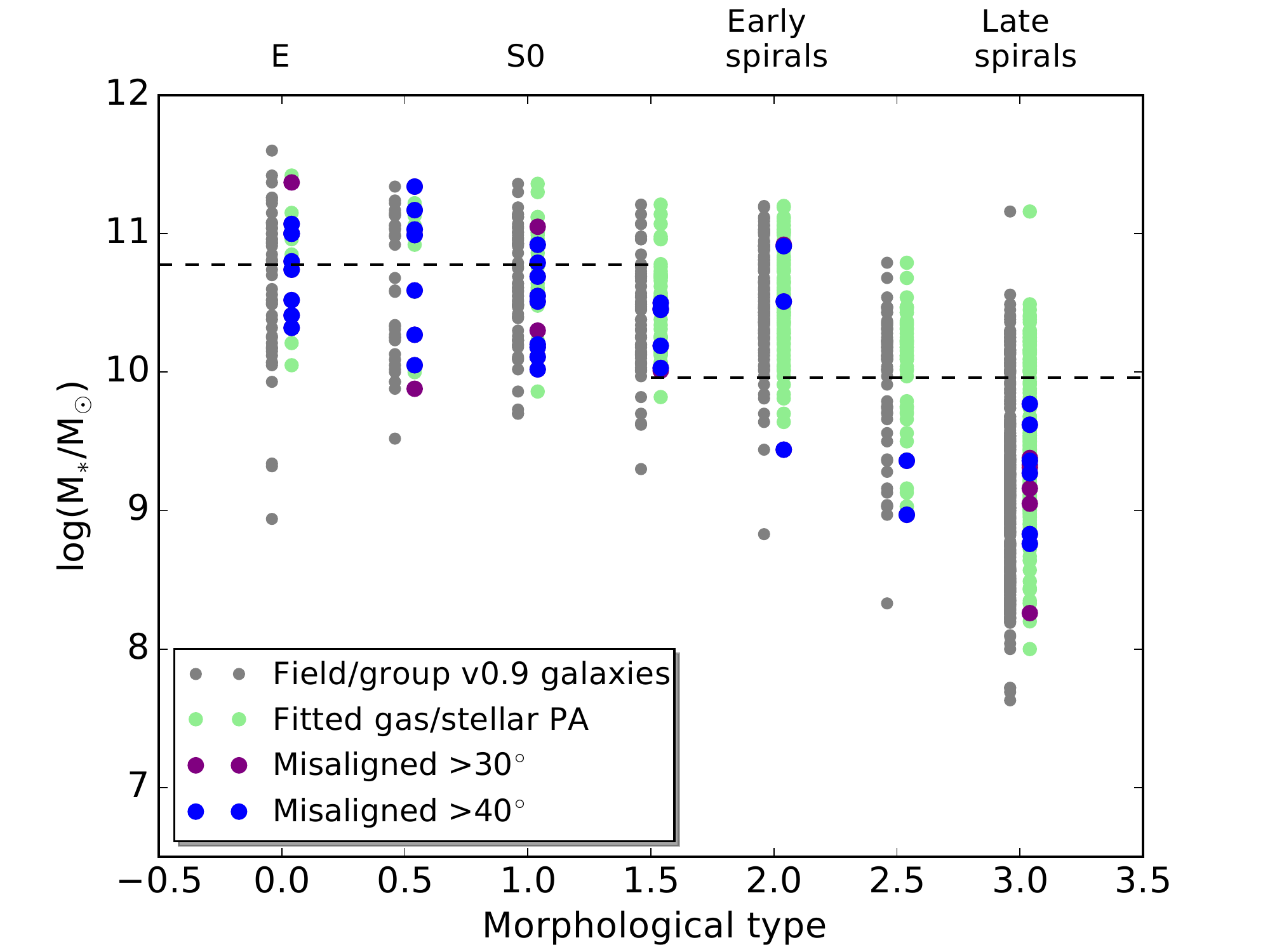}
\end{minipage}%
\begin{minipage}[]{0.5\textwidth}
\includegraphics[width=9.0cm]{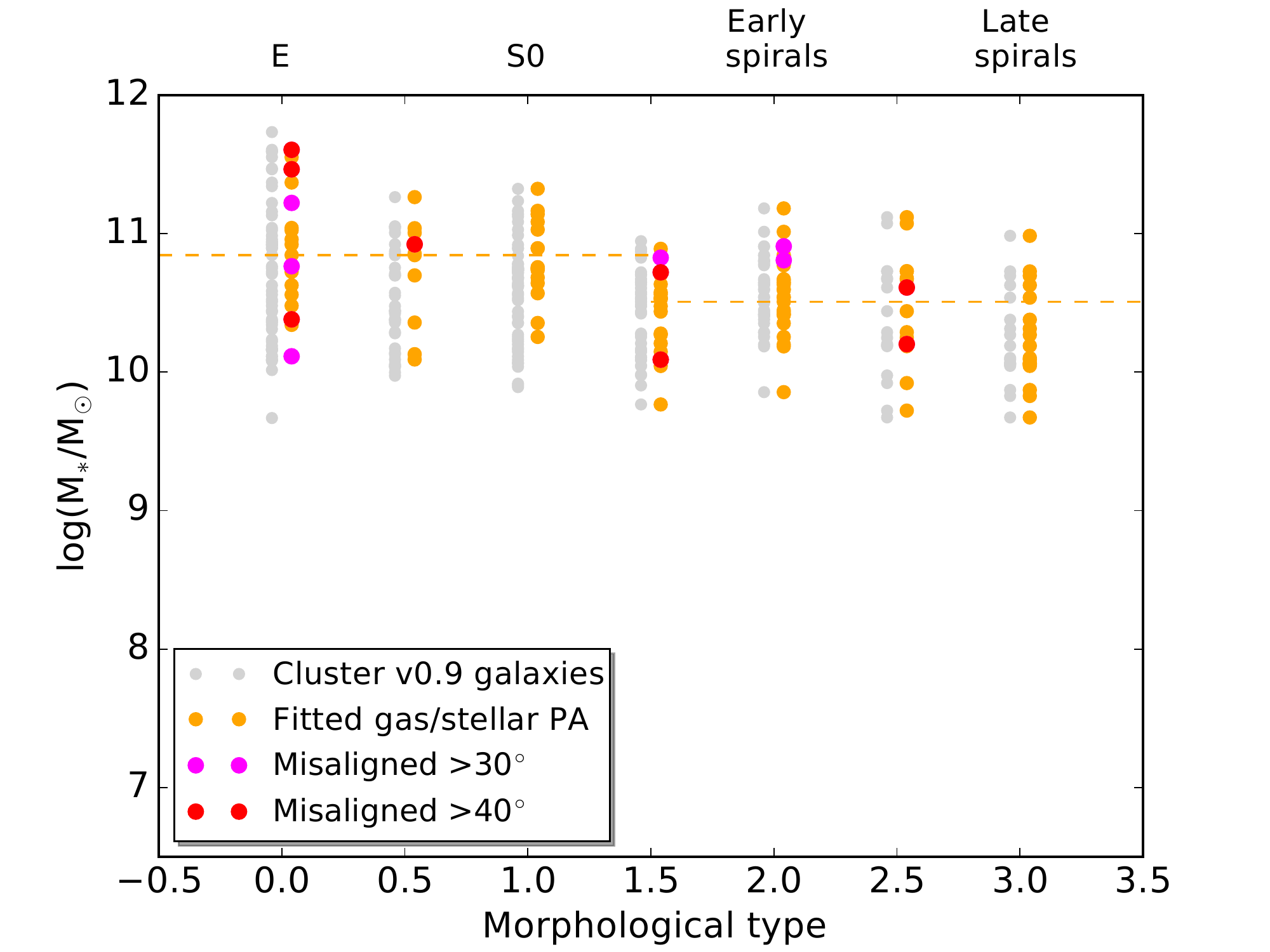}
\end{minipage}%
\vspace*{3mm}
\caption{Stellar mass versus morphology for the field/group (left) and cluster (right) galaxies. Colours are consistent with Fig.~\ref{zvsMstar}; the full v0.9 sample: grey, the galaxies for which both stellar and gas PAs could be fitted (field/group sample: light green; clusters: orange), misaligned (PA$>30^{\circ}$) galaxies (field/group sample: purple; clusters: magenta), misaligned (PA$>40^{\circ}$) galaxies (field/group sample: blue; clusters: red). The grey and orange dashed lines show the median stellar mass for field/group and cluster galaxies respectively, for the ETGs (morphological type $< 1.5$) and LTGs (morphological type $> 1.5$). 
}
\label{Mstarmorph}
\end{figure*}

Fig.~\ref{morphtypehist} (third row) shows the fraction of misalignments compared to stellar mass.  The misalignment fractions in the field/group sample show substantially less variation with stellar mass than with morphology (top row).
The most massive galaxies are not all ETGs. The trend between stellar mass and morphology is shown in Fig.~\ref{Mstarmorph} and highlights that galaxies with stellar masses between $10^{10}$ to $10^{11}\,M_{\odot}$ are dispersed among a wide range of morphological types.  The  $<\sim10^{10.5}\,M_{\odot}$ population is dominated by LTGs but the $>\sim10^{10.5}\,M_{\odot}$ galaxies are a combination of early- and late-type galaxies. For example, the highest stellar mass bin has 54\% ETGs. Therefore the fractions of misaligned galaxies at high stellar mass are not increased in the stellar mass plot because misalignment is driven by morphological type not stellar mass.

The dominant physical driver for misalignment was tested by removing the stellar mass dependence. The morphology, S\'{e}rsic index and colour plots from the left column of Fig.~\ref{morphtypehist} have had their stellar mass dependence removed in the right column. This was done by selecting galaxies weighted by a Gaussian kernel with a standard deviation of 0.1 dex 
around the stellar mass of each misaligned galaxy. This results in a random sample of galaxies with a similar mass. The misalignment fraction for that mass-controlled sample was measured and averaged for each misaligned galaxy within a morphology, S\'{e}rsic index or colour bin. The ratio of the misalignment fractions in the original bins to that of the sample with the same stellar mass then quantified how much more likely are the galaxies to be misaligned in that morphological, S\'{e}rsic index or colour bin, compared to other galaxies with a similar stellar mass. If the dependence was completely due to stellar mass then the right column plots in Fig.~\ref{morphtypehist} would have all points at 1.0. Instead, ETGs in the field/groups sample have $\sim 3$ times the chance of being misaligned than a matched sample in stellar mass, while LTGs are less likely to be misaligned than expected from stellar mass dependence. 

Confirming the morphology trend, high S\'{e}rsic index galaxies are up to $2.5$ times more likely to be misaligned and low S\'{e}rsic index galaxies are less likely to be misaligned than those of a similar stellar mass. The misaligned fraction increases to redder $g-i$ colours in Fig.~\ref{morphtypehist} (fourth row). However, 
the trend in colour is driven by the stellar mass dependence. Therefore, morphology and the shape of the galaxy stellar mass distribution dominates more than stellar mass in the fraction of field/group galaxies misaligned. Physical drivers for this result will be addressed in Section~\ref{MorphMisalign}.

The cluster galaxies have a higher stellar mass distribution than the field/group sample (see Fig.~\ref{zvsMstar}). The stellar mass corrected misalignment distribution in morphology is different for the clusters compared to the field/group sample. This is due to environmentally-driven factors; the lack of gas in cluster ETGs limits the measurement of misaligned gas in that sample, as will be discussed in Section~\ref{clusters}.

\citet{Jin16} using MaNGA, found the fraction of emission-line-detected galaxies that are misaligned increased at higher stellar mass, and peaked at 11\%, in agreement within errors with our results using a 30$^{\circ}$ cut-off for misalignment. Fig.~\ref{morphtypehist} (third row) however, shows a flatter distribution without the peak seen in the MaNGA sample. This discrepancy is not driven by the difference in environment between the SAMI and MaNGA samples because the difference is apparent in the SAMI field/group sample. However, the MaNGA sample shows a significantly larger change in the misalignment fraction between the high and lowest SFR bins, which is a more substantial trend than seen with stellar mass. Based on the findings in Fig.~\ref{morphtypehist}, the strong trend in the MaNGA SFRs may be due to ETGs typically having lower SFR than LTGs. 

 \subsection{Group mass, local environment and misalignment}
\label{GroupResults}

The GAMA-region parent catalogue includes galaxies in densities from the field to groups. Galaxies are classified as field galaxies if they are not identified to have other group members in the G$^3$C (v08) from \citet{robotham11}. 60\% of the field/group galaxies in our sample (for which gas and stellar PAs could be fitted) are in groups, and those groups range in mass from $\sim10^{10}$ to $\sim10^{14.6}\,M_{\odot}$. The largest group masses are similar to the smallest cluster masses for the cluster sample that start from a virial mass of $10^{14.3}\,M_{\odot}$ \citep[see ][]{JB2015}.  

\begin{figure*}
\vspace*{2mm}
\begin{minipage}[]{0.5\textwidth}
\includegraphics[width=7.5cm]{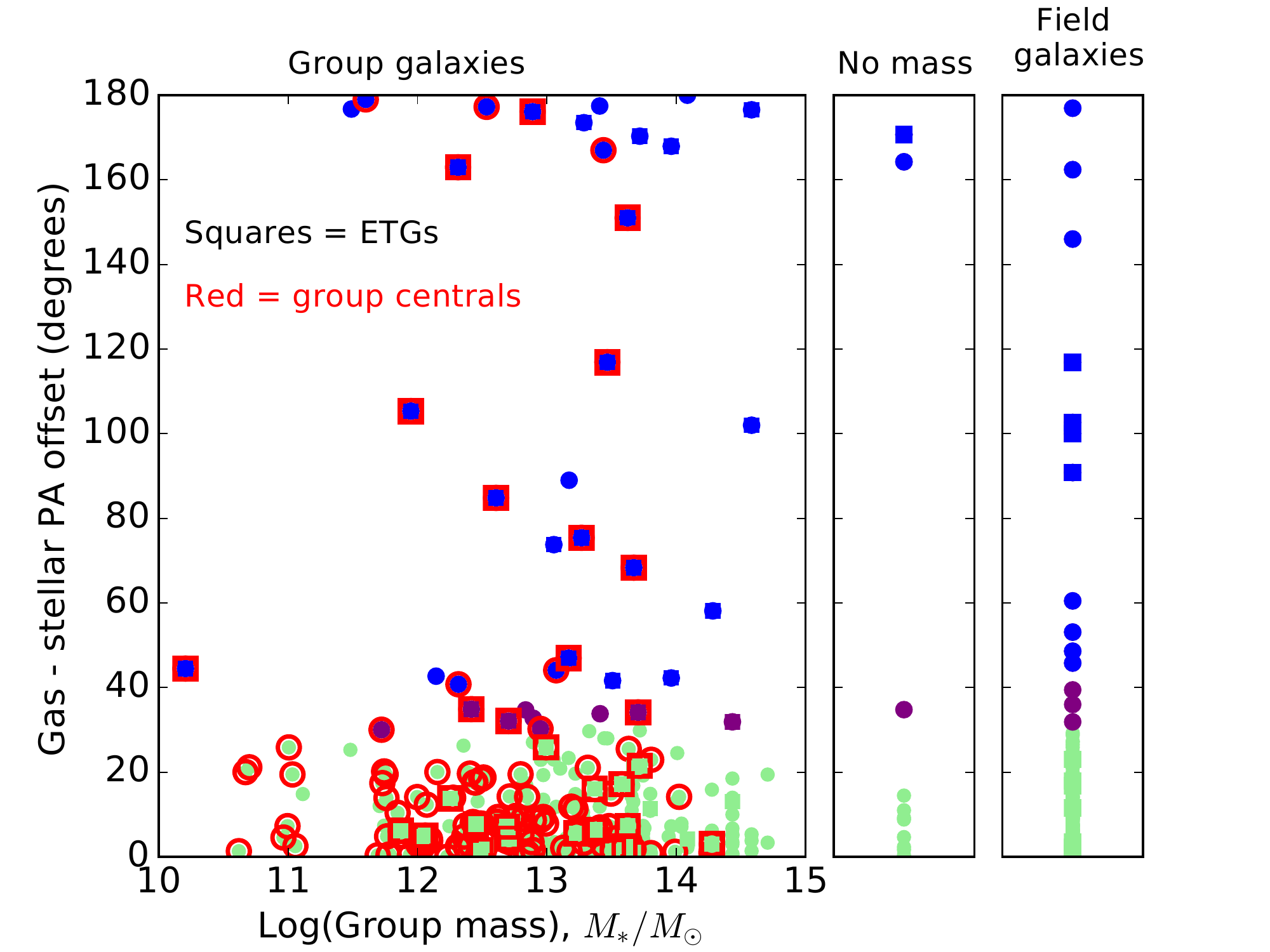} 
\end{minipage}%
\begin{minipage}[]{0.5\textwidth}
\includegraphics[width=6.0cm]{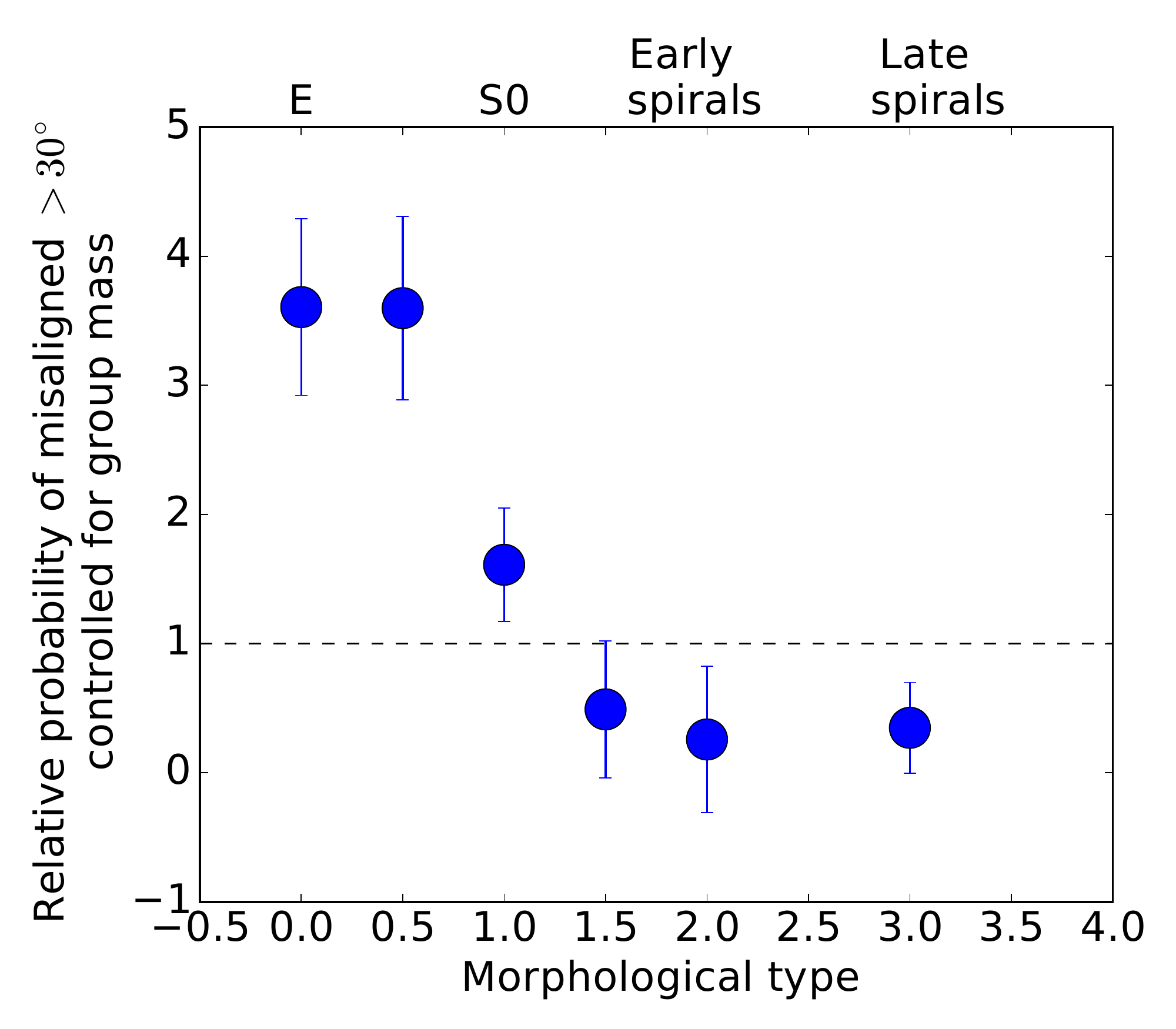} 
\end{minipage}%
\vspace*{1mm}
\begin{minipage}[]{0.5\textwidth} 
\includegraphics[width=7.5cm]{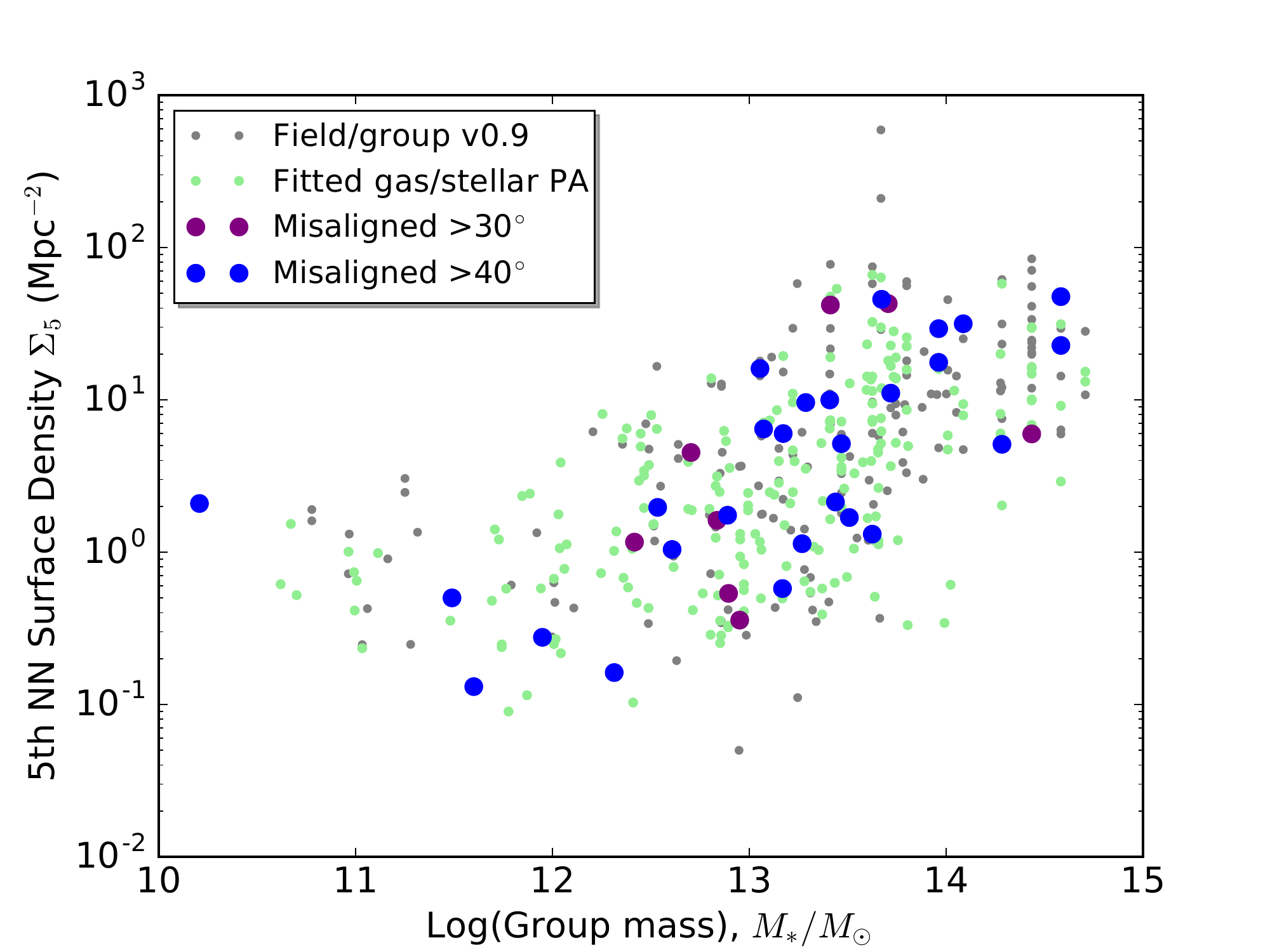}
\end{minipage}%
\begin{minipage}[]{0.5\textwidth} 
\includegraphics[width=6.0cm]{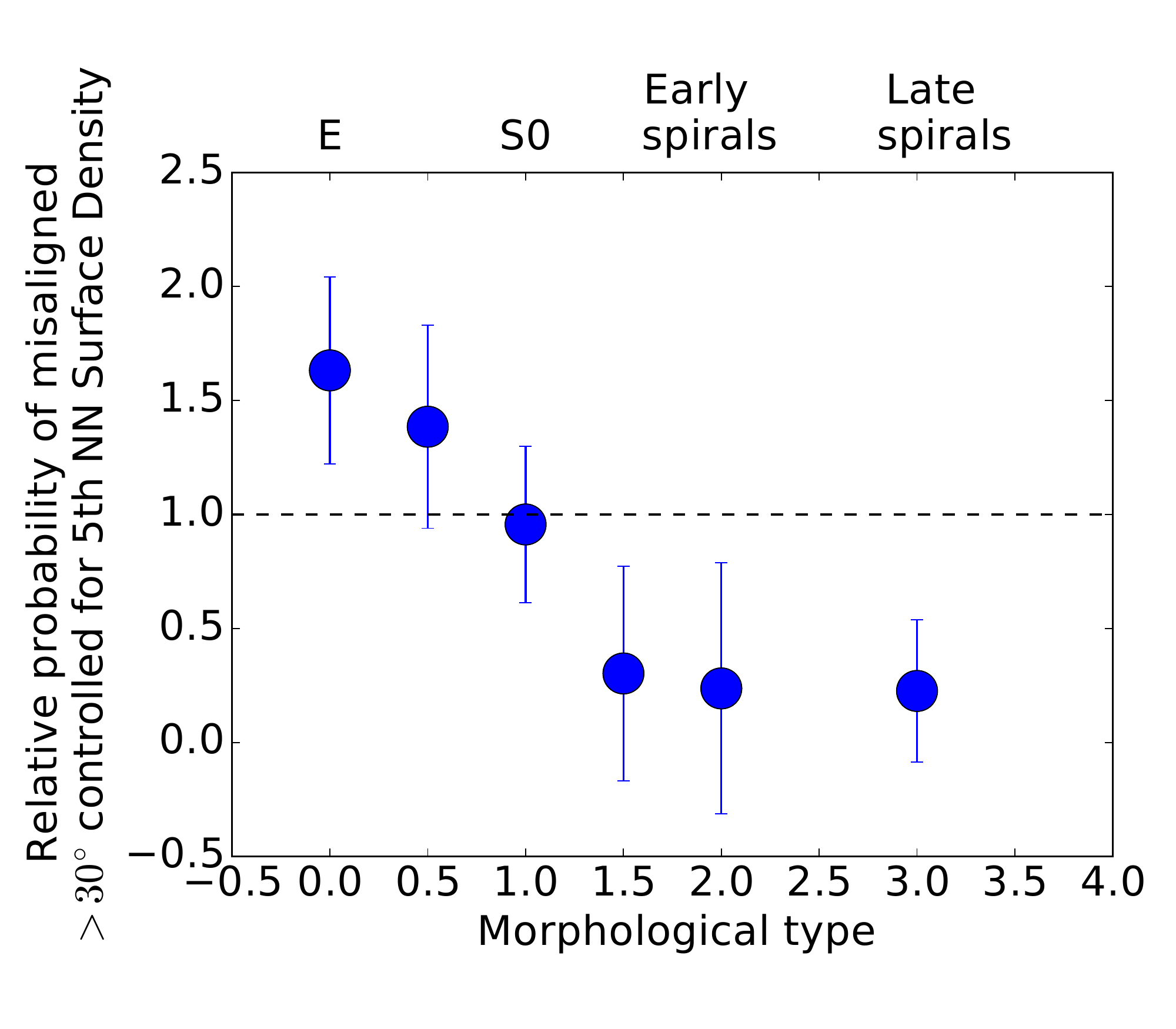}
\end{minipage}%
\vspace*{-2mm}
\caption{Left: Group mass for SAMI galaxies in the field/group samples that are members of groups and have reliable mass measures, versus gas/stellar PA misalignment (top left) and local (5th nearest neighbour surface density) environment (lower left).  In the top left plot, red marks galaxies that are central to their group, and squares are the subset of galaxies that have ETG morphologies (class $<1.5$). The ``no mass" panel are galaxies that are in groups but the group mass is undetermined because the group velocity dispersion was inaccurate \citep[see][for details]{robotham11}, and the ``field galaxies" panel are the field galaxies that are not in groups. 
The legend applies to both left panels; the SAMI v0.9 field/group sample galaxies are marked in grey, and of those, the galaxies for which both stellar and gas PAs could be fit are shown in light green (aligned with PA offset $<30^{\circ}$), purple and blue (misaligned by $>30^{\circ}$ and $>40^{\circ}$ respectively). 
Right: Probability of galaxies in the field/group sample in each morphological type being misaligned by greater than $30^{\circ}$ compared to galaxies with a similar group mass (top right) and 5th nearest neighbour surface density (lower right). These plots are Fig.~\ref{morphtypehist}, top left, controlled for group mass and 5th nearest neighbour surface density by the same method as was used for stellar mass in Section~\ref{Mstar}).  ETGs are more likely (above the dashed line) to be misaligned, while LTGs are less likely (below the dashed line) to be misaligned than galaxies of the same group mass or surface density.
}
\label{Groupmass5thNN}
\end{figure*}

Fig.~\ref{Groupmass5thNN} shows the distribution of group masses compared to the stellar/gas PA offsets and the local density (5th nearest neighbour surface density). The counter-rotating galaxies are found across all group masses from $\sim10^{11.5}$ to $\sim10^{14.6}\,M_{\odot}$.
There is no statistical difference in the distribution of group masses for aligned galaxies compared to misaligned 
(K-S test statistic =0.09-0.18, p-value=0.69-0.99, depending on misalignment definition).
There is also no difference between the distribution of PA offsets for galaxies that are the central galaxy in their group compared to those that are in a group but are not central (K-S test statistic=0.06, p-value=0.97; or statistic=0.26, p-value=0.58 for PA offset $>40^{\circ}$). Therefore the group mass or position within the group does not influence the chance of being misaligned.

However, whether a galaxy lives in a group compared to in the field influences the chance of counter-rotating gas - only $1.6\pm 0.9$\% [$1.0\pm 0.7$\%] of the field galaxies have counter-rotating gas compared to $5.1\pm 1.3$\% [$5.1\pm 1.3$\%] of the galaxies in groups. This is despite equal chance of misaligned (but not counter-rotating) gas between groups and the field galaxies (see Fig.~\ref{Groupmass5thNN} top panel and Table~\ref{Misalign_stats}). The implications of this result will be discussed further in Section~\ref{groups}.

\subsection{Polar rings}
\label{polarrings}

One special case of misalignment are polar ring galaxies, which are singled out in this section.
Polar rings \citep[e.g. AO136-0801 and NGC4650A;][]{Sch83,Ser67}) or polar disks \citep[e.g.][]{Bro08} are star-forming gas rings or disks that have a rotation axis at $\sim90^{\circ}$ to the main stellar disk of the galaxy. 

Polar ring galaxies are relatively rare
and up to a few percent of S0 galaxies are expected to have polar rings \citep[e.g.][]{Mac06}, and therefore this sample is only expect to detect a few.
The increased stability of a polar ring/disk means there should be an increase in the fraction of galaxies with a PA offset around 90$^{\circ}$. Fig.~\ref{PAhistMorph} does in fact show such a peak in the ETGs which is statistically significantly different from a flat distribution of galaxies in the misaligned region of $40<$PA offset$<140^{\circ}$ (K-S test statistic =0.23, p-value 0.032). This peak includes the S0s, and we note that a similar peak was detected in the ATLAS$^{3D}$ sample \citep{Dav16} although not discussed in that paper.   
One example of a polar ring galaxy in our sample is shown in Fig.~\ref{pics} (4th row) in which the on-sky alignment makes it easy to separate the North-South stellar disk rotating into the plane of the sky, surrounded by a clear ring of gas and dust forming stars close to the plane of the sky.

Accounting for polar ring galaxies is important in modelling and understanding the dynamical settling time distribution of galaxies discussed in section~\ref{MorphMisalign}.

\subsection{Distribution of misalignment angles}
\label{distribution}

\begin{figure*}
\begin{minipage}[]{0.5\textwidth}
\includegraphics[width=8.5cm]{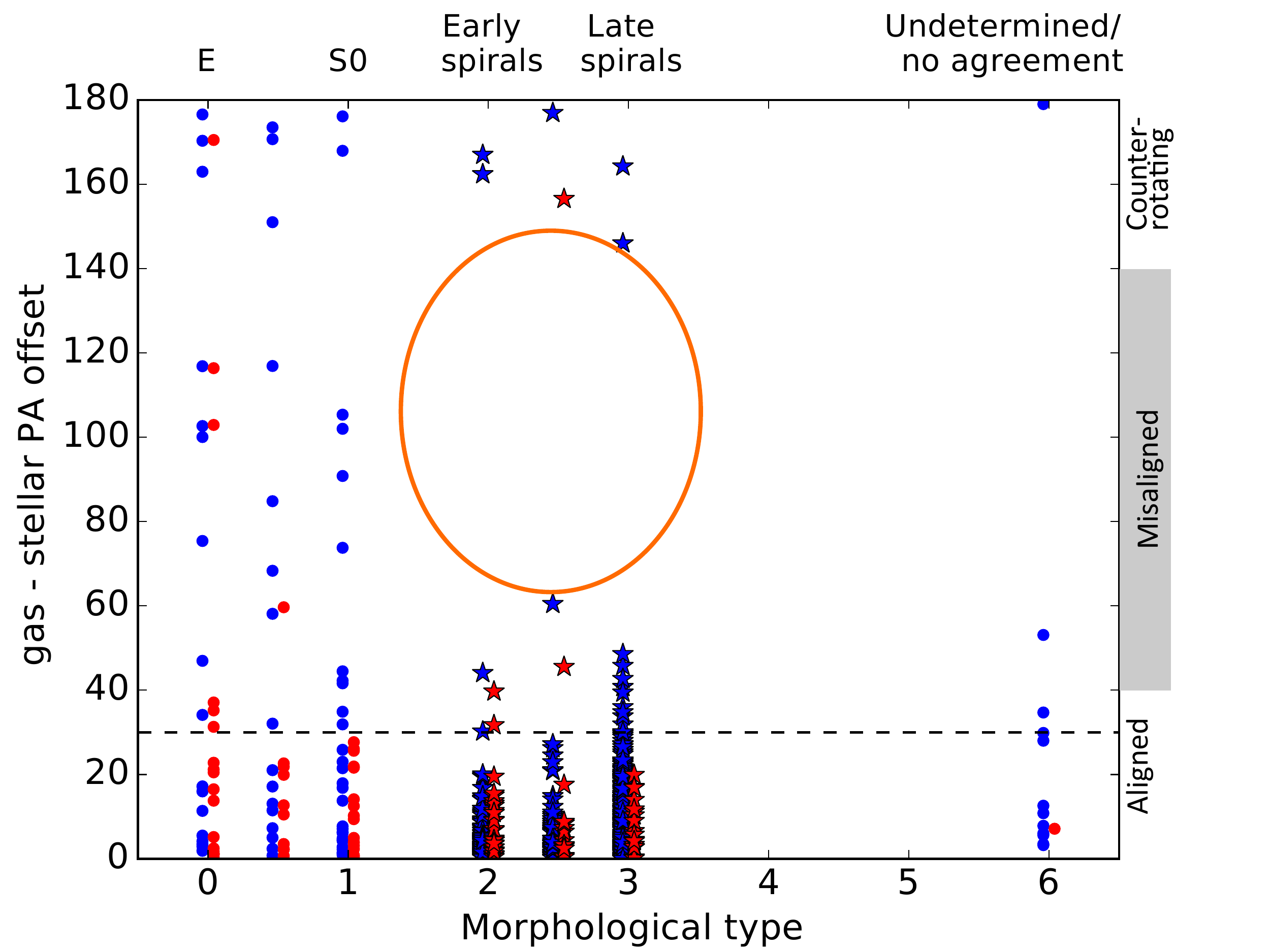} 
\end{minipage}%
\begin{minipage}[]{0.5\textwidth}
\includegraphics[width=9.0cm]{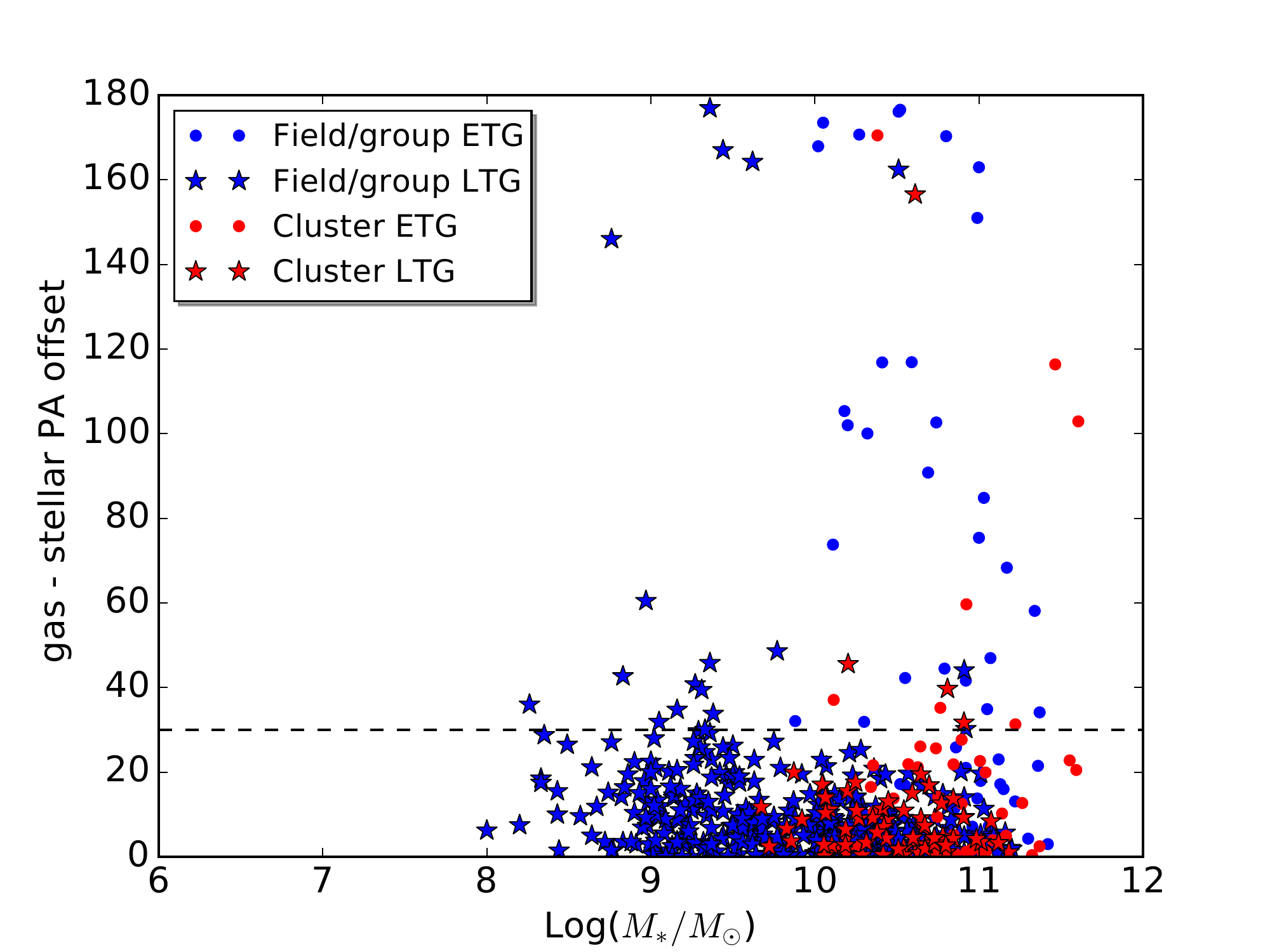} 
\end{minipage}%
\vspace*{3mm}
\caption{Galaxy morphological type (left) and stellar mass (right) versus gas-stellar misalignment angle. 
Morphological types are 0= Elliptical, 1= S0, 2= Early spirals 3=Late spirals, 5 and 6 are galaxies that were undetermined due to complex structure or a lack of consensus among the classification team members. Blended categories are represented between these main bin values (e.g. E/S0 = 0.5) as described in Section~\ref{morph_intro}. Morphological classes $<1.5$ are considered ETGs (dots) and those $>1.5$ are LTGs (stars). Blue points refer to the field/group galaxies while cluster galaxies are in red. The blue and red points in the left plot are offset horizontally from each other for clarity. There is a lack of LTGs with intermediate PA offsets highlighted by the orange ellipse, whereas ETGs populate the full range of PA offsets.
}
\label{morphtype}
\end{figure*}

Differences in the {\it distribution} of the misalignment angles (PA offsets) between populations can indicate different accretion processes beyond what misalignment fractions alone can reveal.

The key result from Fig.~\ref{morphtypehist} showed there is a difference in the misalignment fraction for different morphological types.  It is then notable that there is also a marked difference in the {\it distribution} of those misalignment angles (PA offsets) between ETGs and LTGs. In Fig.~\ref{morphtype} (left) the ETGs are distributed over the full range of PA offsets (aligned, misaligned and counter-rotating), whereas the LTGs clearly have a strong preference to be closer to aligned or counter-rotating, leaving a gap in the PA offsets from $\sim40-140^{\circ}$ (highlighted by the orange ellipse).  Where galaxies sit in this plot of misalignment distribution reveals the physics behind their gas accretion, which will be analysed in Section~\ref{Discussion}.

Section~\ref{Mstar} showed that the PA offset distribution is driven by morphology more than stellar mass which is also highlighted in Fig.~\ref{morphtype}. 
There is a gap at PA offset$\sim40-140^{\circ}$ for low stellar mass galaxies, but 
the bulk of the population lie above a stellar mass of $10^{10}$M$_{\odot}$ in the PA offset$<30^{\circ}$ region which includes both early and late-type galaxies.  The early-type galaxies that are misaligned are all high stellar mass ($>10^{10}$M$_{\odot}$) but the high stellar mass LTGs are never misaligned.

\section{Discussion}
\label{Discussion}
Relaxed stellar systems will have had their last major merger or accretion event at least several gigayears ago. Therefore, misalignment of the gas rotation from the stars in galaxies 
implies there has been new externally accreted gas from the outer halo, filament accretion or from a recent minor merger\footnote{We refer to a {\it minor merger} as a merger in which the bulk stellar rotation of the galaxy is not disrupted, even though a larger gas mass (``wet merger") may be accreted.} or interaction. The global gas rotation PA traces the bulk of the gas, which means gas disks measured to be misaligned have a larger gas mass than was originally in the galaxy before that gas accretion event. The {\it distribution} of the misaligned angles for different galaxy morphologies and environments can give clues to the origin of this accreted gas.

The following discussion will compare physical drivers for the observed distribution of misalignments of ETGs compared to LTGs.
\begin{itemize}
\item 
It will be shown in Section~\ref{MorphMisalign} that the precession of gas disks after accretion due to gravitational dynamical settling, is influenced by ellipticity and $\phi$. However such dynamical settling can only partly account for the misalignments observed and therefore alternative mechanisms must be considered.
\item We then present evidence in Section~\ref{Alignedaccretion} that gas is more likely to be accreted aligned in LTGs and, if accreted misaligned, it will preferentially settle to be aligned rather than counter-rotating due to interaction with existing gas.   Accretion of gas masses substantially smaller than the existing gas disk in gas-rich galaxies are likely to be rapidly disrupted and dissipate into the existing gas disk. 
\item Both group mass and local environment density are found to affect misalignment less than morphology, and mergers are unlikely to be the main source of gas in misaligned field/group galaxies (Section~\ref{Environmentdiscussion}). 
\item In Section~\ref{Environmentdiscussion} we find the cluster environment has a strong impact on misalignment because gas stripping influences the measured gas PA. Misalignment in clusters is then not necessarily due to {\it accretion} of gas (as in the field/group environments), but is due to effects of the cluster medium. 
\end{itemize}

\subsection{Dynamical settling time of accreted gas}
\label{MorphMisalign}

The precession of the gas disk compared to the angular momentum of the stars has been shown in semi-analytic models from \citet[{\sc dark sage};][]{Ste16} to be essential to producing a PA offset distribution in disk galaxies that has both an aligned and counter-rotating peak. The simulations 
predicted that in LTGs with gas precession, the PA offset distribution should have a counter-rotating peak an order of magnitude smaller than the co-rotating peak, and very few galaxies in the misaligned PAs in between. Without gas precession, the distribution has no counter-rotating peak and has a gradual decline in galaxy numbers with PA offset \citep[][Fig. 3]{Ste16}. Our results presented in Fig.~\ref{PAhistMorph} (right) are a direct observational test of those simulations. The simulations without gas precession do not match the observed PA offset distribution and are ruled out by our results. Therefore, precession of the gas disk must be considered in analysis of the physical drivers of misalignment.

Gas introduced to the galaxy 
will have an angular momentum axis which will set the initial PA of rotation, leading it to form a disk. That rotating disk will then be affected by the galaxy's gravitational stellar mass distribution, and after some settling time $t_{s}$ that gas will settle into a stable orbit. Note that this definition of $t_{s}$ is due to the gravitational influence, not the settling due to dissipation of gas that will be discussed in Section~\ref{Alignedaccretion}. After a dynamical settling time $t_{s}$, the PA offset will either be below $40^{\circ}$ (aligned within errors) or above $140^{\circ}$ (counter-rotating within errors). Simulations have shown that counter-rotating gas disks are permanently stable until the gas is consumed in star formation \citep{Osm17}. 

A semi-stable state is a polar-ring with misalignment near $\sim90^{\circ}$ (see Section~\ref{polarrings}). The formation of polar rings or disks has been attributed to gas that is preferentially accreted from either a close passage with another galaxy \citep{Sch83}, or from filamentary cold flows \citep{Mac06, Con06}, with perpendicular rotation to the stellar disk.  Simulations show that merging and interactions can transform late-type disk into S0s with polar-rings which persist for many Gyrs \citep{Bek98,Bou03} and therefore most polar rings are found in S0 galaxies. 
Polar rings have a very slow dynamical settling time due to the high angle between gas and stellar disks. Cosmological evolution and dynamical friction simulations as well as stellar ages in polar rings have indicated that stability depends on how close the ring is to $90^{\circ}$, and within $<40^{\circ}$ of polar a ring will remain stable for at least $1.6-3$ Gyr \citep[depending on the model:][]{Sch83,Mac06,Cox06, Gal02}.

\begin{figure}
\includegraphics[width=9.0cm]{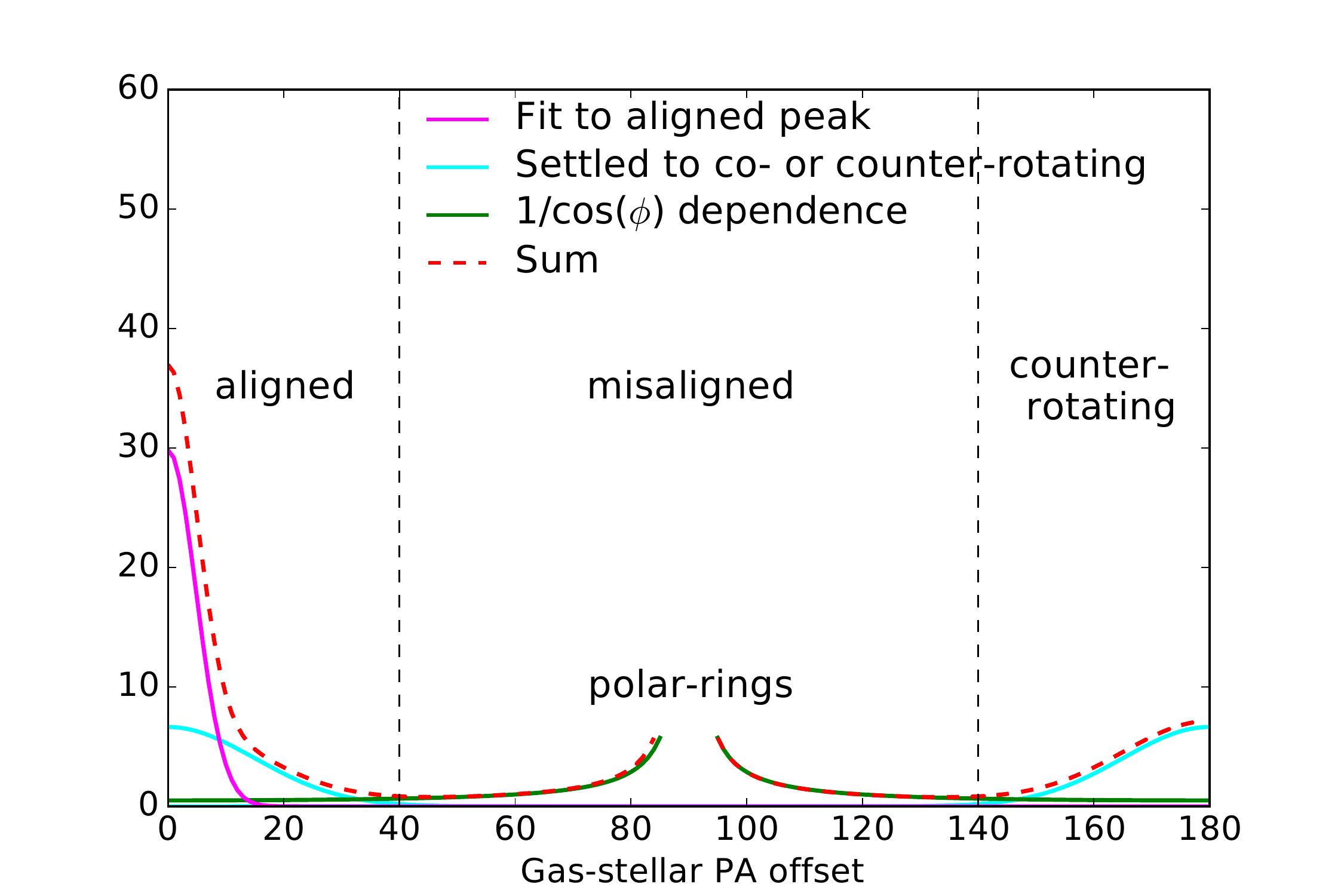} 
\caption{Cartoon illustrating the proposed contributions to the distribution of PA offset values observed. We refer to the region with $0 - 40^{\circ}$ as {\it aligned}, $40 - 140^{\circ}$ as {\it misaligned} and $140 - 180^{\circ}$ as {\it counter-rotating}. Galaxies in the misaligned region are assumed to have gas disks that are in the process of settling to be either aligned or counter-rotating. The magenta curve is a fit to the aligned peak in the LTG distribution (see Fig.~\ref{PAhistMorph}) showing the apparent distribution of galaxies with good S/N and settled, stable, aligned gas and stellar rotation.  Cyan curves represent galaxies that have settled from misaligned to aligned or counter-rotating in the basic case that equal numbers settle both ways. Note that in the discussion it is argued that more galaxies will settle to aligned, but the relative contributions of the magenta and cyan curves in the aligned region cannot be disentangled from the observations. The histograms in Fig.~\ref{PAhistMorph} have  additional galaxies between $\sim20-40^{\circ}$ than is predicted by these curves, suggesting a contribution from the galaxies with higher PA errors and/or more galaxies settle to aligned than misaligned. The $\frac{1}{cos\phi}$ dependence of Eq.~\ref{td} is represented by the green curve. The dashed red line is the sum of all curves. The y axis scaling on these curves was arbitrarily set to mimic the distribution in Fig.~\ref{PAhistMorph} (left) and designed only to give a picture of the concepts in the discussion in Section~\ref{MorphMisalign}. 
}
\label{Toymodel}
\end{figure}

Galaxies with misalignments between $40 - 140^{\circ}$ have gas disks that are not stable and are assumed to have recently entered the galaxy but not yet settled into a co-rotating or counter-rotating configuration. The angles defining boundaries allow for conservative errors in the misalignments as discussed in Section~\ref{nature}. However, there is not a clear boundary, therefore the following statistics are also considered in a more relaxed range of $30 - 150^{\circ}$ to match the literature and for each number given below, the corresponding value with this broader misalignment range is given in square brackets. 
To streamline the following discussion, the PA offset ranges of $0 - 40^{\circ}$, $40 - 140^{\circ}$ and $140 - 180^{\circ}$ are called {\it aligned}, {\it misaligned} and {\it counter-rotating} respectively. Fig.~\ref{Toymodel} shows these regions in a cartoon that illustrates the contributions to the misalignment distribution referred to in the following discussion. 

To investigate how the dynamical torquing or settling of the misaligned gas is dependent on the galaxy stellar mass shape, the distribution of misalignment angles between ETGs and LTGs seen in Figs.~\ref{morphtype} and ~\ref{morphtypehist} will be compared.
ETGs have $26\pm 6$\% [$32\pm 6$\%] with misaligned gas, and for LTGs it is much less with $1.7\pm 0.7$\% [$4\pm 1$\%].   However, if it was assumed that accreted gas can come in equally from any angle, then accounting for galaxies
that have been accreted aligned or counter-rotating by chance (increase of $180^{\circ}/100^{\circ}$ [$180^{\circ}/120^{\circ}$]), gives $47\pm 10$\% [$48\pm 10$\%] for ETGs and $3\pm 1$\% [$7\pm 2$\%] for LTGs.
This misalignment fraction alone would only give a robust measure of gas accretion event rate (number of accretions / settling time) if gas was equally likely to be accreted at any angle, and always had the same dynamical settling time. However, this is not the case, and we now discuss the physical processes that drive dynamical settling time and how that is affected by the galaxy stellar mass shape.

We define dynamical settling time as the time for the angular momentum axis of a misaligned rotating disk to torque towards that of the galaxy's stellar mass distribution until it aligns or anti-aligns. Within the central $\sim6-7$\,kpc seen by SAMI, we assume that the stellar density is higher than the density of the dark matter and therefore the stellar mass distribution will impart a gravitational torque on the gas disk. 
In Appendix 2, the time taken for that rotating disk to precess onto the galaxy plane is shown to be 
\begin{equation}
t_{s} \propto \frac{1}{\dot{g}} \propto \frac{R}{V_{rot}\, (2\epsilon-\epsilon^2) \, cos(\phi)}
\label{td}
\end{equation}

Eq.~\ref{td} highlights several intuitive aspects of dynamically settling disks. First, in a regularly rotating disk, the gradient of $V_{rot}/R$ ($d$V/$d$R) is steep in the centre of the galaxy until the radius at which the rotation curve flattens out.
Therefore, the dynamical time for gas in the disk at lower radius, to precess on to the plane of the galaxy's stellar mass distribution, will be less than gas at higher radius leading to a radial warp in the disk over time. Such warps have been well studied in nearby galaxies such as Centaurus A \citep[e.g.][]{JBH87, Nic92, Spa96, Lak83}.
Secondly, as the ellipticity value ($\epsilon$) increases, the dynamical settling time decreases. The angular momentum axis of an accreted gas disk will precess towards that of the existing stellar mass disk more rapidly for LTGs than for ETGs with low $\epsilon$ values.
Thirdly, gas accreted with an angular momentum more highly inclined to that of the galaxy, and hence with larger $\phi$ (up to 90$^{\circ}$), will have a longer dynamical time than gas accreted at a small angle. This is one reason for the stability of polar ring/disk galaxies. We now consider how the ellipticity, R/V and $\phi$  terms in this equation contribute to the distribution of misalignments we observe.

\subsubsection{
Does the impact of ellipticity on dynamical settling time drive the morphology-misalignment distribution? }

Eq.~\ref{td} has a dependence on ellipticity. Is that term large enough that the PA offset distributions of ETGs and LTGs in Fig.~\ref{morphtype} (left) are driven by the dynamical settling time of the gas once it has been accreted?  

Intrinsic ellipticity results in a difference in settling time in Eq.~\ref{td} of at most $\sim2.7$ times slower for ETGs. That assumes the intrinsic ellipticities of ETGs and LTGs are at the limits of sensible values of 0.2 and 0.8 respectively \citep[e.g.][]{Bin81, Lam92, Fra92, Men08}. The actual ellipticities will vary and not all be at those limit values \citep{Fos17}, reducing the impact of ellipticity on dynamical settling time. If ellipticity was the main driver for the difference in PA offsets between morphological types there should be $<2.7$ times as many misaligned ETGs. 
The impacts of both the R/V and ellipticity terms in Eq.~\ref{td} can be assessed for galaxies for which we have measurements of R/V in Appendix 2. The ratio of R/(V*(2$\epsilon-\epsilon^2$)) in Eq.~\ref{td} should be $1.4\pm0.1$ times higher for ETGs than LTGs (where the error is the relative error not including the inherent error introduced in the assumptions used in Appendix 2).
However, {\it if there is no morphological dependence on R/V} then the dynamical settling time in Eq.~\ref{td} would be $2.7$ times slower for ETGs than LTGs.  The corresponding observed ratio of gas that is misaligned for early to late types is $15\pm7$ [$8.2\pm2.6$].  
Despite the uncertainties in the assumptions for R/V and ellipticity, both the predicted ratio from Eq.~\ref{td} and observed ratios show that there are and should be a higher fraction of ETGs that are misaligned. However, this observed ratio ($15\pm7$) is substantially higher than predicted by Eq.~\ref{td}($<2.7$). Therefore the elliptical shape of the galaxy stellar mass distribution affects the time it takes for the gas disk to torque towards the stellar disk, but this is not the only effect driving the observed statistics.

\subsubsection{Impact of gas angular momentum angle on dynamical settling time}

Is the lack of misaligned LTGs (marked by the orange ellipse in Fig.~\ref{morphtype}; left) due to gas in LTGs {\it settling more rapidly} towards aligned/counter-rotating angles, or is it due to LTGs {\it not accreting} gas at the misaligned angles in the orange ellipse in the first place? A toy model for the misalignment distribution resulting from the dynamical settling time equation will be used to understand the difference in misalignment distributions between ETGs and LTGs. 

Based on Eq.\ref{td}, the misalignment fractions should have a dependence on the angle $\phi$ of the gas disk angular momentum to that of the stellar disk. The impact of $\phi$ is that if gas can be accreted from any angle then the distribution of misaligned PAs should not be flat. Instead it will be peaked towards  $\phi=90^{\circ}$ because galaxies closer to that PA offset value will have a longer dynamical settling time. This contribution to the PA offset distribution is illustrated in the cartoon in Fig.~\ref{Toymodel}. An additional peak at counter-rotating PAs is then the sum of the probability of galaxies that have gas accreted by chance within the counter-rotating range (green line at $>140^{\circ}$), plus those galaxies that, after a dynamical settling time, have settled from misaligned to counter-rotating (cyan).  In both of those cases the counter-rotating galaxies indicate externally acquired gas. It is notable that in the aligned peak ($<40^{\circ}$), galaxies with externally accreted gas that have settled to be aligned cannot be separated from those with gas generated internally.

If galaxies accrete gas at any angle then this toy model based on dynamical settling time predicts that any galaxies with $\sim40^{\circ}<PA<\sim150^{\circ}$ should have a distribution peaking around $90^{\circ}$. This is clearly the case for the ETGs but not for LTGs (Fig.~\ref{PAhistMorph}). We showed above that the ellipticity term was insufficient to remove all LTGs from that region, and therefore if gas was accreted at any angle into LTGs the lack of a peak around $90^{\circ}$ in LTG means that unlike ETGs, gas is simply {\it not accreted at all angles in LTGs}. 
These differences in the ETG and LTG misalignment distributions must be driven by additional physical processes. The next subsection focusses on the preference for gas to be accreted at certain angles.

\subsection{Preferentially aligned accretion in LTGs}
\label{Alignedaccretion}

The distribution of LTGs at misaligned angles does not fit the model of the gravitational dynamical settling time in which the $\phi$ term dictates there should be more galaxies towards $90^{\circ}$. Nor can the counter-rotating galaxies then be justified as having been accreted by chance as counter-rotating, or settled to counter-rotating in the same time-frame as the ETGs. Instead, the lack of misaligned galaxies points to either a more rapid settling time than described in Eq.\ref{td}, or the galaxies were never misaligned. If a faster dynamical settling time was the only reason for the lack of late-type misaligned galaxies, then the expectation is that they would be settling both to aligned and misaligned symmetrically, unless more rapid dissipation occurs due to interactions with existing gas in the galaxy. Accretion of gas with a very much smaller mass than in the existing disk would be expected to dissipate onto the existing disk fast and would not have its rotation measured because it is not the dominant gas disk. However, if a galaxy accretes a larger gas mass than already in the disk, viscous forces between gas in the disk and the incoming gas may influence that gas to more likely settle to be aligned rather than counter-rotating.

The fraction of galaxies settling to co-rotating is substantially higher than the fraction settling to counter-rotating in the LTG sample. There are clearly more LTGs at low misaligned PA offsets than at high misaligned values. $83\pm19$\% [$86\pm13$\%] of the misaligned LTGs are very close to co-rotating (between $40-50^{\circ}$ [$30-50^{\circ}$]). The percentage is much lower in ETGs being $25\pm11$\% [$40\pm10$\%].
We note that the galaxies in that PA offset range are not dominated by scatter from the strong aligned peak, because a Gaussian fit to that peak (shown in Fig.~\ref{Toymodel}, magenta curve) has minimal contribution at PA offsets greater than $40^{\circ}$. 
{\it Therefore that gas has a predominance to be accreted with an angular momentum aligned or close to aligned to that of the stellar disk in LTGs or is torqued towards aligned by the influence of existing gas.}

This result supports the halo accretion models of \citet{Dan15} in which in late-type disk galaxies, the accreted gas angular momentum was shown to torque rapidly, within one orbital period, towards the stellar angular momentum. This happens at $\sim0.1-0.3$ virial radii, outside the radius typically observed by SAMI  
\citep[see also][]{Cev16}. 
\citet{Wel17} simulations also demonstrated how galaxy mass distributions flatten into disks when fed by smooth accretion with gas that has coherent angular momentum, and this is crucial to the formation of LTGs. 
Even gas accreted from minor mergers or interactions may have a tendency to be aligned in LTGs as the merging galaxy can torque towards alignment with the disk before merging \citep{Wel15}. 
\citet{Sal12} show from the GIMIC simulations that 
disk galaxies result when gas accreted over time has the same spin as the in situ gas, resulting in a disk in which the spin is enhanced with time. 
Accretion is more likely to be aligned in LTGs then simply because disks have been built from aligned accretion. \citet{Sal12} find the opposite is the case for ETGs. ETG spheroids form from multiple misaligned accretions of gas over time, that each form stars which in combination average over the spin of the galaxy. 

Simulations have shown that smooth accretion of gas onto ETGs is more likely than mergers to account for the  fraction of misaligned galaxies observed. 
\citet{Lag15} used the {\sc galform} model of galaxy formation, set in the cold dark matter framework and coupled it with a Monte-Carlo simulation to follow the angular momenta flips driven by matter accretion onto haloes and galaxies. They found that mergers alone could account for only $2-5$\% of misalignments between stars and ionised gas, but the addition of smooth accretion bolstered the misalignment fraction (defined in their case as PA offset $>30^{\circ}$) to $\sim 46$\%, similar to the value measured in our field/group sample of $45\pm6$\%. This implies that misaligned gas in ETGs is not predominately from mergers in order to reach the misalignment fractions found in Fig.~\ref{morphtype} and the fractions that were found in the ATLAS$^{3D}$ ETG sample in \citet{Dav13}.  The  ATLAS$^{3D}$ ETG misalignment distribution is remarkably similar to our SAMI ETG PA offsets, and their data also showed some evidence for a peak near $90^{\circ}$ although the paper did not comment on that. In their toy model to match this distribution, they needed to invoke an exceptionally long dynamical settling time of 80 dynamical times, which they attributed to smooth accretion. Accounting for the $\phi$ dependence presented above may somewhat reduce the settling time required in their model to reproduce the correct PA offset distribution, but long settling times still support smooth accretion rather than mergers as the dominant accretion origin.

Therefore, the fraction of counter-rotating and heavily misaligned galaxies plus the gravitational dynamical settling time for LTGs is inconsistent with accretion of gas at any angle within the radius seen by SAMI. The skewed distribution of LTGs towards aligned is due to preferentially aligned accretion or gas dissipation, not rapid gravitational dynamical settling of gas disks, which is in agreement with simulations discussed above. This is not surprising as the flattened stellar mass distribution in LTGs is a product of the aligned accretion of matter over time. This predominance of aligned accretion and interactions with existing gas in LTGs drives the difference in the misalignment fraction compared to ETGs more than the dynamical settling time.  Both effects (the ellipticity in $t_{s}$ and preferentially aligned accretion) are connected to the galaxy's stellar mass distribution (morphology) and are responsible for the observed dependence of misalignment on morphology and S\'{e}rsic index rather than stellar mass or colour.

 \subsubsection{Origin of counter-rotating gas in LTGs and the relative mass of the accreted gas}
 \label{CounterRot}
 
The results presented so far support simulations in which gas in LTGs has a predominance to be accreted close to aligned to the stellar angular momentum. There is nevertheless a small 1.4\% of the LTGs in the field/cluster sample that have counter-rotating gas, which raises the question of what unusual scenarios could lead to these galaxies not following the rest of the population. 
There are particular scenarios for mergers and filament accretion that would be expected to be uncommon but will lead to counter-rotation.

Counter-rotating gas must have previously been externally accreted and then settled into this stable orbit. 
If gas is closer to counter-rotating when accreted, then as it gets pulled onto the disk, it will experience viscous forces and/or shocks with the gas already in the disk.
Both \citet{Jin16} and \citet{Che16} argue that evidence for triggering of star formation in the centre of blue star forming galaxies with counter-rotating gas/stars suggests that the incoming gas rapidly loses angular momentum due to interaction with existing gas and falls to the centre of the galaxy. We note that in the case of their data and ours, the gas rotation measured however, remains the dominant gas mass rotating in an ordered disk. Therefore the angular momentum loss can not yet be large enough that the incoming gas has been disrupted by the existing gas disk or the latter would dominate the gas rotation.

If the accreted gas mass is small compared to the in situ gas, it may be expected to briefly distort the gas dynamics but then be disrupted and eventually align to have the same angular momentum axis as the stars. For example, simulations by \citet{Tha96,Tha98} have shown that it is not possible to form a counter-rotating disk in a LTG through a minor merger with a dwarf galaxy. Alternatively, if the incoming gas is plentiful from a gas-rich system, then it will dominate the rotation if the gas mass is much larger than the existing gas. In that case it can form a counter-rotating disk. 
From these basic arguments, the origin of gas in our observed LTGs with counter-rotating systems may be the unlikely combination of mergers or interactions with firstly, an accretion angle closer to counter-rotating and secondly, a larger gas mass than in the host galaxy.  
The probability of such an accretion scenario would be dependent on environment, and is discussed further in Section~\ref{groups}.

Alternatively, \citet{Alg14} simulate a scenario in which filament accretion can result in counter-rotating gas and stars. They investigate an isolated disk galaxy in cosmological simulations in which accretion of gas from two separate filaments over the life of a disk galaxy result in gas being introduced counter-rotating to the original stellar disk.  
No merger was involved, which highlights the difficulty in using misalignment rates as a tracer of galaxy merger rates. However, they note that very specific `V-shaped' configuration of two filaments and specific timing of the accretion from each filament is required, which again limits the chance of counter-rotation from filament accretion.
Therefore, it is neither surprising that some of our observed LTGs are counter-rotating, nor that the fraction of them is very low as very particular scenarios are likely to be required to result in counter-rotating systems.

 \subsection{The influence of environment on misalignment}
 \label{Environmentdiscussion}

\subsubsection{Group mass}
\label{groups}

In Section~\ref{GroupResults} it was shown that there is no influence of group mass on the misalignment fractions and that morphology remains the biggest driver of misalignment in group galaxies (see Fig. ~\ref{Groupmass5thNN}, top right). Therefore if mergers are more likely within larger group masses, then mergers are not the dominant driver of accreted gas in these galaxies. 

It was notable in Section~\ref{GroupResults} that group galaxies are more than three times as likely to have {\it counter-rotating} stars/gas compared to field galaxies (see Fig.~\ref{Groupmass5thNN} top left panel and Table~\ref{Misalign_stats}). Therefore in field galaxies it is less likely for externally accreted gas to settle to be counter-rotating rather than aligned. 
Galaxies classified here as field galaxies are not necessarily isolated galaxies, they may have smaller satellites not detected down to the limits of the G$^3$C \citep[see][for details of group classification]{robotham11}. Any accreted gas from such a satellite is likely to be small. In addition gas available for smooth accretion from the halo is expected to be less in field galaxies than in groups \citep[see for example, ][]{Mul10}.
If the mass of incoming gas is lower compared to the existing gas disk, it will be easily disrupted and end up co-rotating with the existing gas disk as discussed in Section~\ref{Alignedaccretion}, not counter-rotating \citep{Tha96,Tha98}. Therefore the lower counter-rotating fraction found in the field sample compared to the groups suggests accreted gas masses are lower in the field galaxy environments.

The accretion of small gas masses in the field sample may have led to all of the misaligned ETGs having $80<$PA offset$<120^{\circ}$ (Fig.~\ref{Groupmass5thNN}) (squares in top right panel)). ETGs on average have less gas mass than disk galaxies and the accreted gas is therefore a larger fraction of the existing gas. The reduced dissipation of accreted gas in ETGs will leave the gas misaligned for longer.

 \subsubsection{Local environment density}
 \label{LocEnv}

Galaxies with a higher local density may be expected to have an increased chance of mergers or accreting gas from neighbouring galaxies \citep[e.g.][]{vDok99}. 
One test of whether gas has accreted from mergers rather than halo or filament accretion is a comparison of the misalignment rates in different local environments.
5th nearest neighbour surface densities \citep[][]{Bro13,Bro17} were used as a measure of the local environment around each galaxy.
The misaligned and counter-rotating galaxies are a representative sample of all galaxies in both the distribution of group mass (K-S test statistic=0.10, p-value=0.95) and in 5th nearest neighbour surface density (K-S test statistic=0.13, p-value=0.77) (see Fig.~\ref{Groupmass5thNN}). 
By controlling the misalignment fractions in morphology bins by 5th nearest neighbour surface density in Fig.~\ref{Groupmass5thNN} (lower right),  we have shown that morphology drives the misalignment trend rather than the local density.   Therefore, if galaxies have a higher chance of mergers and interactions in a higher local density then mergers are not the main driver for misalignment.

\subsubsection{Cluster environment}
\label{clusters}

\citet{Dav11} found a significantly lower misalignment fraction in ETGs in the Virgo cluster compared with field/group galaxies from ATLAS$^{3D}$. We similarly find a lower fraction, but the physical mechanisms in the the two environments are not the same. The main influences on misalignment statistics in the clusters are: (a) the inability to measure gas in ETGs biases the included sample, and (b) the cluster environment causes apparent misalignment due to gas {\it stripping} rather than {\it accretion}.

As a galaxy falls through the intra-cluster medium (ICM), cold gas may be depleted by ram pressure stripping or shock heating \citep{Gun72, Bek99, Sch01, Vol01, Cen14b, Pen15}. 
Hot gas in the halo of a galaxy can also be removed \citep{Lar80}, limiting that potential reservoir of gas which could have fuelled external accretion. 
Gas in the process of being stripped will have a gradient in the gas velocity in the direction of the infall through the ICM. This may then be measured as misalignment when it is in fact not due to a misaligned rotating gas disk.

The cluster environment reduces the chance of measuring ETG gas, removing the trend with misaligned fraction in ETGs compared to LTGs as seen in Fig.~\ref{morphtypehist}. This is despite the higher stellar mass distribution of the cluster sample compared to the field/group sample (see Fig.~\ref{zvsMstar}). The observed statistics are then dominated not by whether the gas is misaligned, but whether the gas rotation remains measurable. In {\it non-cluster galaxies}, 
$43\pm 4$\% of the ETGs and $63\pm 2$\% of the LTGs had both stellar and gas PAs measurable. However 
in {\it clusters}, only $30\pm 4$\% have both stellar and gas rotation measurable, but $89\pm 4$\% of LTGs had measurable gas and stellar rotation PAs. The chance of measuring misaligned gas in clusters is therefore less than in the field/group galaxies since misalignment is dominant in ETGs.

\begin{figure*}
\centering
\includegraphics[width=18.0cm]{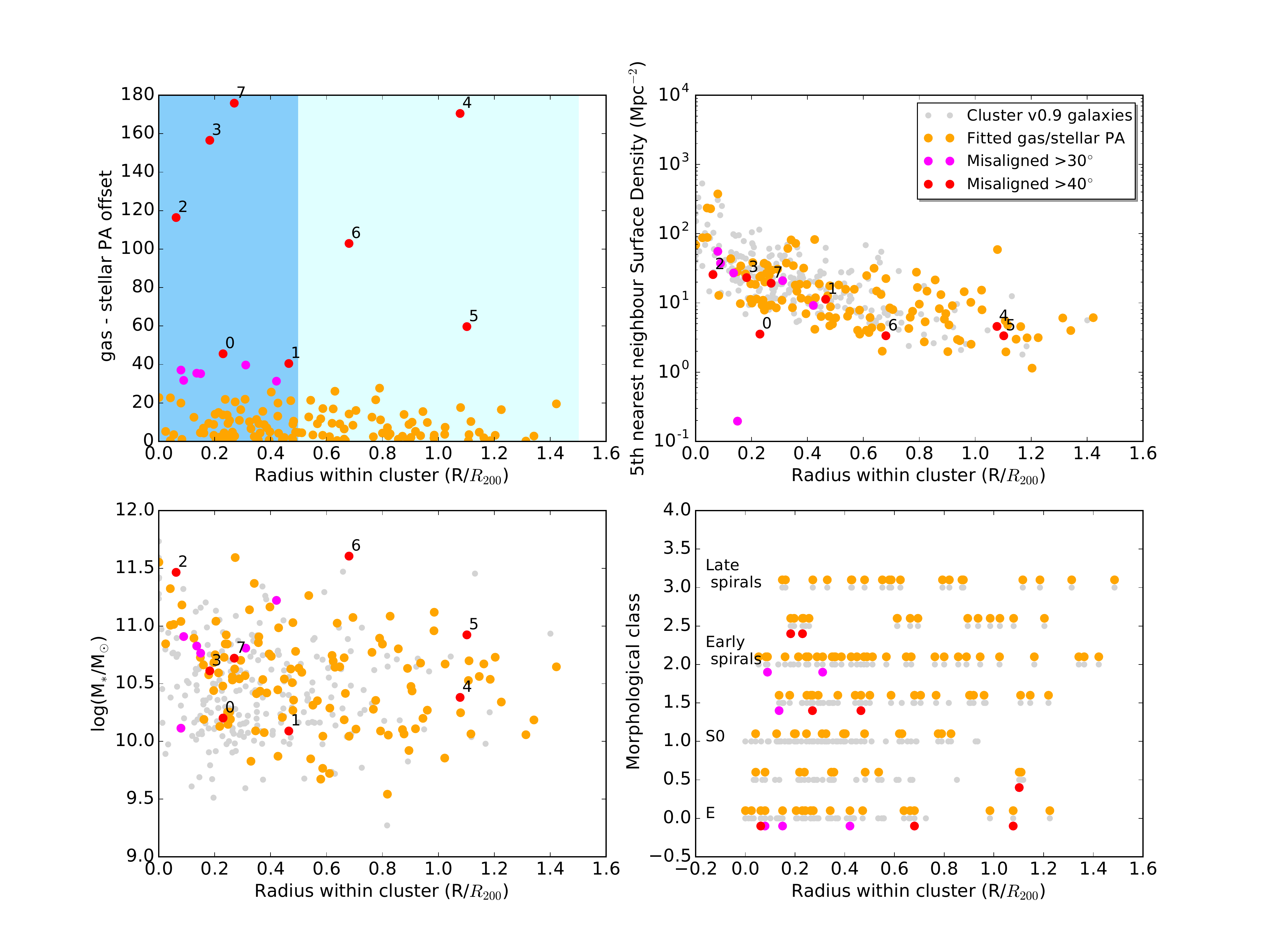}
\vspace*{3mm}
\caption{Top left: Misalignment versus radius within the cluster as a fraction of R$_{200}$. 
Top right:  5th nearest neighbour surface density versus radius within the cluster for galaxies in the 8 clusters. Lower left: Stellar mass versus radius within the cluster. Lower right: Morphological class versus radius within the cluster. The latter 3 panels include the original v0.9 SAMI sample in grey, and all the galaxies for which each of the properties and PA offset can be measured in orange. Those misaligned by $>30^{\circ}$ are magenta and $>40^{\circ}$ are red. 
}
\label{cluster_radius}
\end{figure*}

\begin{figure}
\begin{minipage}[]{0.22\textwidth}
\includegraphics[width=4.5cm]{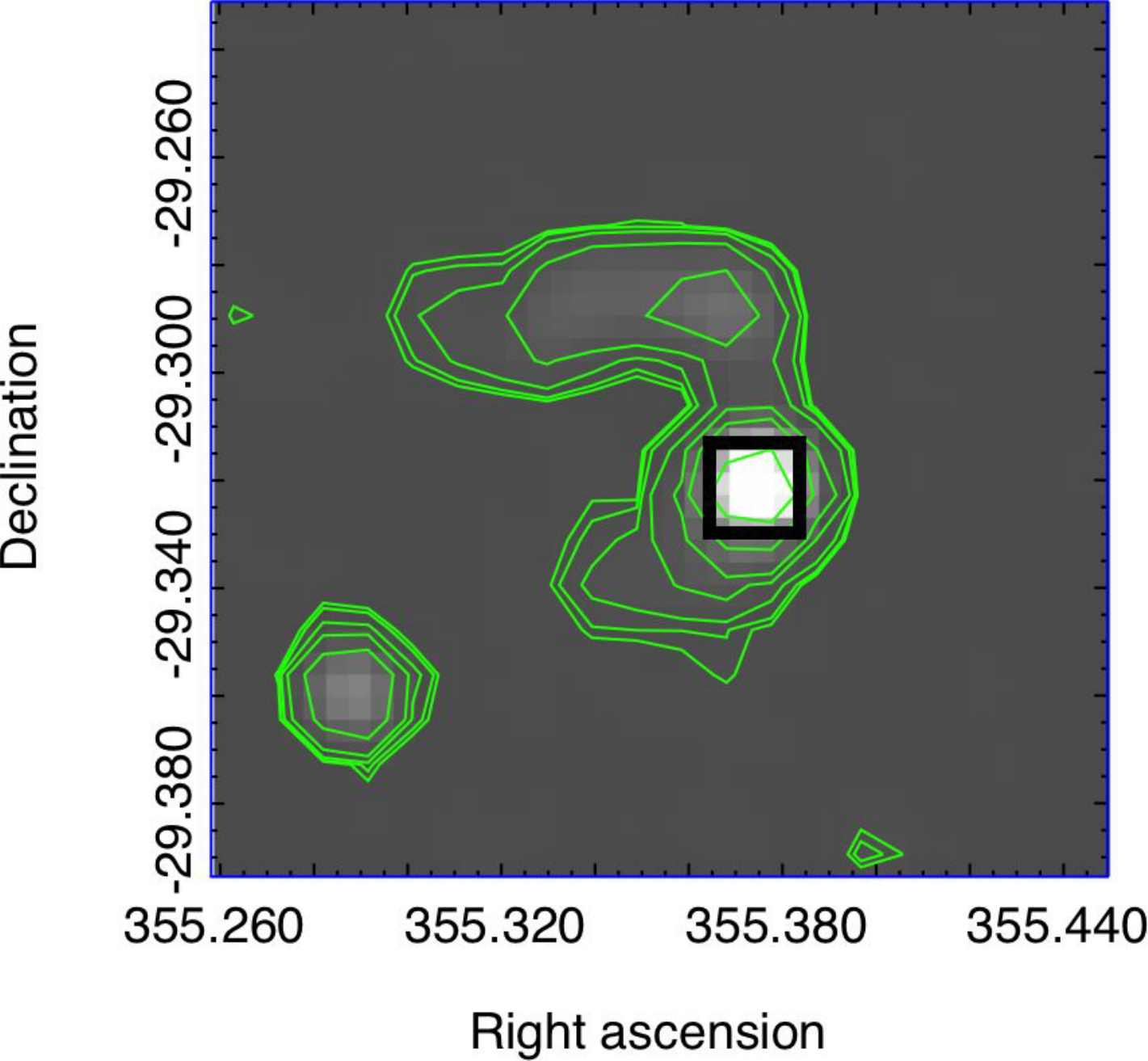} 
\end{minipage}%
\hspace*{5mm}
\begin{minipage}[]{0.22\textwidth}
\includegraphics[width=3.0cm]{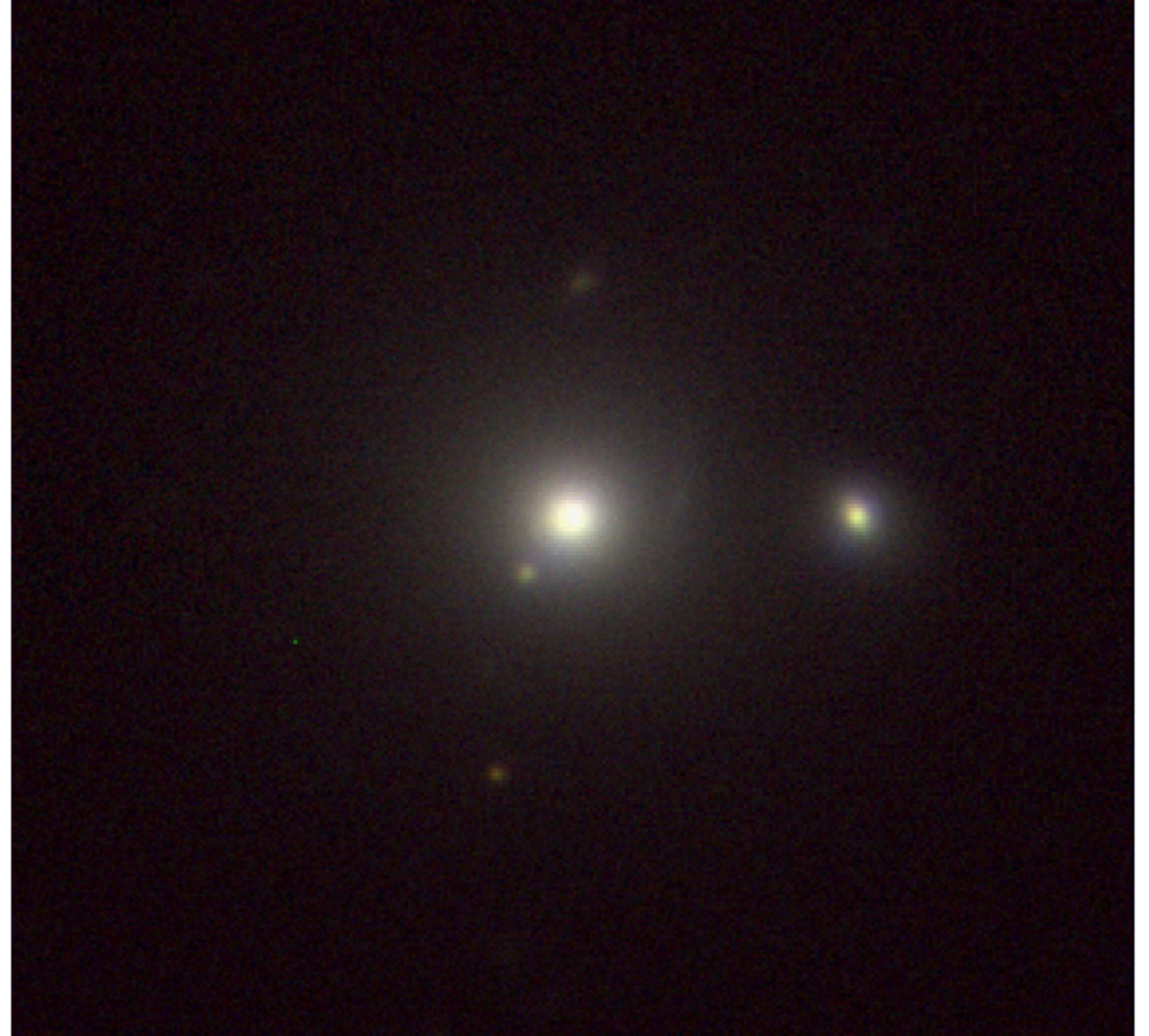}
\end{minipage}%
\vspace*{1mm}
\begin{minipage}[]{0.25\textwidth}
\includegraphics[width=4.0cm]{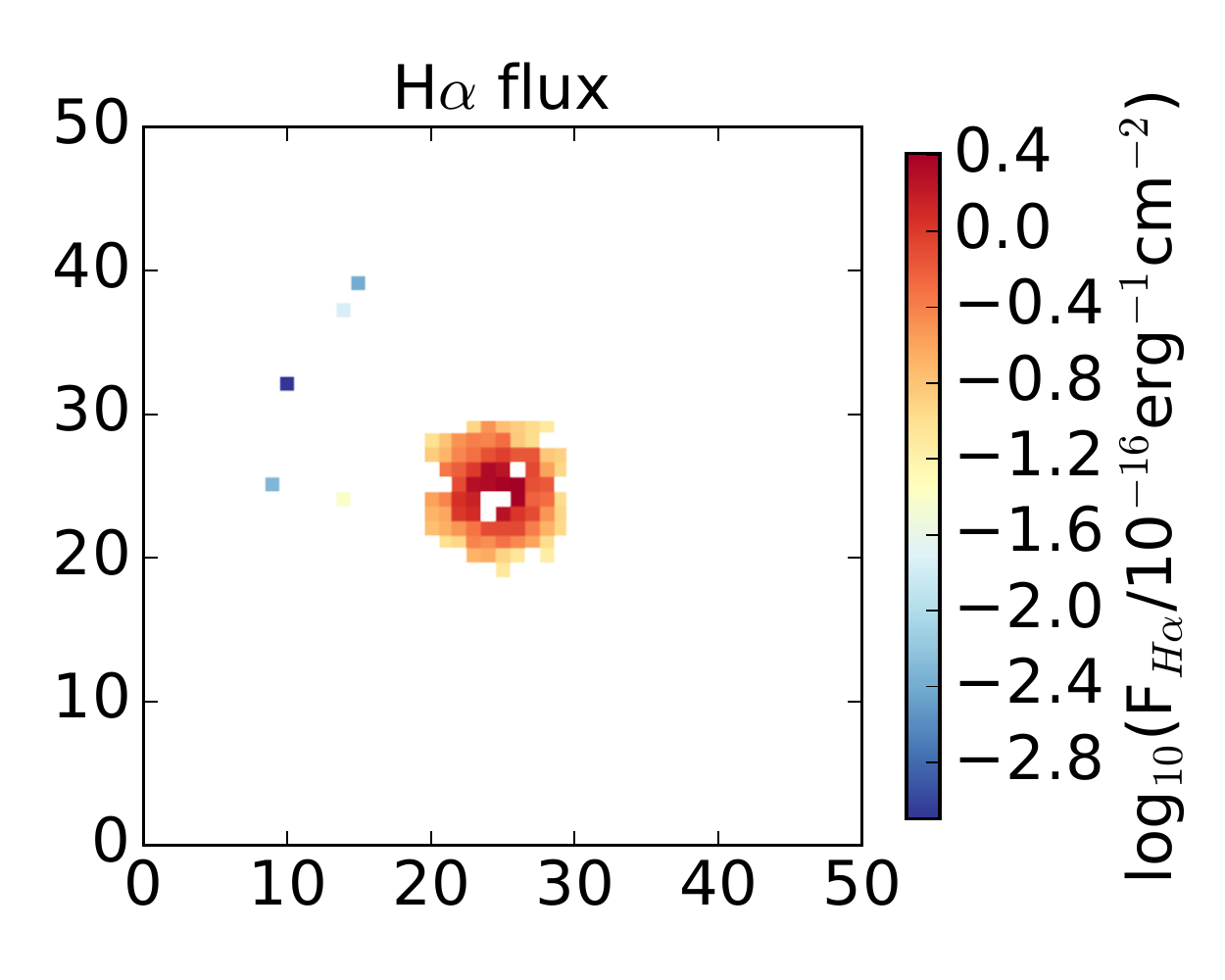}
\end{minipage}%
\begin{minipage}[]{0.25\textwidth}
\includegraphics[width=4.0cm]{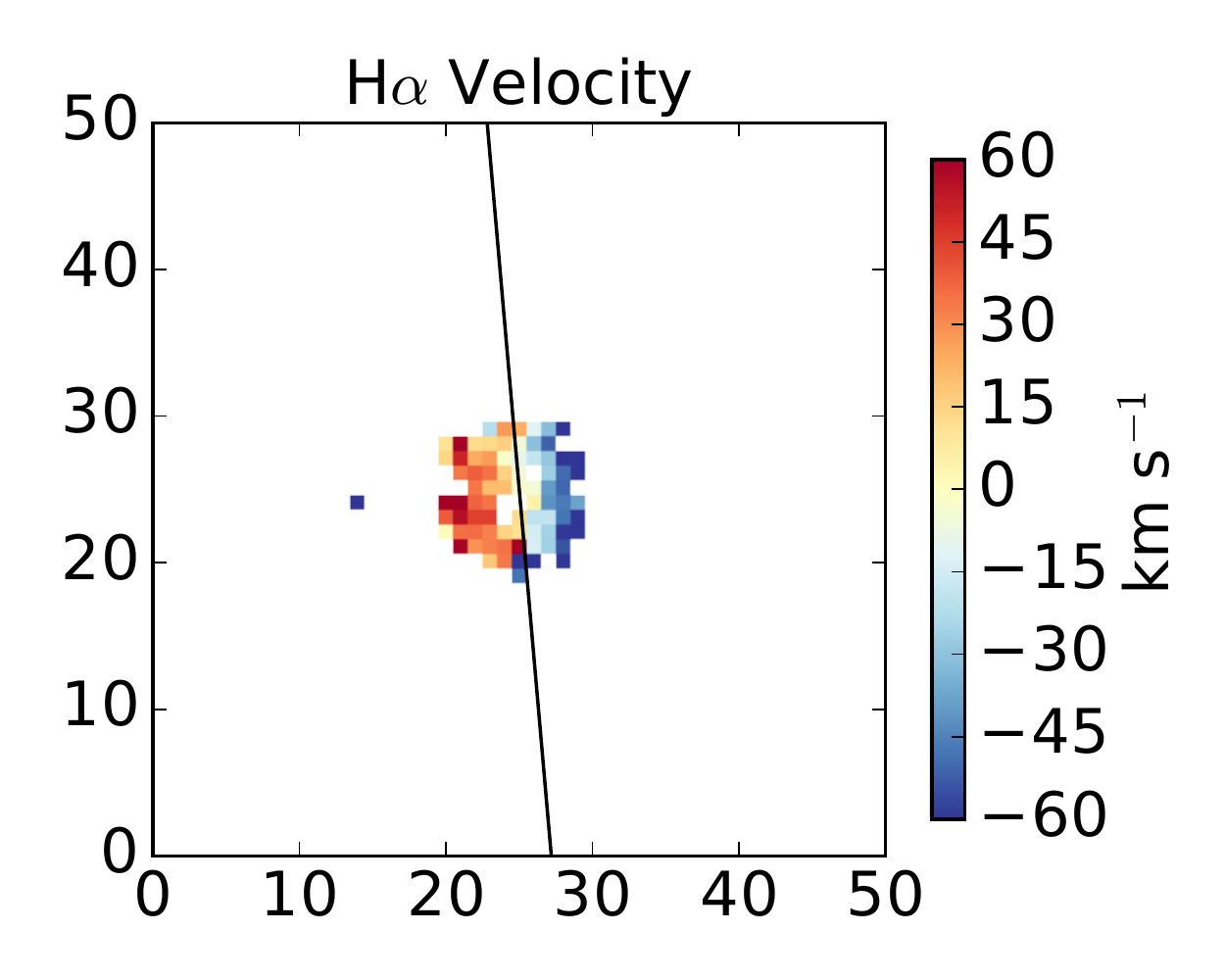}
\end{minipage}%
\vspace*{1mm}
\begin{minipage}[]{0.25\textwidth}
\includegraphics[width=4.0cm]{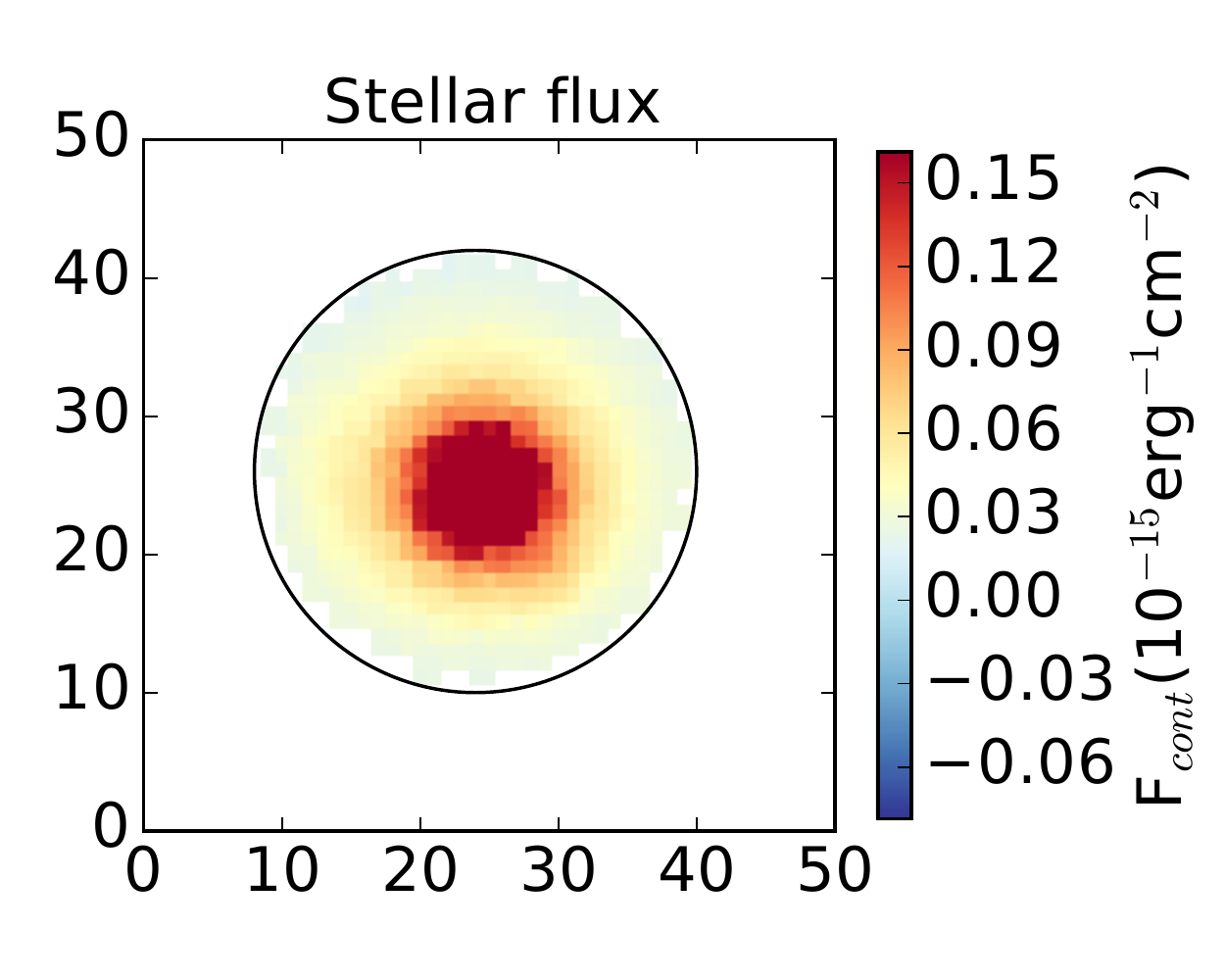}
\end{minipage}%
\begin{minipage}[]{0.25\textwidth}
\includegraphics[width=4.0cm]{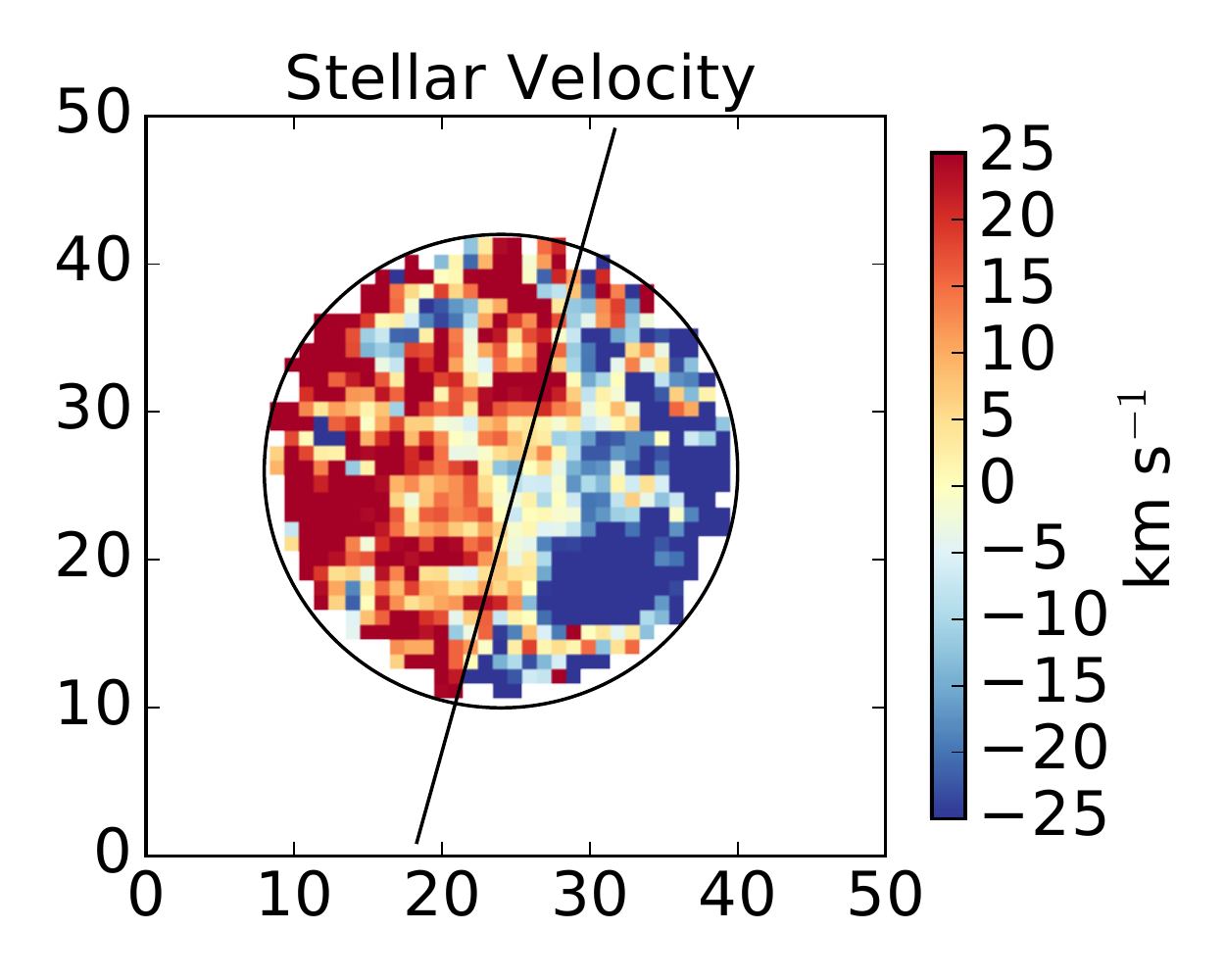}
\end{minipage}%
\vspace*{1mm}
\caption{Example of cluster galaxy where the gradient in the gas dynamics is not likely to be due to rotation in a gas disk. Instead, this source has a wide-angled tail morphology at 1.4\,GHz shown in the NVSS log contours (top left), which is evidence that it is moving rapidly through the ICM (apparent motion is to the right in this image). The VLT Survey Telescope's ATLAS (VST/ATLAS) survey false-colour g, r, i image (top right) fits within the black square in the radio image. The remaining panels show the H$\alpha$ and stellar flux and velocity.
There is very little gas left and it is compact in the centre of the galaxy, but the small amount that has not been stripped has a clear gradient in the gas that is aligned with the expected gas velocity gradient if the gas is being ram pressure stripped. 
}
\label{clusterpics}
\end{figure}

Here are highlighted examples of misaligned cluster galaxies in Fig.~\ref{cluster_radius} in which ram pressure stripping or cluster processes cause misalignment with no gas accretion involved:
\begin{itemize}
\item Number 2:  
is an ETG with
high mass ($10^{11.47}$\Msol). The low apparent cluster radius and very large velocity of 1764\,km\,s$^{-1}$ (cluster velocity dispersion $\sigma_{200} = 840$\,km\,s$^{-1}$)
suggests it is likely to be close to the cluster centre. Dynamical friction can result in massive galaxies preferentially falling to the centre of the cluster where they may then rapidly merge with existing galaxies \citep{McG09,Coo05}.  
It has a giant narrow-angled tail source imaged with the VLA at 4.9GHz \citep{Fer99}, confirming the influence of the intra-cluster medium on this galaxy as it falls in to the cluster. The stellar rotation is regular. Notably, the little gas that it has, has a one-sided velocity gradient 
in the direction of motion indicated by the radio structure.
The velocity gradient in the gas is therefore likely to {\it not} be due to rotation of the gas but a result of ram pressure stripping.
There are several galaxies with narrow-angled tail radio structures like this in the sample, and one ETG example is shown in Fig.~\ref{clusterpics}.  
There are at least four other large wide- and narrow-angled tail radio sources that did not make it into our sample because there was insufficient gas to measure a PA. 
Therefore, misalignment in cluster galaxies is not necessarily an indicator of the rate of externally acquired gas, but in these cases is instead due to gas stripping distorting the dynamics of existing gas within in-falling galaxies.
\item Number 3: 
This galaxy is an edge-on spiral galaxy falling face-on into the cluster. It is also relatively high mass ($10^{10.61}$\Msol) and while it has no radio emission, it has evidence of ram pressure stripping because the gas is offset to one side of the disk, and broadening of the lines indicate the H$\alpha$ may be shock excited by the ram-pressure.
\item Number 7 
is a LTG with counter-rotating gas, indicating that it has acquired external gas that has had time to settle into a counter-rotating rotation without being stripped. 
If it is on first approach the gas may have been accreted outside the cluster and settled to counter-rotating during the infall time (typically $>1$Gyr). The gas may have been retained because the galaxy is high mass ($10^{10.72}$\Msol), low velocity (472\,km\,s$^{-1}$), in a low virial mass cluster and falling in edge-on which may shield the gas \citep{Moo99}.
\end{itemize}

We have shown with this small population that the misalignment of gas in our cluster galaxies has different drivers to that in the field/group sample. The main origin of misaligned gas in the centre of clusters appears to be dynamical friction of gas-rich galaxies, which puts large galaxies in the centre of a cluster where the galaxy will have its dynamics impacted by ram pressure.
A larger sample of cluster galaxies is required to further decipher the impacts of the cluster environment on the origins of gas, and this will be investigated further in a later paper with the full SAMI Galaxy Survey cluster sample.

\section{Summary}
\label{Conclusion}

The PA offset between the gas and stellar rotation in a sample of 1213 SAMI Galaxy Survey Galaxies was investigated, resulting in 618 galaxies with fitted gas and stellar PAs. A strong correlation was identified between gas/stellar misalignment fraction and galaxy morphology (and S\'{e}rsic index) in which misaligned galaxies are predominantly ETGs.  The correlation with stellar mass, colour or local environment was not as significant. 
The expected dependence of dynamical settling time of misaligned gas {\it after} accretion due to torques towards the stellar disk  was derived and is dependent on the intrinsic ellipticity and the initial angular momentum of the incoming gas. Dynamical settling time is calculated to be longer in ETGs but not by enough to account for the observed misaligned galaxy distribution. 

Analysis of the distribution of PA offsets in ETGs and LTGs demonstrates that the fraction of galaxies with misaligned stellar and gas rotations is {\it not} an indicator of the galaxy merger rate or the fraction of galaxies with externally accreted gas. The distributions instead support simulations in which gas can be accreted from any angle in ETGs but in LTGs there is a strong predominance for external gas to be accreted with an angular momentum axis aligned with that of the stars, or torqued towards existing gas. 

Based on the comparison of galaxies in field and group environments, we propose that counter-rotating galaxies can result from accretion of gas more massive than that in the original galaxy disk, but are unlikely if the accreted gas mass is smaller.
The main source of accreted gas is proposed to be from halo or filament accretion rather than mergers, based on both groups and 5th nearest neighbour surface density misalignment results, plus a comparison of misaligned ETG and LTG rates to predictions from simulations.
 
In cluster environments the cluster-driven processes have a larger impact on misaligned gas and stars than the galaxy morphology or stellar mass. The small sample of misaligned cluster galaxies are influenced by gas-stripping processes such that the gas velocity gradient may not indicate a rotating disk. 

Since the overall fractions of misaligned galaxies are only $\sim11$\% in all environments, and the angular momentum axis of accreted star-forming gas is intricately linked to large-scale structure, a very much larger sample would be required to further subdivide morphological types into local and global environments to disentangle the drivers of misalignment and their link to gas origins as proposed in this paper. Such a statistically large sample will be possible with the Hector Galaxy Survey \citep{Bry16,JBH15,Law14}, beginning in 2019.

\subsection*{Acknowledgements}

The SAMI Galaxy Survey is based on observations made at the Anglo-Australian Telescope. The Sydney-AAO Multi-object Integral-field spectrograph (SAMI) was developed jointly by the University of Sydney and the Australian Astronomical Observatory.  Initial seed funding came from Bland-Hawthorn's ARC Federation Fellowship (2008-13). The SAMI input catalogue is based on data taken from the Sloan Digital Sky Survey, the GAMA Survey and the VST ATLAS Survey. The SAMI Galaxy Survey is funded by the Australian Research Council Centre of Excellence for All-sky Astrophysics (CAASTRO), through project number CE110001020, and other participating institutions. The SAMI Galaxy Survey website is http://sami-survey.org/ .

Parts of this research were supported by the Australian Research Council Centre of Excellence for All Sky Astrophysics in 3 Dimensions (ASTRO 3D), through project number CE170100013.

We would like to thank the Australia Astronomical Observatory and University of Sydney instrumentation groups for their support and dedication to making the SAMI instrument.
The SAMI survey has greatly benefitted from the excellent technical support offered by the AAO in Sydney and by site staff at the Anglo-Australian Telescope.

GAMA is a joint European-Australasian project based around a spectroscopic campaign using the Anglo-Australian Telescope. The GAMA input catalogue is based on data taken from the Sloan Digital Sky Survey and the UKIRT Infrared Deep Sky Survey. Complementary imaging of the GAMA regions is being obtained by a number of independent survey programs including GALEX MIS, VST KiDS, VISTA VIKING, WISE, Herschel-ATLAS, GMRT and ASKAP providing UV to radio coverage. GAMA is funded by the STFC (UK), the ARC (Australia), the AAO, and the participating institutions. The GAMA website is: http://www.gama-survey.org/.

JJB acknowledges support of an Australian Research Council Future Fellowship (FT180100231). SMC acknowledges the support of an Australian
Research Council Future Fellowship (FT100100457). 
LC acknowledges support under the Australian Research Council's Discovery 
Projects funding scheme (DP130100664) and Future Fellowship (FT180100066) funded by the Australian Government.  
Support for AMM is provided by NASA through Hubble Fellowship grant \#HST-HF2-51377 awarded by the Space Telescope Science Institute, which is operated by the Association of Universities for Research in Astronomy, Inc., for NASA, under contract NAS5-26555. NS acknowledges support from a University of Sydney Postdoctoral Research Fellowship. SB acknowledges the funding support from the Australian Research Council through a Future Fellowship (FT140101166). JvdS is funded under Bland-Hawthorn's ARC Laureate Fellowship (FL140100278). MSO acknowledges the funding support from the Australian Research Council through a Future Fellowship (FT140100255). C.F. acknowledges funding provided by the Australian Research Council (Discovery Projects DP150104329 and DP170100603, and Future Fellowship FT180100495), and the Australia-Germany Joint Research Cooperation Scheme (UA-DAAD).

Based on data products (VST/ATLAS) from observations made with ESO Telescopes at the La Silla Paranal Observatory under programme ID 177.A-3011(A-J).

This research has made use of the NASA/IPAC Extragalactic
Database (NED), which is operated by the Jet
Propulsion Laboratory, California Institute of Technology,
under contract with the National Aeronautics and Space
Administration. 

Funding for SDSS-III has been provided by the Alfred P. Sloan Foundation, the Participating Institutions, the National Science Foundation, and the U.S. Department of Energy Office of Science. The SDSS-III web site is http://www.sdss3.org/.

We thank the referee for suggestions that improved the clarity and flow of this paper.

\section*{Appendix 1}
\label{App1}
\begin{figure}
\begin{minipage}[]{0.5\textwidth}
\includegraphics[width=5.3cm]{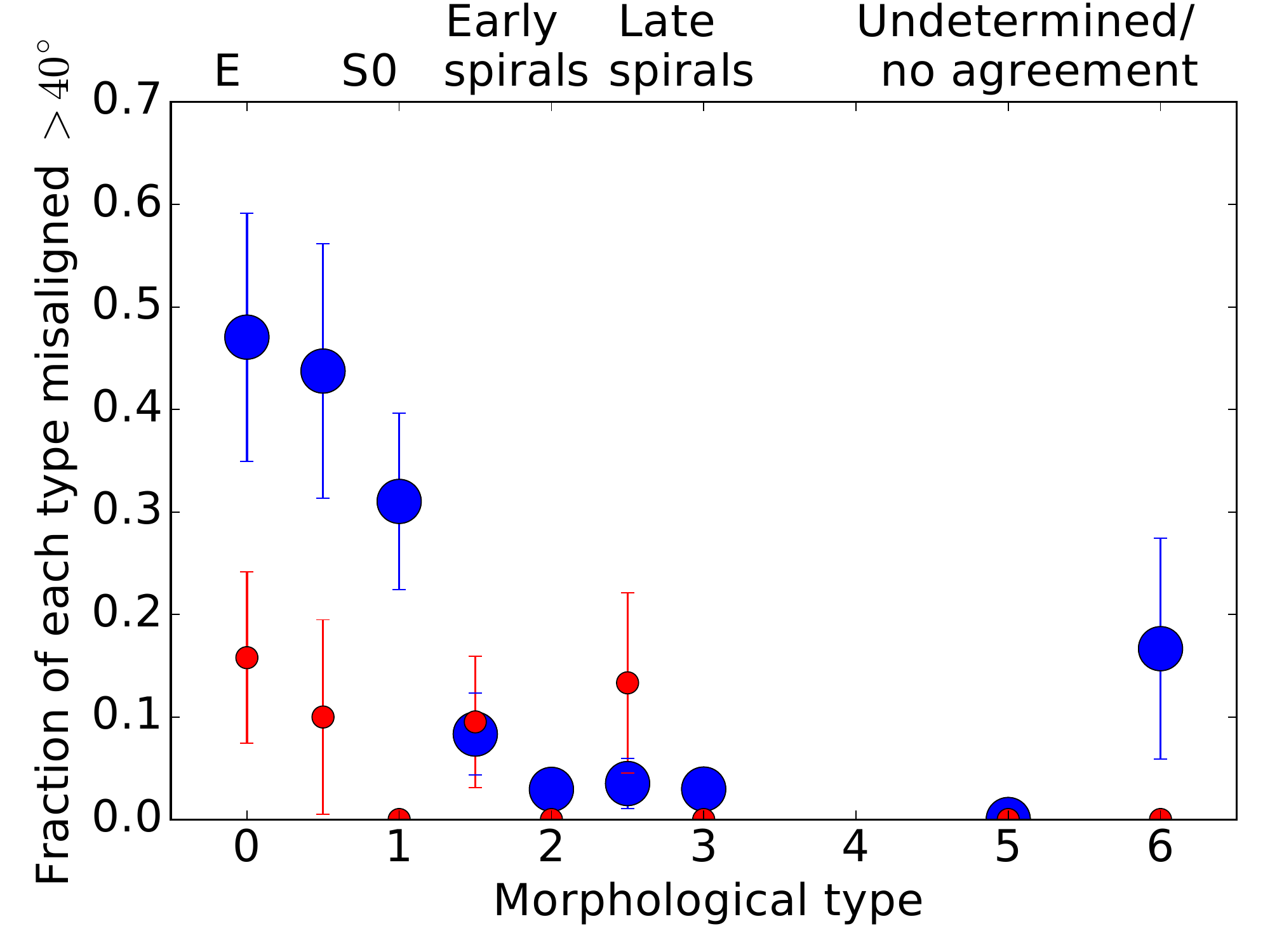}
\end{minipage}%
\vspace*{0.1mm}
\begin{minipage}[]{0.5\textwidth}
\includegraphics[width=5.3cm]{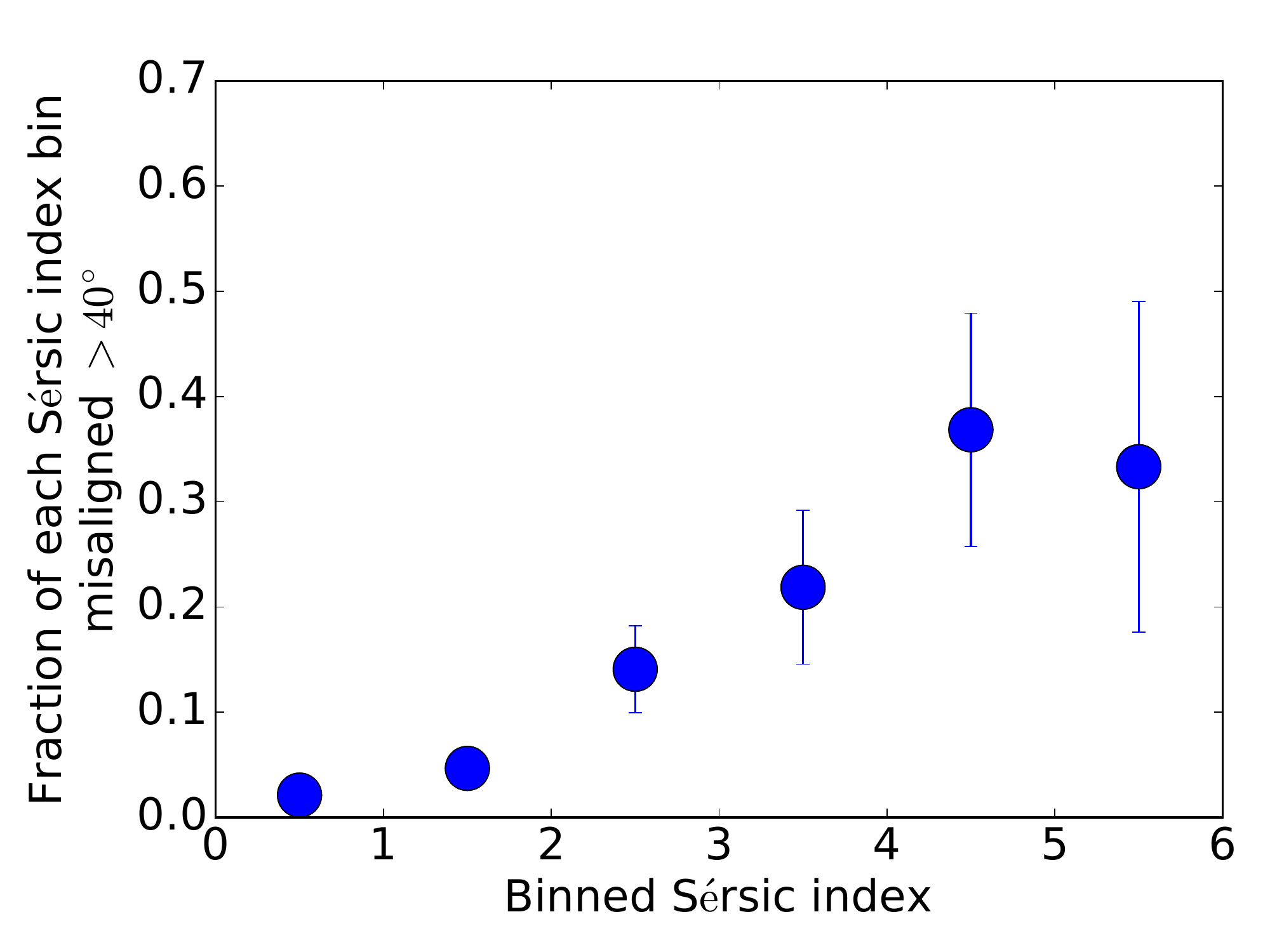}
\end{minipage}%
\vspace*{0.1mm}
\begin{minipage}[]{0.5\textwidth}
\includegraphics[width=5.3cm]{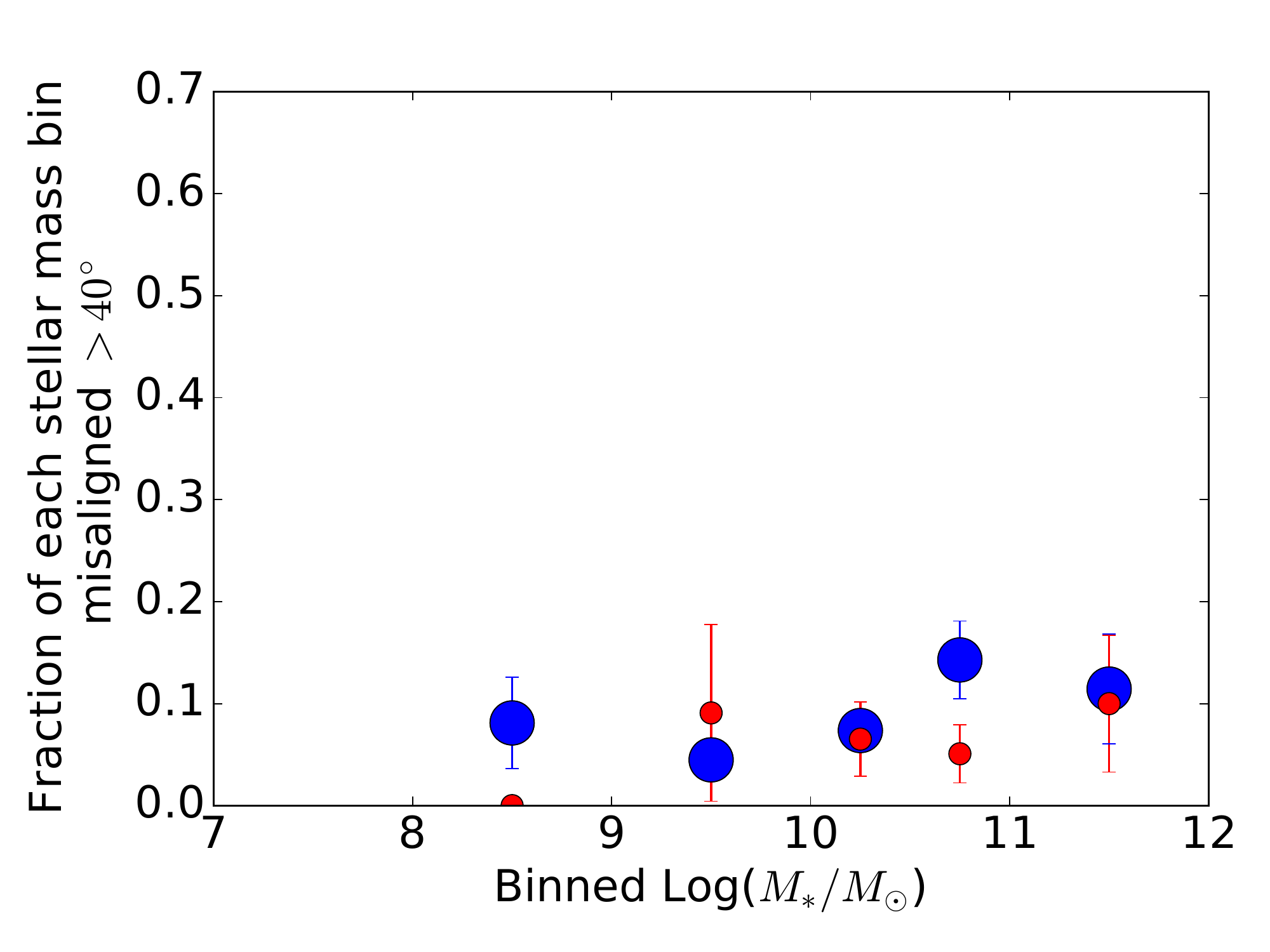}
\end{minipage}%
\vspace*{0.1mm}
\begin{minipage}[]{0.5\textwidth}
\includegraphics[width=5.3cm]{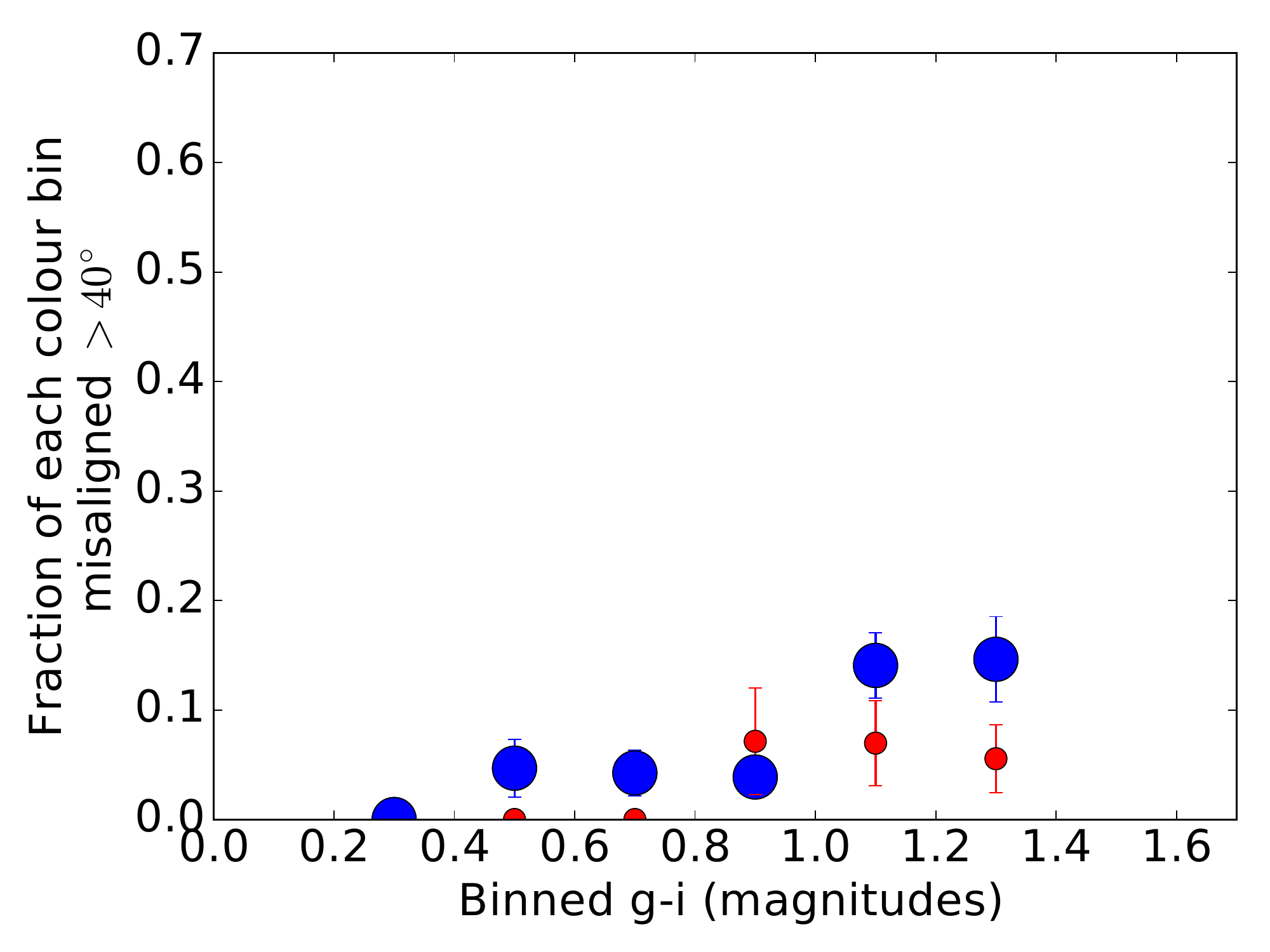}
\end{minipage}%
\caption{Fraction of galaxies that are misaligned by greater than $40^{\circ}$ matching the plots shown for PA offset$>30^{\circ}$ in Fig.~\ref{morphtypehist}.  The trends found when the PA offset cut-off was set to $>30^{\circ}$, also apply with the stricter cut-off of PA$>30^{\circ}$ shown here. The colours and annotations are as described in Fig.~\ref{morphtypehist}.
}
\label{morphtypehistAppx}
\end{figure}

Figure ~\ref{morphtypehistAppx} shows the equivalent plots to the left column of Fig.~\ref{morphtypehist}, but with a PA offset cut off of $40^{\circ}$. Similar trends are found for both the $30^{\circ}$ and $40^{\circ}$ cut-offs.

\section*{Appendix 2 - Derivation of dynamical settling time equation}
\label{App2}
In our calculation, the dynamical settling time $t_{s}$, depends on (a) the initial misalignment $\phi$ of the disk from the stellar plane, (b) the mass, $M_{sm}$, of the galaxy within the semi-major axis, $A$, of the galaxy's stellar mass distribution (not the semi-major axis of the gas disk), (c) the angular velocity of the gas disk $\omega=V_{rot}$/R for rotation velocity $V_{rot}$ at disk radius R, and (d) the shape of the galaxy's stellar mass distribution given by the 
ellipticity, defined as  $\epsilon = 1 - (\frac{C}{A})$ in the axisymmetric case where $A=B$. A and C are intrinsic or de-projected major and minor axes respectively, as opposed to the projected ellipticity, $\epsilon_{proj}=1 - \frac{b}{a}$ which is shown in Fig.~\ref{Re_ellip}. The inclination, i,  of the angular momentum vector to the line of sight is given by
\begin{equation}
cos^2i = \frac{[(\frac{b}{a})^2 - (\frac{C}{A})^2]}{[1-(\frac{C}{A})^2]}.
\label{incl}
\end{equation}
for apparent axis lengths of a and b. 

A particle in an ellipsoidal gravitational potential, rotating at $\phi$ degrees from the major axis rotation angle, will have a precession rate of that orbit given by classical mechanics \citep{Toh82} as
\begin{equation}
\dot{g} =-\frac{3}{2} \omega J_{2} cos(\phi) =-\frac{3 V_{rot} J_{2} cos(\phi) }{2 R}  
\label{gdot}
\end{equation}
where $J_{2}$ is the dimensionless second coefficient of the quadrupole moment of the gravitational potential given by
\begin{equation}
J_{2} = \frac{C_{i}-0.5(B_{i}+A_{i})}{M_{P} A_{r}^{2}}
\label{J2}
\end{equation}
where $A_{r}$ is the radius of the enclosed mass $M_{P}$ within the galaxy's gravitational stellar mass distribution and $A_{r}$=A for $M_{P}=M_{sm}$, the mass within the semi-major axis radius, and $A_{i}$, $B_{i}$ and $C_{i}$ are the moments of inertia about the three axes of the distribution shape, and are related to the intrinsic axis lengths by
\begin{equation}
\begin{aligned}
A_{i}= \frac{M_{P}}{5}(B^{2} + C^{2})\\
B_{i}= \frac{M_{P}}{5}(A^{2} + C^{2})\\
C_{i}= \frac{M_{P}}{5}(A^{2} + B^{2}).
\end{aligned}
\label{inertia}
\end{equation}

In a triaxial shape $A \neq B \neq C$, but in a generalised ellipsoid $A_{i}$=$B_{i}$ and hence A=B. Therefore,
\begin{equation}
J_{2} = \frac{A^{2} - C^{2}}{5A_{r}^{2}}
\end{equation}
If $A_{r}$=A, then
\begin{equation}
J_{2} = -\frac{2\epsilon-\epsilon^2}{5}
\label{J2e}
\end{equation}
The time taken for that rotating disk to precess onto the galaxy plane is then 
\begin{equation}
t_{s} \propto \frac{1}{\dot{g}} \propto \frac{R}{V_{rot}\, (2\epsilon-\epsilon^2) \, cos(\phi)}
\label{tdApp}
\end{equation}

\subsection*{Comparison of the R/V term between ETGs and LTGs}

\begin{figure}
\centering
\includegraphics[width=9.0cm]{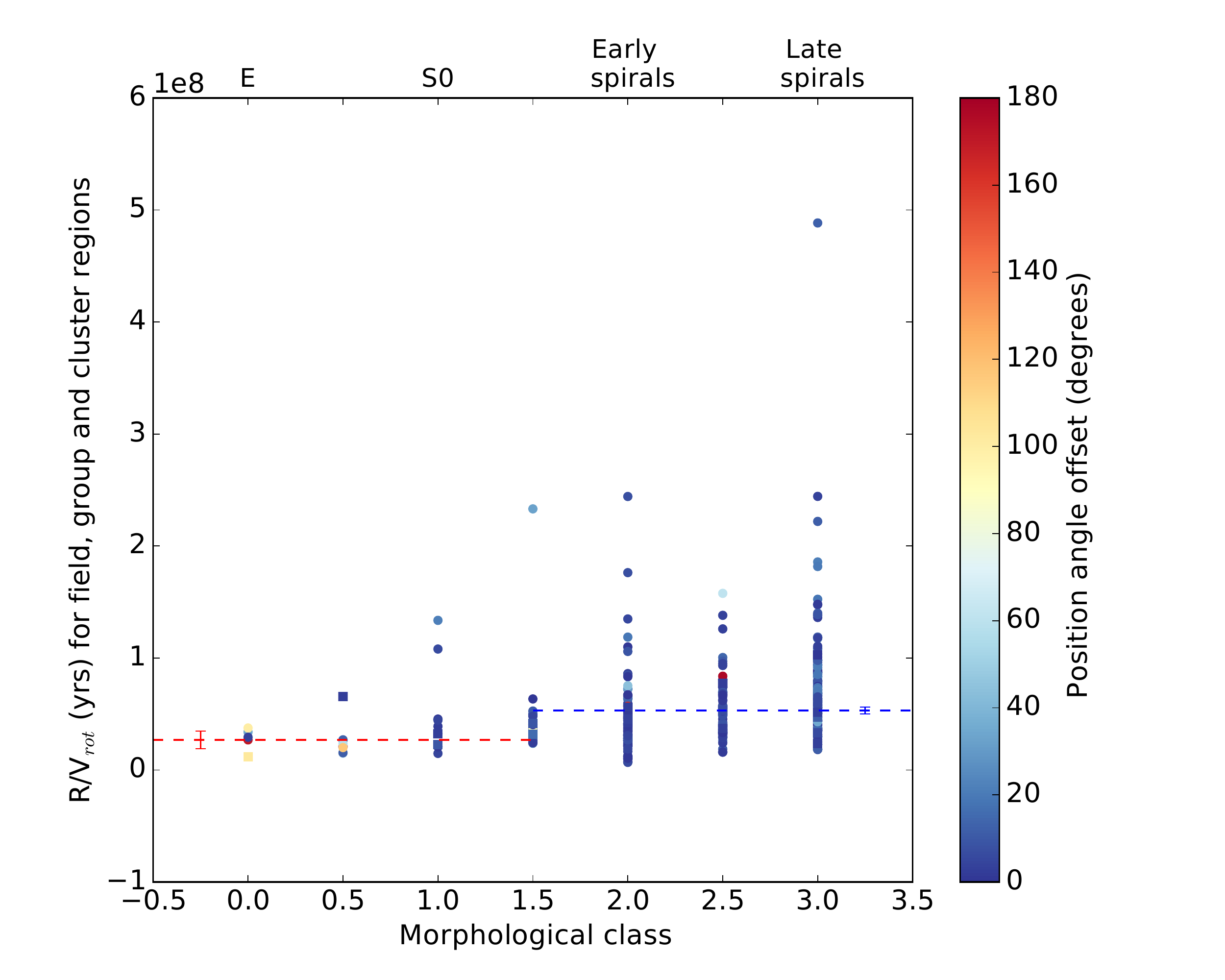}
\caption{Morphological type versus R/V (top) 
as defined in Eq.~\ref{td}, colour coded by gas/stellar PA offset. The dashed lines mark the median R/V values for ETGs (red) and LTGs (blue). The error bars on the dashed lines indicate the standard error of the median. The rotation velocities are measured using the method in \citet{Blo17b} and are measured at 2.2R$_{e}$ which is then used for the R value. 
}
\label{tdynMorph}
\end{figure}

The $R/V$ term in Eq.~\ref{tdApp} is similar for ETGs and LTGs at the radius where the velocity has flattened out to a maximum value. While not strictly equal, there is a substantial overlap in measured values because scatter is dominated by inclination corrections (that are particularly difficult for ETGs), and differences in measurement methods for $V_{rot}$.  
\citet{She09} showed only a weak dependence of rotational velocity on morphological type within spiral galaxies, confirming that the bulge has little effect and their typical values for spirals were $V_{max}\sim200$\,km/s. ETGs measured in the ATLAS$^{3D}$ sample in \citet{Dav13}, had non-inclination-corrected rotation velocities ranging between $\sim50-200$\,km/s, and therefore \citet{Dav16} adopted a value or $V_{rot} \sim 200$\, km/s for ETGs.

$R/V$ at maximum V can be measured for at least a subset of the SAMI v0.9 sample. The $V_{rot}$ values for the field and group SAMI galaxies from \citet{Blo17a, Blo17b} have been extended to include measurements for the cluster galaxies using the same method. Galaxies are fitted using kinemetry then an arctan form is fitted to the 2-d rotation curve from the first kinematic moment in order to find the velocity at 2.2 effective radii where the rotation curves are assumed to have flattened \citep[see][for details]{Blo17b}. The median values of $R/V$ shown in Fig.~\ref{tdynMorph} are within a factor of two (ratio of $0.51\pm0.04$) for the ETG and LTG samples. However, there are only 23 ETGs shown because there are many ETGs for which this method does not give a measure of rotation velocity because the intensity-weighted line-of-sight velocity (1st order moment) map from kinemetry is not well fitted if there is small amount of H-alpha gas or if that gas has patchy distribution. In those cases the bulk rotation needed to fit the PA is clear but the kinemetry fit fails. 

\bsp	
\label{lastpage}

\begin{thebibliography}{}
\bibitem[Algorry et al.(2014)]{Alg14} Algorry D. G., Navarro J. F., Abadi M. G., Sales L.V., Steinmetz M., Piontek F., 2014, MNRAS, 437, 3596
\bibitem[Allen et al.(2015)]{All15} Allen J. et al., 2015, MNRAS, 446, 1567.
\bibitem[Baldry et al.(2010)]{Bal10} Baldry I. K., et al., 2010, MNRAS, 404, 86
\bibitem[Barrera-Ballesteros et al.(2014)]{Bar14} Barrera-Ballesteros J. K., et al., 2014, A\&A 568, A70 
\bibitem[Barrera-Ballesteros et al.(2015)]{Bar15} Barrera-Ballesteros J. K., et al., 2015, A\&A 582, 21
\bibitem[Bauermeister, Blitz \& Ma(2010)]{Bau10} Bauermeister A., Blitz L., Ma C., 2010, ApJ, 717, 323
\bibitem[Bekki(1998)]{Bek98} Bekki K., 1998, ApJ, 499, 635
\bibitem[Bekki(1999)]{Bek99} Bekki, K. 1999, ApJL, 510, L15
A\&A, 292, 13
\bibitem[Binney \& de Vaucouleurs(1981)]{Bin81} Binney J., de Vaucouleurs G., 1981, MNRAS, 194,679
\bibitem[Bland, Taylor \& Atherton(1987)]{JBH87} Bland J., Taylor K., Atherton P.D., 1987, MNRAS, 228, 595 
\bibitem[Bland-Hawthorn et al.(2011)]{JBH2011} Bland-Hawthorn J. et al., 2011, Optics Express, 19, 2649
\bibitem[Bland-Hawthorn(2015)]{JBH15} Bland-Hawthorn J., 2015, in Ziegler B. L., Combes F., Dannerbauer H., Verdugo M., eds, IAU Symposium Vol. 309, Galaxies in 3D across the Universe. pp 21Ð28 (arXiv:1410.3838), doi:10.1017/S1743921314009247
\bibitem[Bloom et al.(2017a)]{Blo17a} Bloom J. V., et al., 2017, MNRAS, 465, 123
\bibitem[Bloom et al.(2017b)]{Blo17b} Bloom J. V., et al., 2017, MNRAS, in press
\bibitem[Bournaud \& Combes(2003)]{Bou03} Bournaud F., Combes F., 2003, A\&A, 401, 817
\bibitem[Brook et al.(2008)]{Bro08} Brook et al. ApJ, 689, 678, 2008
\bibitem[Brooks et al.(2009)]{Bro09} Brooks A.~M., Governato F., Quinn T., Brook C.~B., Wadsley J., 2009, ApJ, 694, 396
\bibitem[Brough et al.(2013)]{Bro13} Brough S., et al., 2013, MNRAS, 435, 2903
\bibitem[Brough et al.(2017)]{Bro17} Brough S., et al., 2017, arXiv:1704.01169
\bibitem[Bryant et al.(2011)]{JB2011} Bryant J. J., O'Byrne J. W., Bland-Hawthorn J., Leon- Saval S. G., 2011, MNRAS, 415, 2173
\bibitem[Bryant et al.(2012)]{JB2012} Bryant J. J., et al., 2012, McLean I. S., Ramsay S. K., Takami H., eds., in Ground-based and Airborne Instrumentation for Astronomy IV. Proceedings of the SPIE, Volume 8446, article id. 84460X.
\bibitem[Bryant et al.(2014)]{JB2014} Bryant J. J., Bland-Hawthorn J., Fogarty L. M. R., Lawrence J. S., Croom S. M., 2014, MNRAS, 438, 869
\bibitem[Bryant et al.(2015)]{JB2015} Bryant J. J., et al., 2015, MNRAS, 447, 2857
\bibitem[Bryant et al.(2016)]{Bry16} Bryant J. J., et al., 2016, Proc. SPIE 9908-52, arXiv:1608.03921
\bibitem[Bundy et al.(2015)]{Bun15} Bundy K. et al., 2015, ApJ, 798, 7
\bibitem[Bureau \& Chung(2006)]{Bur06} Bureau M., Chung A., 2006, MNRAS, 366, 182
\bibitem[Cappellari \& Emsellem(2004)]{Cap04} Cappellari M., Emsellem E., 2004, PASP, 116, 138
\bibitem[Cappellari et al.(2007)]{Cap07} Cappellari M. et al., 2007, MNRAS, 379, 418
\bibitem[Cappellari et al.(2011)]{Cap11} Cappellari M. et al., 2011a, MNRAS, 413, 813
\bibitem[Cen(2014)]{Cen14b} Cen R., 2014, ApJ, 781, 38
\bibitem[Ceverino et al.(2016)]{Cev16} Ceverino D., Almeida J. S., Tu\~n\'on C. M., Dekel A., Elmegreen B. G., Elmegreen D. M., Primack J., 2016, MNRAS, 457, 2605
\bibitem[Chen et al.(2016)]{Che16} Chen, Y.-M. et al., 2016, Nat. Commun., 7, 13269
\bibitem[Chung et al.(2012)]{Chu12} Chung A., Bureau M., van Gorkom J. H., Koribalski B., 2012, MNRAS, 422, 1083
\bibitem[Connors et al.(2006)]{Con06} Connors T. W., Kawata D., Bailin J., Tumlinson J., Gibson B. K., 2006, ApJ, 646, L53
\bibitem[Cooray \& Milosavlijevic(2005)]{Coo05} Cooray A., Milosavlijevis M., 2005, ApJL, 627, 85
\bibitem[CortŽs, Kenney \& Hardy(2015)]{Cor15} CortŽs J. R., Kenney J. D. P., Hardy E., 2015, ApJS, 216, 9
\bibitem[Cortese et al.(2016)]{Cor16} Cortese L., et al., 2016, MNRAS, in press.
\bibitem[Cox, Sparke \& van Moorsel(2006)]{Cox06} Cox A. L., Sparke L. S., van Moorsel G. 2006, AJ, 131, 828
\bibitem[Croom, Saunders \& Heald(2004)]{Cro04} Croom S., Saunders W., Heald R., 2004, AAONw, 106, 12
\bibitem[Croom et al.(2012)]{Cro12} Croom S., et al., 2012, MNRAS 421, 872 
\bibitem[Danovich et al.(2015)]{Dan15} Danovich M., Dekel A., Hahn O., Ceverino D., Primack J., 2015, MNRAS 449, 2087 
\bibitem[Davis et al.(2011)]{Dav11} Davis T. A., 2011, MNRAS, 417, 882
\bibitem[Davis et al.(2013)]{Dav13} Davis T. A., 2013, MNRAS, 429, 534
\bibitem[Davis et al.(2016)]{Dav16} Davis T. A., 2016, MNRAS, 457, 272
\bibitem[Dressler(1980)]{Dre80} Dressler A., 1980, ApJ, 236, 351
\bibitem[Driver et al.(2011)]{Dri11} Driver S. P. et al., 2011, MNRAS, 413, 971
\bibitem[Dubois et al.(2014)]{Dub14} Dubois Y.,  et al, 2014, MNRAS, 444, 1453
\bibitem[Emsellem et al.(2011)]{Ems11} Emsellem E., et al. 2011, MNRAS, 414, 888
\bibitem[Engel et al.(2010)]{Eng10} Engel, H., et al. 2010, A\&A, 524, A56
\bibitem[Epinat et al.(2008)]{Epi08} Epinat B., et al. 2008, MNRAS, 388, 500
\bibitem[Fathi et al.(2009)]{Fat09} Fathi K., et al. 2009, ApJ, 704, 1657
\bibitem[Feretti et al.(1999)]{Fer99} Feretti L., Dallacasa D., Govoni F., Giovannini G., Taylor G. B., Klein U., 1999, A\&A, 344, 472
\bibitem[Foster et al.(2017)]{Fos17} Foster C., et al., 2017, MNRAS, 472, 966
\bibitem[Fogarty et al.(2014)]{Fog2014} Fogarty L. M. R. et al., 2014, MNRAS, 443, 485
\bibitem[Franx \& De Zeeuw(1992)]{Fra92} Franx M., De Zeeuw T., 1992, ApJ, 392, L47
\bibitem[Gallagher et al.(2002)]{Gal02} Gallagher J. S., Sparke L. S., Matthews L. D., Frattare L. M., English J., Kinney A. L., Iodice E.,  Arnaboldi M., 2002, ApJ, 568, 199
\bibitem[Ganda et al.(2006)]{Gan06} Ganda K., Falc\'{o}n-Barroso J., Peletier R. F., Cappellari M., Emsellem E., McDermis R. M., de Zeeuw P. T., Carollo C. M., 2006, MNRAS, 367, 46
\bibitem[Green et al.(2017)]{Gre17} Green A., et al., 2017, MNRAS, in press.
\bibitem[Gunn \& Gott(1972)]{Gun72} Gunn J. E.,  Gott J. R. 1972, ApJ, 176, 1
\bibitem[Hill et al.(2011)]{Hil11} Hill D., et al., 2011, MNRAS, 412, 765
\bibitem[Hinshaw et al.(2009)]{Hin09}Hinshaw G. et al.,  2009, ApJS, 180, 225
\bibitem[Ho et al.(2016)]{Ho16a} Ho I. et al., 2016, MNRAS, 457, 1257
\bibitem[Ho et al.(2016)]{Ho16b} Ho I. et al., 2016, Ap\&SS, 361, 280 
\bibitem[Houck \&Bregman(1990)]{Hou90} Houck J. C., Bregman J. N., 1990, ApJ, 352, 506
\bibitem[Jin et al.(2016)]{Jin16} Jin Y., et al., 2016, MNRAS, 463, 913
\bibitem[Johnson et al.(2016)]{Joh16} Johnson H. L., Harrison C. M., Swinbank A. M., Bower R. G., Smail Ian, Koyama Y., Geach J. E., 2106 MNRAS 460, 1059
\bibitem[Kannappan \& Fabricant(2001)]{Kan01} Kannappan S. J., Fabricant D. G., 2001, ApJ, 121, 140
\bibitem[Kelvin et al.(2012)]{Kel12} Kelvin L.S., et al., 2012, MNRAS, 421, 1007
\bibitem[Kennicutt, Tamblyn \& Congdon(1994)]{Ken94} Kennicutt R. C., Tamblyn P., Congdon C. E. ,1994, ApJ, 435, 22
\bibitem[Kere{\v s} et al.(2005)]{Ker05} Kere{\v s} D., Katz N., Weinberg D.~H., 
Dav{\'e} R., 2005, MNRAS, 363, 2
\bibitem[Kere{\v s} et al.(2009)]{Ker09} Kere{\v s} D., Katz N., Fardal M., Dave R., Weinberg D. H. 2009, MNRAS, 395, 160
\bibitem[Krajnovi\'{c} et al.(2006)]{Kra06} Krajnovi\'{c} D., et al., 2006, MNRAS 366, 787
\bibitem[Krajnovi\'{c} et al.(2008)]{Kra08} Krajnovi\'{c} D. et al., 2008, MNRAS, 390, 93
\bibitem[Krajnovi\'{c} et al.(2011)]{Kra11} Krajnovi\'{c} D., et al., 2011, MNRAS 414, 2923
\bibitem[Lagos et al.(2014)]{Lag14} Lagos C. d. P., Davis T. A., Lacey C. G., Zwaan M. A., Baugh C. M.,
Gonzalez-Perez V., Padilla N. D., 2014, MNRAS, 443, 1002
\bibitem[Lagos et al.(2015)]{Lag15} Lagos C. del P., Padilla N. D., Davis T. A., Lacey C. G., Baugh C. M., Gonzalez-Perez V., Zwaan M. A., Contreras S., 2015, MNRAS, 448, 1271
\bibitem[Lake \& Norman(1983)]{Lak83} Lake G., Norman C., 1983, ApJ, 270, 51
\bibitem[Lambas, Maddox \& Loveday(1992)]{Lam92} Lambas D. G., Maddox S. J., Loveday J., 1992, MNRAS, 258, 404
\bibitem[Larson, Tinsley \& Caldwell(1980)]{Lar80} Larson R. B., Tinsley B. M., Caldwell C. N., 1980, ApJ, 237, 692
\bibitem[Lawrence et al.(2014)]{Law14} Lawrence J. S., et al., 2014, in Ground-based and Airborne Instrumentation for Astronomy V. p. 91476Y, doi:10.1117/12.2055734
\bibitem[Liske et al.(2015)]{Lis15} Liske J., et al., 2015, MNRAS, 452, 2087
\bibitem[Maccio, Moore \& Stadel(2006)]{Mac06} Macci\`{o} A.V., Moore B., Stadel J., 2006, ApJ, 636, L25
\bibitem[Martinsson et al.(2013)]{Martinsson13} Martinsson T. P. K., Verheijen M. A. W., Westfall K. B., Bershady M. A., Schechtman-Rook A., Andersen D. R., Swaters R. A., 2013, A\&A, 557, 130
\bibitem[Massey et al.(2007)]{Mas07}   Massey R., et al. 2007, Nature, 445, 286
\bibitem[McGee et al.(2009)]{McG09} McGee S. L., Balogh M. L., Bower R. G., Font A. S., McCarthy I. G., 2009, MNRAS, 400, 937
\bibitem[Mendez-Abreu et al.(2008)]{Men08} Mendez-Abreu J., Aguerri J. A. L., Corsini E. M., Simonneau E., 2008, A\&A, 478, 353
\bibitem[Moore et al.(1999)]{Moo99} Moore B., Lake G., Quinn T., Stadel J., 1999, MNRAS, 304, 465
\bibitem[Mulchaey \& Jeltema(2010)]{Mul10} Mulchaey J. S., Jeltema T. E., 2010, ApJ, 715, 1
\bibitem[Nicholson, Bland-Hawthorn \& Taylor(1992)]{Nic92} Nicholson R.A., Bland-Hawthorn J., Taylor K., 1992, ApJ, 387, 503
\bibitem[Ocvirk, Pichon \& Teyssier(2008)]{Ocv08} Ocvirk P., Pichon C., Teyssier R., 2008, MNRAS, 390, 1326
\bibitem[Osman \& Bekki(2017)]{Osm17} Osman O., Bekki K., 2017, MNRAS, 471, L87
\bibitem[Owers et al.(2017)]{Owe17} Owers M. S. et al., 2017, MNRAS, 468, 1824
\bibitem[Peng, Maiolino \& Cochrane(2015)]{Pen15} Peng Y., Maiolino R., Cochrane R., 2015, Nature, 521, 192
\bibitem[Pizzella et al.(2004)]{Piz04} Pizzella A., Corsini E. M., Vega Beltr\'{a}n J. C., Bertola F., 2004, A\&A, 424, 447
\bibitem[Putman et al.(2012)]{Put12}  Putman M. E., Peek J. E. G., Joung M. R., 2012, ARA\&A, 50, 491
\bibitem[Robotham et al.(2011)]{robotham11} Robotham A.~S.~G., et al., 2011, MNRAS, 416, 2640
\bibitem[S\'{a}nchez Almeida et al.(2014)]{San14} S\'{a}nchez Almeida J., Elmegreen B. G., Mu\~{n}oz-Tu\~{n}\'{o}n C., Elmegreen D. M., 2014, A\&ARv, 22, 71
\bibitem[S\'{a}nchez-Bl\'{a}zquez et al.(2006)]{San06} S\'{a}nchez-Bl\'{a}zquez P., et al., 2006, MNRAS, 371, 703
\bibitem[S\'{a}nchez et al.(2012)]{San12} S\'{a}nchez S. F., et al., 2012, A\&A, 538, A8
\bibitem[Sancisi et al.(2008)]{San08} Sancisi R., Fraternali F., Oosterloo T., van der Hulst T., 2008, A\&ARv, 15, 189
\bibitem[Sales et al.(2012)]{Sal12} Sales L. V., Navarro J. F., Theuns T., Schaye J., White S. D. M., Frenk C.
S., Crain R. A., Dalla Vecchia C., 2012, MNRAS, 423, 1544 
\bibitem[Sarzi et al.(2006)]{Sar06} Sarzi et al., 2006, MNRAS, 266, 1151
\bibitem[Saunders et al.(2004)]{Sau04} Saunders, W., et al. 2004, in A. F. M. Moorwood, M. Iye, eds, SPIE Conf. Ser. Vol. 5492, Ground-based Instrumentation for Astronomy,  p389Ð400
\bibitem[Scannapieco et al.(2009)]{Sca09} Scannapieco C., White S. D. M., Springel V., Tissera P. B., 2009, MNRAS, 396, 696
\bibitem[Schulz \& Struck(2001)]{Sch01} Schulz S., Struck C., 2001, MNRAS, 328, 185
\bibitem[Schweizer, Whitmore \& Rubin(1983)]{Sch83} Schweizer F., Whitmore B.C., Rubin V.C., 1983, AJ, 88, 909
\bibitem[Serra et al.(2012)]{Ser12} Serra P., et al., 2012, MNRAS, 422, 1835
\bibitem[Serra et al.(2014)]{Ser14} Serra P., et al., 2014, MNRAS, 444, 3388
\bibitem[S\'{e}rsic(1967)]{Ser67} S\'{e}rsic J.L., 1967,  Z. Astrophys., 67, 306
\bibitem[Shapiro \& Field(1976)]{Sha76} Shapiro P. R., Field G. B., 1976, ApJ, 205, 762
\bibitem[Sharp \& Birchall(2010)]{Sha10} Sharp R., Birchall M. N., 2010, PASA, 27, 91
\bibitem[Sharp et al.(2015)]{Sha15} Sharp R., et al. 2015, MNRAS, 446, 1551
\bibitem[Sharp et al.(2006)]{Sha06} Sharp R., et al., 2006, in McLean I.S., Iye W., eds, SPIE Conf. Ser. Vol. 6269, id. 62690G, in Ground based and Airbourne Instrumentation for Astronomy
\bibitem[Shen et al.(2009)]{She09} Shen S., Wang C., Chang R., Shao Z., Hou J., Shu C., 2009, ApJ, 705, 1496
\bibitem[Smith et al.(2004)]{Smi04} Smith, G. A., Saunders, W., Bridges, T., et al. 2004, in A. F. M. Moorwood, M. Iye, eds, SPIE Conf. Ser. Vol. 5492, Ground-based Instrumentation for Astronomy,  p410Ð420
\bibitem[Sparke(1996)]{Spa96} Sparke L. S., 1996, ApJ, 473, 810
\bibitem[Stevens, Croton \& Mutch(2016)]{Ste16} Stevens A. R. H., Croton D. J., Mutch S. J., 2016, MNRAS, 461, 859
\bibitem[Tacconi et al.(2010)]{Tac10} Tacconi L. J., et al., 2010, Nature, 463, 781
\bibitem[Taylor et al.(2011)]{Tay11} Taylor E. N. et al., 2011, MNRAS, 418, 1587
\bibitem[Thakar \& Ryden(1996)]{Tha96} Thakar A. R., Ryden B. S., 1996, ApJ, 461, 55
\bibitem[Thakar \& Ryden(1998)]{Tha98} Thakar A. R., Ryden B. S., 1998, ApJ, 506, 93
\bibitem[Tohline, Simonson \& Caldwell(1982)]{Toh82} Tohline J. E., Simonson G. F., Caldwell N., 1982, ApJ, 252, 92
\bibitem[van de Sande et al.(2017)]{vdS17} van de Sande J., et al., 2017, ApJ, 835, 104
\bibitem[van de Voort et al.(2015)]{vdV15} van de Voort F., Davis T. A., Kere\v s D., Quataert E., Faucher-Gigu\`{e}re C., Hopkins P. F., 2015, MNRAS, 451, 3269
\bibitem[van Dokkum et al.(1999)]{vDok99} van Dokkum P. G., Franx M., Fabricant D., Kelson D. D., Illingworth G. D., 1999, ApJ, 520, 95
\bibitem[Vollmer et al.(2001)]{Vol01} Vollmer B., Cayatte V., Balkowski C., Duschl W. J., 2001, ApJ, 561, 708
\bibitem[Walcher et al.(2014)]{Wal14} Walcher C. J. et al., 2014, A\&A, 569, 1
\bibitem[Welker et al.(2015)]{Wel15} Welker C., Dubois Y., Pichon C., Devriendt J., Chisari E. N., 2015, eprint arXiv:1512.00400.
\bibitem[Welker et al.(2017)]{Wel17} Welker C., Dubois Y., Devriendt J., Pichon C.,  Kaviraj S., Peirani S., 2017, MNRAS, 465, 1241
\bibitem[Wild et al.(2014)]{wil14} Wild, V., et al. 2014, A\&A, 567, A132
\bibitem[Young et al.(2000)]{You14} Young et al., 2014, MNRAS, 444, 3408

\end{thebibliography}
\end{document}